\documentclass[traditabstract,longauth]{aa} 
\usepackage{graphicx}
\usepackage{txfonts}
\usepackage{natbib}
\usepackage{longtable}

\newcommand{\HI}{{\rm H~{\sc i}}} 
\newcommand{\CIV}{{\rm C~{\sc iv}}} 
\newcommand{\CIII}{{\rm C~{\sc iii]}}} 
\newcommand{\AlIII}{{\rm Al~{\sc iii}}}
\newcommand{\SiIV}{{\rm Si~{\sc iv}}} 
\newcommand{\MgII}{{\rm Mg~{\sc ii}}}

\newcommand{\lya}{Lyman-$\alpha$}

\def\Sec#1{Section~\ref{s:#1}}

\def\Eq#1{Eq.~\ref{eq:#1}}
\def\Fig#1{Fig.~\ref{fig:#1}}
\def\Tab#1{Table~\ref{t:#1}}

\defcitealias{hewett2010}{HW10}

\begin{document}
   \title{The Sloan Digital Sky Survey quasar catalog: ninth data release}

   \author{Isabelle P\^aris
          \inst{1,2}
          \and
          Patrick Petitjean
          \inst{1}
          \and
          \'Eric Aubourg
          \inst{3}
          \and
          Stephen Bailey
          \inst{4}
          \and
          Nicholas P. Ross
          \inst{4}
          \and
          Adam D. Myers
          \inst{5,6}
          \and
          Michael A. Strauss
          \inst{7}
          \and
          Scott F. Anderson
          \inst{8}
          \and
          Eduard Arnau
          \inst{9}
          \and
          Julian Bautista
          \inst{3}
          \and
          Dmitry Bizyaev
          \inst{10} 
          \and
          Adam S. Bolton
          \inst{11}
		  \and
		  Jo Bovy\thanks{Hubble fellow}
		  \inst{12}
          \and
          William N. Brandt
          \inst{13,14}
		  \and
		  Howard Brewington
		  \inst{10} 
		  \and
		  Joel R. Browstein
		  \inst{11}
		  \and
		  Nicolas Busca
		  \inst{3}
		  \and
		  Daniel Capellupo
		  \inst{15,16}
		  \and
		  William Carithers
		  \inst{4}
		  \and
		  Rupert A.C. Croft
		  \inst{17}
		  \and
		  Kyle Dawson
		  \inst{11}
		  \and
		  Timoth\'ee Delubac
		  \inst{18}          
          \and
          Garrett Ebelke
          \inst{10} 
		  \and
		  Daniel J. Eisenstein
		  \inst{19}
          \and
          Philip Engelke
          \inst{20}
          \and
          Xiaohui Fan
          \inst{21}
          \and
          Nur Filiz Ak
          \inst{13,14,22}
          \and
          Hayley Finley
          \inst{1}
          \and
          Andreu Font-Ribera
          \inst{4,23}
          \and
          Jian Ge
          \inst{15}
          \and
          Robert R. Gibson
          \inst{8}
          \and
          Patrick B. Hall
          \inst{24}
         \and
          Fred Hamann
          \inst{15}
          \and
          Joseph F. Hennawi
          \inst{6}
          \and
          Shirley Ho
          \inst{17}
          \and
          David W. Hogg
          \inst{25}
          \and
          \v{Z}eljko Ivezi\'{c}
          \inst{8}
          \and
		  Linhua Jiang
		  \inst{21}
		  \and
		  Amy E. Kimball
		  \inst{8,26}
		  \and
		  David Kirkby
		  \inst{27}      
          \and
          Jessica A. Kirkpatrick
          \inst{4}    
         \and
         Khee-Gan Lee
         \inst{6,28}
         \and
         Jean-Marc Le Goff
         \inst{18}
         \and
          Britt Lundgren
          \inst{20}
          \and
          Chelsea L. MacLeod
          \inst{9}
          \and
          Elena  Malanushenko
          \inst{10} 
          \and
          Viktor Malanushenko
          \inst{10} 
          \and
          Claudia Maraston
          \inst{29}
          \and
          Ian D. McGreer
          \inst{21}
          \and
          Richard G. McMahon
          \inst{30}
          \and
          Jordi Miralda-Escud\'e
          \inst{9,31}
          \and
          Demitri Muna
          \inst{32}
          \and
          Pasquier Noterdaeme
          \inst{1}
	      \and
	      Daniel Oravetz
	      \inst{10} 
	      \and   
          Nathalie Palanque-Delabrouille
          \inst{18}
          \and
          Kaike Pan
          \inst{10} 
          \and
          Isma\"el Perez-Fournon
          \inst{33,34}
          \and
          Matthew M. Pieri
          \inst{29}
          \and
          Gordon T. Richards
          \inst{35}
          \and
          Emmanuel Rollinde
         \inst{1}
          \and
          Erin S. Sheldon
          \inst{36}
          \and
          David J. Schlegel
          \inst{4}
          \and
          Donald P. Schneider 
          \inst{13,14}
          \and
          Anze Slosar
          \inst{36}
          \and
          Alaina Shelden
          \inst{10} 
          \and
          Yue Shen
          \inst{19}
          \and
          Audrey Simmons
          \inst{10} 
          \and
          Stephanie Snedden
          \inst{10} 
          \and
          Nao Suzuki
          \inst{4,37}
          \and
          Jeremy Tinker
          \inst{32}
          \and
          Matteo Viel 
         \inst{38,39}
          \and
          Benjamin A. Weaver
          \inst{32}
          \and
          David H. Weinberg
          \inst{40}
          \and
          Martin White
          \inst{4}
          \and
          W. Michael Wood-Vasey
          \inst{41}
          \and
          Christophe Y\`eche
          \inst{18}     
 }

   \institute{UPMC-CNRS, UMR7095, Institut d'Astrophysique de Paris, F-75014, Paris, France,  \email{paris@iap.fr}
         \and
            Departamento de Astronom\'ia, Universidad de Chile, Casilla 36-D, Santiago, Chile
         \and
         	APC, Astroparticule et Cosmologie, Uniiversit\'e Paris Diderot, CNRS/IN2P3, CEA/Irfu, Observatoire de Paris, Sorbonne Paris Cit\'e, 10, rue Alice Domon \& L\'eonie Duquet, 75205 Paris Cedex 13, France
         \and
        	Lawrence Berkeley National Lab, 1 Cyclotron Rd, Berkeley CA, 94720, USA
         \and
            Department of Physics and Astronomy, University of Wyoming, Laramie, WY 82071, USA 
         \and
            Max-Planck-Institut f\"ur Astronomie, K\"onigstuhl 17, D-69117 Heidelberg, Germany
         \and
            Princeton University Observatory, Peyton Hall, Princeton, NJ 08544, USA
         \and
            University of Washington, Dept. of Astronomy, Box 351580, Seattle, WA 98195, USA
		 \and
	     	Institut de Ci\`encies del Cosmos (IEEC/UB), Barcelona, Catalonia
		 \and
		 	Apache Point Observatory, P.O. Box 59, Sunspot, NM 88349-0059, USA
         \and
            Department of Physics and Astronomy, University of Utah,UT, USA
         \and
            Institute for Advanced Study, Einstein Drive, Princeton, NJ 08540, USA
         \and
            Department of Astronomy and Astrophysics, The Pennsylvania State University, University Park, PA 16802, USA
		 \and
		 	Institute for Gravitation and the Cosmos, The Pennsylvania State University, University Park, PA 16802, USA
         \and
            Department of Astronomy, University of Florida, Gainesville, FL 32611-2055, USA
		 \and
		 	School of Physics and Astronomy, Tel Aviv University, Tel Aviv 69978, Israel
		 \and
		 	Carnegie Mellon University, Physics Department, 5000 Forbes Ave, Pittsburgh, PA 15213, USA
         \and
            CEA, Centre de Saclay, Irfu/SPP, 91191, Gif-sur-Yvette, France
         \and
            Harvard-Smithsonian Center for Astrophysics, 60 Garden St., MS\#20, Cambridge, MA 02138 USA 
	     \and
            Department of Astronomy, Yale University, New Haven, CT06511, USA
         \and
            Steward Observatory, University of Arizona, 933 North Cherry Avenue, Tucson, AZ 85721
         \and
          	Faculty of Sciences, Department of Astronomy and Space Sciences, Erciyes University, 38039, Kayseri, Turkey
		 \and
		 	Institute of Theoretical Physics, University of Zurich, 8057 Zurich, Switzerland
    	 \and
        	Department of Physics and Astronomy, York University, Toronto, ON M3J1P3, Canada 
		 \and
		 	Center for Cosmology and Particle Physics, Department of Physics, New York University, 4 Washington Place, New York, NY 10003, USA
		 \and
			National Radio Astronomy Observatory, 520 Edgemont Rd., Charlottesville, VA, 22903, USA
		 \and
		 	Department of Physics and Astronomy, UC Irvine, 4129 Frederick Reines Hall, Irvine, CA 92697-4575, USA
	     \and
	     	Department of Astrophysical sciences, Princeton university, Princeton 08544, USA
		 \and
		 	Institute of Cosmology and Gravitation, University of Portsmouth, Dennis Sciama building, Portsmouth P01 3FX, UK
		 \and
            Institute of Astronomy, University of Cambridge, Madingley Road, Cambridge CB3 0HA, UK
		 \and
		 	Instituci\'o Catalana de Recerca i Estudis Avan\c cats, Catalonia
         \and
         	Center for Cosmology and Particle Physics, Department of Physics, New York University, New York, NY 10003 USA
         \and
            Instituto de Astrofísica de Canarias (IAC), E-38200 La Laguna, Tenerife, Spain
		 \and
		 	Departamento de Astrofisica, Universidad de La Laguna (ULL), E-38205 La Laguna, Tenerife, Spain
         \and
            Department of Physics, Drexel University, Philadelphia, PA 19104, USA
         \and
            5 Brookhaven National Laboratory, Blgd 510, Upton, NY 11375, USA
	     \and
	     	Department of Physics, University of California Berkeley, Berkeley, CA 94720, USA
    	 \and
INAF - Osservatorio Astronomico di Trieste, Via G.B. Tiepolo 11 
         \and
INFN/National Institute for Nuclear Physics, Via Valerio 2, I-34127 Trieste, Italy
         \and
         Astronomy Department and Center for Cosmology and AstroParticle Physics, Ohio State University, 140 West 18th Avenue, Columbus, OH 43210, USA
		 \and
		 PITT PACC, Department of Physics and Astronomy, University of Pittsburgh, Pittsburgh, PA 15260, USA
         }

   \date{Received xxx; accepted xxx}

  \abstract{We present the Data Release 9 Quasar (DR9Q) catalog from the Baryon Oscillation Spectroscopic Survey (BOSS) 
of the Sloan Digital Sky Survey III.
The catalog includes all BOSS objects that were targeted as quasar 
candidates  during the survey, are spectrocopically confirmed as quasars via visual inspection,
 have luminosities $M_{\rm i}$[z=2]~$<$~$-$20.5 (in a $\Lambda$CDM cosmology with $H_0$ = 70 km~s$^{-1}$~Mpc$^{-1}$, 
$\Omega_{\rm M}$ = 0.3, and $\Omega_{\Lambda}$ = 0.7) and either display at least one emission line with 
full width at half maximum (FWHM) larger than 500~km~s$^{-1}$
or, if not, have interesting/complex absorption features. It includes as well, known quasars (mostly from SDSS-I and II) 
that were reobserved by BOSS.
This catalog contains 87,822 quasars (78,086 are new discoveries) detected over 3,275~deg$^{2}$ with robust identification and redshift 
measured by a combination of principal component eigenspectra newly derived from a training set of 8,632 spectra from SDSS-DR7.
The number of quasars with $z>2.15$ (61,931) is $\sim$2.8 times larger than the number of $z>2.15$ quasars previously known.
Redshifts and FWHMs are provided for the strongest emission lines (C~{\sc iv}, C~{\sc iii}], Mg~{\sc ii}). 
The catalog  identifies 7,533  broad absorption line quasars  and gives their characteristics.
For each object the catalog presents 
five-band (\textit{u}, \textit{g}, \textit{r}, \textit{i}, \textit{z}) CCD-based photometry with typical accuracy of 0.03 mag, 
and information on the morphology and selection method. 
The catalog also contains X-ray, ultraviolet, near-infrared, and radio emission properties of the quasars, 
when available, from other large-area surveys. 
The calibrated digital spectra cover the wavelength region 3,600-10,500~\AA\ at 
a spectral resolution in the range 1,300~$<$~$R$~$<$~2,500; the spectra can be retrieved from the SDSS Catalog Archive Server.
We also provide a supplemental list of an additional 949 quasars that have been identified,
among  galaxy targets of the BOSS or among quasar targets after DR9 was frozen. 
}

   \maketitle
 
%
\section{Introduction}
\label{s:Introduction}
Since their discovery \citep{schmidt1963}, interest in quasars has 
grown steadily, both because of their peculiar properties and because of their importance 
for cosmology and galaxy evolution.  
Many catalogs have gathered together increasing numbers
of quasars either from heterogeneous samples
\citep[see][ and references therein]{hewitt1993,veron2006} 
or from large surveys, most importantly: the Large Bright
Quasar Survey \citep[LBQS,][]{morris1991,hewett1995}; the 2dF Quasar
Redshift Survey \citep[2QZ;][]{boyle2000,croom2001} and
the successive releases of the
Sloan Digital Sky Survey \citep[SDSS,][]{york2000} Quasar
Catalogs \citep[e.g.,][ for DR7]{schneider2010}.  

This paper describes the first quasar catalog of the Baryon Oscillation Spectroscopic Survey \citep[BOSS,][]{schlegel2007,dawson2012}.
BOSS is the main dark time legacy survey of the third 
stage of the Sloan Digital Sky Survey \citep[SDSS-III,][]{eisenstein2011}. It is based on the 
ninth data release of the SDSS \citep{DR9}.
BOSS is a five-year program to obtain spectra of 1.5 million of galaxies and over 150,000 $z>2.15$ quasars.
The main goal of the survey is to detect the characteristic scale imprinted by baryon acoustic oscillations 
(BAO) in the early universe from the spatial distribution of both luminous red galaxies at $z\sim 0.7$ and
H~{\sc i} absorption lines in the intergalactic medium (IGM) at $z\sim 2.5$. 
BOSS uses the same imaging data as in SDSS-I and II, with an extension in the South Galactic Cap (SGC).

The  BAO clustering measurements in the IGM require a quasar catalog of maximal purity and accurate redshifts.
Indeed the spectra of any non-quasar object, especially at high signal-to-noise ratio, will dilute
the signal and/or increase the noise in the clustering measurement. The automated processing of the spectra
\citep{bolton2012} is sophisticated, but is not perfect. 
The identification of the objects and their redshifts have therefore to be certified before any analysis is performed. 
The present catalog, henceforth denoted DR9Q catalog, contains 87,822 quasars identified among the objects targeted as quasar
candidates over an area of 3,275 ${\rm deg^2}$ surveyed during the first two years of BOSS operations.
We also give a supplemental list of quasars identified among galaxy targets.
This catalog keeps the tradition of producing quasar catalogs \citep{schneider2002,schneider2003,schneider2005,schneider2007,schneider2010} 
from SDSS-I and II \citep{york2000}. The final version of the SDSS-II quasar catalog \citep{schneider2010} based on the 
seventh SDSS data release \citep{DR7} contains 105,783 objects 
mostly at $z<2$ \citep[see ][ for their properties]{shen2011}.
Note that the DR9Q catalog does not contain all DR7 quasars but only those DR7 quasars that 
were reobserved during the two first years of BOSS\footnote{All known $z>2.15$ quasars in BOSS footprint are
being reobserved to obtain spectra of uniformly high SNR in the Lyman-$\alpha$ forest and to enable variability studies}. 
High redshift ($z>2$) quasar continua together with pixel masks, improved noise estimates,  and other products designed to aid in the 
BAO-Lyman-$\alpha$ clustering analysis will be released in \cite{lee2012}.

The selection of candidates and observations are summarized in \Sec{survey}.
We describe the visual inspection of all targets in \Sec{Construction_Catalog},
present accurate redshifts for the quasars in \Sec{AutoZ} and
describe the detection and measurement of broad absorption lines (BALs) in \Sec{BAL}. 
The catalog is described in \Sec{Catalog_description}. We give a  catalog summary in \Sec{DR9summary} and
comment on the supplemental lists of quasars in \Sec{VAC6}.
We conclude in \Sec{Summary}.

In the following we will use a $\Lambda$CDM cosmology with $H_0$ = 70 ${\rm km \  s^{-1} \ Mpc^{-1}}$, 
$\Omega_{\rm M}$ = 0.3, and $\Omega_{\Lambda}$ = 0.7 \citep{Spergel2003}.

Most of the objects in the catalog show at least an emission line with FWHM~$>$~500 ${\rm km \ s^{-1}}$
in their spectra. However, there are a few exceptions: a few objects
have emission lines with smaller FWHM due to noise or dust obscuration (Type II quasars)
others have very weak emission 
lines but are identified as quasars because of the presence of the Lyman-$\alpha$ forest
\citep{Diamond-Stanic2009}.
We will call a quasar an object with a luminosity
$M_{\rm i}$[z=2]~$<$~$-$20.5 and either displaying at least one emission line 
with FWHM greater than 500~km~s$^{-1}$ or, if not, having interesting/complex absorption features.
This definition is slightly different from the one used in SDSS-DR7. The change in absolute
magnitude is to include a few low-$z$ objects in the catalog. Because BOSS is targeting $z>2.15$ quasars,
the median absolute luminosity is higher in BOSS than in SDSS-DR7. All BOSS
objects with $z>2$ qualify for the SDSS-DR7 definition: FWHM~$>$~1000~km~s$^{-1}$ and $M_{\rm i}$[z=0]~$<$~$-$22
(adopting the same cosmology and $\alpha_{\nu}$~=~$-0.5$).
 In the following, all magnitudes will be PSF magnitudes.
 
%
%
\section{Survey outline}
\label{s:survey}

%
In order to measure the BAO scale in the Lyman-$\alpha$ forest 
at $z \sim 2.5$, BOSS aims to obtain spectra of over 150,000 quasars in the redshift range 
$2.15 \leq z \leq 3.5$, where at least part of the \lya\ forest lies in the BOSS spectral range.
The measurement of clustering in the IGM is independent of the properties of 
background quasars. 
Therefore the quasar sample does not need to be uniform and a variety of selection methods are used 
to increase the surface density of high redshift quasars \citep{ross2012}. 
Some quasars with $z<2$ will be targeted in the course of
specific ancillary science programs or as a consequence of imperfect high-redshift quasar selection.

To detect the BAO signal, a surface density of 15 quasars with $z \geq 2.15$ per square degree is required \citep{mcdonald2007}. 
For comparison, SDSS-I/II targeted about $\sim14,000$ $z\geq 2.15$ quasars over the full survey,
e.g. $\sim$8,400~deg$^2$ \citep{schneider2010}, leading to a 
surface density of $\sim$2 quasars per square degree in the redshift range of interest for BOSS.
To reach the BAO quasar density requirement implies targeting to fainter magnitudes than 
SDSS-I/II. The BOSS limiting magnitude for target selection is $r \leq 21.85$ or $g \leq 22$ 
\citep{ross2012}, while $z \geq 3$ quasars were selected  
to be brighter than $i \sim 20.2$ in SDSS-I/II \citep{richards2002}.

	\subsection{Imaging data}
BOSS uses the same imaging data as that of the original SDSS-I/II survey, with an extension in the SGC. 
These data were gathered using a dedicated 2.5 m wide-field telescope \citep{gunn2006}
to collect light for a camera with 30 2k$\times$2k CCDs \citep{gunn1998} over five broad bands - \textit{ugriz} \citep{fukugita1996}; 
this camera has imaged 14,555 unique square degrees of the sky, including $\sim$7,500 deg$^2$ in the NGC and 
$\sim$3,100 deg$^2$ in the SGC \citep{DR8}. The imaging data were taken on dark photometric nights
of good seeing \citep{hogg2001}. Objects were detected and their properties were measured \citep{lupton2001,stoughton2002}
 and calibrated photometrically 
\citep{smith2002,ivezic2004,tucker2006,padmanabhan2008}, and astrometrically \citep{pier2003}.
%

%
	\subsection{Target selection} 
The target selection of quasar candidates is crucial for the goals of the quasar BOSS survey. 
On average 40 fibers per square degree are allocated by the survey to the quasar project. 
The surface density of $z \geq 2.15$ quasars to the BOSS magnitude limit is approximately 28 per deg$^2$ \citep[see][]{palanque2012}. 
Thus, recovering these quasars from 40 targets per square degree in single-epoch
SDSS imaging is challenging because photometric errors are significant at this depth and because the quasar locus
(in \textit{ugriz}) crosses the stellar locus at $z \sim 2.7$ \citep{fan1999,richards2002,ross2012}. 
All objects classified as point-sources in the imaging data and brighter than 
either $r=21.85$ or $g=22$ (or both, magnitudes dereddened for Galactic extinction) are  passed through the various 
quasar target selection algorithms. The quasar target selection for the first two years of BOSS operation is fully described 
in \cite{ross2012}. We briefly summarize here the key steps.

The target selection algorithm is designed to maximize the number of quasars useful for the \lya\ forest analyses and reach 
the requirement of 15  deg$^{-2}$ quasars with $z \geq 2.15$. 
Several target selection methods are therefore combined
and data in other wavelength bands are used when available.
At the same time, in order to use the quasars themselves for statistical studies, such as the quasar luminosity function or clustering analyses \citep[e.g.][]{white2012}, 
part of the sample must be uniformly selected.
Thus, the BOSS quasar target selection is split in two parts:
\begin{itemize}
	\item About half of the targets are selected as part of the so-called ``CORE" sample using a single uniform target selection algorithm.
	The likelihood method \citep{kirkpatrick2011} was adopted for the CORE selection during the first year of observations. 
	Starting with the second year of operation, it was replaced by the extreme deconvolution method \citep[XDQSO;][]{bovy2011} 
which better takes photometric errors into account. 
	\item Most of the remaining quasar candidates are selected as part of the so-called ``BONUS" sample through a combination of
 four methods: the Non-Parametric Bayesian Classification and Kernel Density Estimator \cite[KDE;][]{richards2004,richards2009}, 
the likelihood method \citep{kirkpatrick2011}, a neural network \citep{yeche2010} and the XDQSO method
\citep[][objects for lower likelihood 
than in the CORE sample, over a slightly expanded redshift range, and incorporating data from 
UKIDSS; \cite{lawrence2007}; from GALEX; \cite{martin2005}; and, where available, from coadded 
imaging in overlapping SDSS runs]{bovy2011,Bovy2012}. The outputs of all of these BONUS methods are combined using a neural network
\end{itemize}

Point-sources that match FIRST \citep{becker1995} and that are not blue in $u - g$ (which would be characteristic of $z < 2$ quasars)
are also always included.  
In addition, previously known $z > 2.15$ quasars (mostly from SDSS I/II) were also re-targeted for several reasons: (i) the BOSS wavelength
range is more extended than in previous surveys; (ii)  BOSS spectra have usually higher signal-to-noise ratio (SNR) than SDSS spectra \citep{DR9};
(iii) the two epoch data will allow spectral variability studies. 
This sample is selected using the SDSS-DR7 quasar catalog \citep{schneider2010}, 
the 2dF QSO Redshift Survey \citep[2QZ;][]{Croom2004}, the 2dF-SDSS LRG and QSO Survey \citep[2SLAQ;][]{Croom2009}, 
the AAT-UKIDSS-SDSS (AUS) survey, and the MMT-BOSS pilot survey \citep[Appendix C in][]{ross2012}. 
Quasars observed at high spectral resolution by UVES-VLT and HIRES-Keck were also included in the sample.
Finally, BOSS includes targeting of a number of ancillary programs, some designed specifically to target quasars 
\citep[e.g., the variability programs;][]{palanque2011, macleod2012}. 
The corresponding programs include:\par\noindent
$\bullet$ Reddened Quasars: Quasar candidates with high intrinsic reddening.\par\noindent
$\bullet$ No Quasar Left Behind: Bright variable quasars on Stripe 82.\par\noindent
$\bullet$ Variability-Selected Quasars: Variable quasars on Stripe82, focused on $z>2.15$.\par\noindent
$\bullet$ K-band Limited Sample of Quasars: Quasars selected from SDSS and UKIDSS K photometry.\par\noindent
$\bullet$ High-Energy Blazars and Optical Counterpars of Gamma-Ray Sources: Fermi sources, plus blazar candidates from radio and X-ray.\par\noindent
$\bullet$ Remarkable X-ray Source Populations: XMM-Newton and Chandra sources with optical counterparts.\par\noindent
$\bullet$ BAL Quasar Variability Survey: Known BALs from SDSS-I/II.\par\noindent
$\bullet$ Variable Quasar Absorption: Known Narrow-line absorption quasars from SDSS-I/II.\par\noindent
$\bullet$ Double-Lobed Radio Quasars: Point sources lying between pairs of FIRST radio sources.\par\noindent
$\bullet$ High-Redshift Quasars: Candidates at $z>3.5$ in overlap between scanlines.\par\noindent
$\bullet$ High-Redshift Quasars from SDSS and UKIDSS: Candidates at $z>5.5$ from SDSS and UKIDSS photometry.\par\noindent
$\bullet$ Previously Known Quasars with $1.8 < z < 2.15$: Reobserved to constrain metal absorption in the Ly$\alpha$ forest.\par\noindent
$\bullet$ Variable Quasars: selected from repeat observations in overlaps of SDSS imaging runs.\par\noindent
These programs are described in detail in the Appendix and Tables~6 and 7 of \cite{dawson2012}.

%
\subsection{Spectroscopy}

Because BOSS was designed to observe targets two magnitudes fainter than the original SDSS spectroscopic 
targets, substantial upgrades to the SDSS spectrographs were required and prepared during the first year of SDSS-III
\citep{Smee2012}. 
New CCDs were installed in both red and blue arms, with much higher quantum efficiencies both at the reddest and bluest 
wavelengths. These are larger format CCDs with smaller pixels, that match the upgrade of the fiber system from 640 fibers with 3 arcsec optical
diameter to 1,000 fibers (500 per spectrograph) with 2 arcsec diameter. The larger number of fibers alone improves
survey efficiency by 50\%, and because BOSS observes point sources (quasar targets) and distant galaxies in the
sky-dominated regime the smaller fibers yield somewhat higher SNR spectra in typical APO seeing,
though they place stiffer demands on guiding accuracy and differential refraction. The original diffraction gratings 
were replaced with higher throughput, volume-phase holographic (VPH) transmission gratings, and other 
optical elements were also replaced or recoated to improve throughput. The spectral resolution varies 
from  $\sim$1,300 at 3,600~\AA~ to 2,500 at 10,000~\AA~
The instrument is described in detail in \cite{Smee2012} and the BOSS survey is explained 
in \cite{dawson2012}.

\begin{figure}[htbp]
	\centering{\includegraphics[width=45mm,angle=-90]{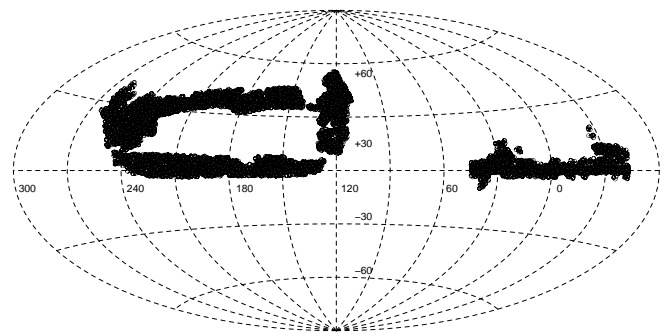}}
	\caption{
	The space distribution in equatorial coordinates of the SDSS-III DR9 data release quasars.	 
	}
	\label{fig:SkyCoverage}
\end{figure}

\begin{figure}[htbp]
	\centering{\includegraphics[width=75mm]{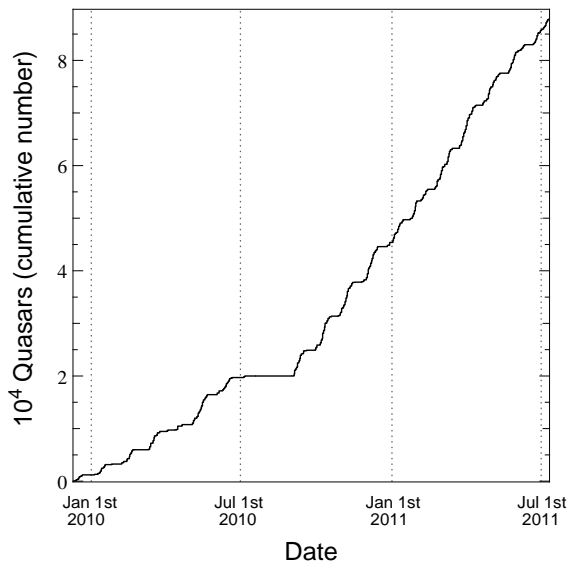}}
	\caption{ 
	Cumulative number of quasars as a function of observation date during the first two years of the survey.	
Horizontal times are due to the
yearly summer shutdown during monsoon rains (summer 2010 at MJD = 55400) and the monthly bright time when
BOSS does not observe.
	}
	\label{fig:ProgressPlot}
\end{figure}

BOSS spectroscopic observations are taken in a series of at least three 15-minute exposures.
Additional exposures are taken until the squared signal-to-noise ratio per pixel, (SNR)$^2$, reaches the survey-quality threshold for each CCD.
These thresholds are ${\rm (SNR)^2} \geq 22$ at $i$-band magnitude 21 for the red camera
and ${\rm (SNR)^2} \geq 10$ at $g$-band magnitude 22 for the blue camera (extinction corrected magnitudes). 
Recall that the pixels are co-added, linear in log~$\lambda$ with sampling from 0.82 to 2.39~\AA\ 
over the wavelength range from 3,610 to 10,140~\AA. 
The current spectroscopic reduction pipeline for BOSS spectra is described  
in \cite{bolton2012}.
SDSS-III uses plates with 1000 spectra each, more than one plate can cover a tile
\citep{dawson2012}. 819 plates were observed between December 2009 and July 2011.
Some have been observed multiple times. In total, 87,822 unique quasars have been spectroscopically confirmed
based on our visual inspection. \Fig{SkyCoverage} shows the observed area in the sky.
The total area covered by the SDSS-DR9 is 3,275 ${\rm deg^2}$.
\Fig{ProgressPlot} displays the cumulative number of quasars as a function of the observation date.

%
%
%

\begin{figure}[htbp]
	\centering{\includegraphics[width=.8\linewidth]{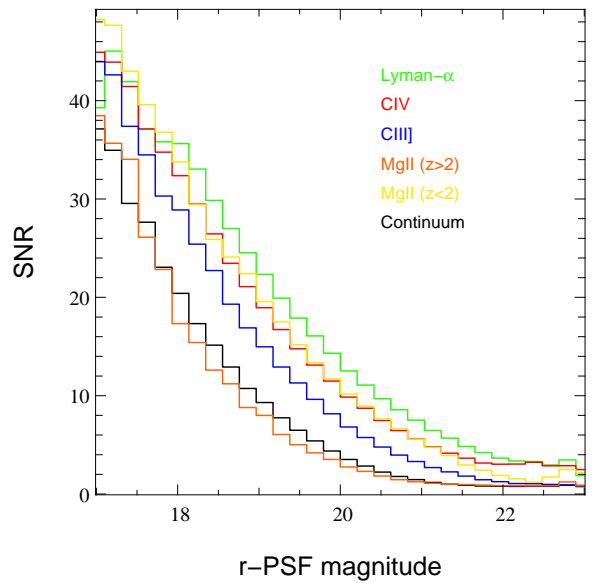}}
	\caption{
		Median observed SNR per pixel at the top of the \lya\ (green), \CIV\ (red), \CIII\ complex (blue), \MgII\ 
at $z > 2$ (orange) and \MgII\ at $z < 2$ (yellow) emission lines and in emission-line free regions (black) versus
$r$-PSF magnitude (corrected for Galactic extinction).
Two redshift ranges are considered for \MgII\ because the emission line is redshifted in regions
of the spectra with very different characteristics.
		At $r \sim 22$, the median SNR per pixel at the top of the \lya\ and \CIV\ emission lines is about 4;
sufficient to identify most of the quasars.
Outside of the emission-line regions, at the same magnitude, the SNR per pixel
is about unity. 
	}
	\label{fig:SNR_emission}
\end{figure}

As $z>2$ quasars are usually identified by the presence of strong Lyman-$\alpha$ and \CIV\ emission
lines, we determine the SNR effectively achieved at the position of these lines. 
The median SNR per pixel at the position of various emission lines  (\lya , \CIV , \CIII\ complex and \MgII )
and in the continuum are shown in \Fig{SNR_emission}. 
While the SNR per pixel in regions free of emission lines (black histogram) drops to be equal to $\sim$ 1 at 
$r \sim 22$, the SNR at the top of the \lya\ (green histogram) and \CIV\ (red histogram) emission 
lines stays above about 4, allowing the identification of a fair fraction
of these objects at this magnitude.

In order to classify the object, each spectrum is fit by the BOSS pipeline\footnote{The software used is called idlspec2d and is publicly 
available. The current version is v5\_4\_45.
Details can be found at http://www.sdss3.org/dr9/software, 
\cite{bolton2012}}  with a library of star templates,
a PCA decomposition of galaxy spectra and a PCA decomposition of quasar spectra.
Each class of templates is fit over a range of redshifts:
galaxies from $z=-0.01$ to $1.00$
quasars from $z=0.0033$ to $7.00$;
and stars from $z=-0.004$ to $0.004$ ($\pm 1200$~{km/s}).
The combination of redshift and template with the overall best fit (in terms of the lowest reduced chi-squared) is adopted 
as the pipeline classification ({\tt CLASS}) and redshift measurement ({\tt Z}~$\pm$~{\tt Z\_ERR}).
A warning bitmask ({\tt ZWARNING}) is set to indicate
poor wavelength coverage, negative star template fits, 
broken/dropped fibers, fibers assigned to mesure sky background,
and fits which are within $\Delta \chi^2/\rm{dof} = 0.01$ of the next
best fit (comparing only fits with a velocity difference of 1,000 ${\rm km \ s^{-1}}$).
A {\tt ZWARNING} equals to zero indicates a robust classification with no pipeline-identified problem \citep{DR8,bolton2012}.

The classifications by the BOSS pipeline are not perfect however and visual inspection is required. 
Most misclassified spectra have low SNR.
At SNR per pixel~$\sim$~2, some objects are fit equally well by a star and a 
quasar template. Even if the object is correctly identified as a quasar, the redshift
can be erroneous, because one line is misidentified; the
most common case is Mg~{\sc ii}$\lambda$2800  is misidentified as Lyman-$\alpha$. 
But this can be also because of a strong absorption feature (e.g. a damped Lyman-$\alpha$ system, DLA,  or a BAL) spoils the 
profile of an emission line and the pipeline is unable to recover it.


\subsection{Calibration warnings}
\subsubsection{Excess flux in the blue}
The BOSS spectra often show excess light at the blue end (a similar problem was
found in SDSS-DR7 spectra; P\^aris et al., 2011).

\begin{figure}[htbp]
	\centering{\includegraphics[width=55mm,angle=-90]{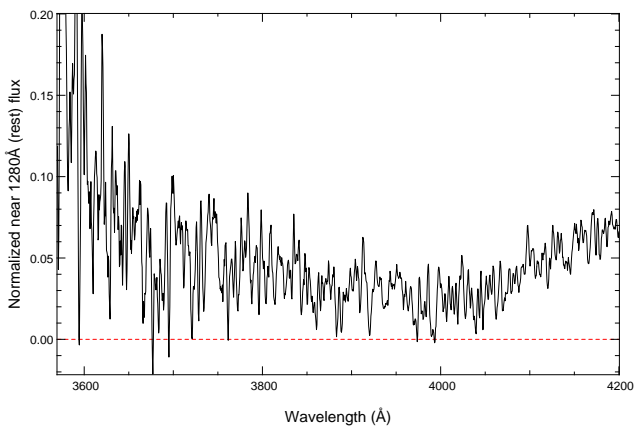}}
	\caption{
Stack of DR9 BOSS spectra where a damped
Lyman-$\alpha$ system is seen at an absorpstion redshift higher than 3.385 with a column
density $N$(H~{\sc i})~$\geq$~10$^{20.5}$~cm$^{-2}$. The spectra are normalized to unity
near 1280~\AA~ in the quasar rest frame. Owing to the presence of
the DLA, the flux is expected to be zero at observed wavelengths
below $\sim$4100~\AA~(e.g. below the Lyman limit of all DLAs). This is not the case in the very blue part
of the spectrum ($\lambda_{\rm obs} \leq 4000 $\AA) where the mean observed flux appears to increase
(spuriously).
}
\label{fig:StackLLS}
\end{figure}

To quantify this problem we selected spectra where a damped Lyman-$\alpha$ system (DLA)  
is observed with aborption redshift greater than 3.385 and with
a column density $N$(H~{\sc i})~$\geq$~10$^{20.5}$~cm$^{-2}$. There are
402 such quasars in the sample. In these spectra, 
and because of the presence of the DLA, the flux is expected to be zero
at $\lambda_{\rm obs}$~$\leq$~4100~\AA~ (e.g. below the Lyman limit of all DLAs). 
When stacking the selected lines of sight
(\Fig{StackLLS}), we note instead that the flux increases for wavelengths
below 4000~\AA. The excess light at
 $\lambda_{\rm obs} \sim$~3600~\AA~ is 10\% of the flux at 
$\lambda_{\rm rest}$~=~1280\,\AA~ where the spectra are normalized.
This problem can affect the analysis of the Lyman-$\alpha$ forest \citep[see e.g.][]{font2012b}
 and is probably a consequence of imperfect sky subtraction \citep{dawson2012}.
It will be corrected in a future version of the pipeline.

\subsubsection{Spectrophotometric calibration}
To maximize the flux in the blue part of the quasar spectra, where the Lyman-$\alpha$ forest lies, it was decided
to offset the position of the quasar target fibers to compensate for 
atmospheric refraction and different focus in the blue \citep{dawson2012}. These offsets were not applied to the standard stars.
The current pipeline flux calibration does not take these fiber offsets into account, 
therefore the spectrophotometry of the main QSO targets (e.g. not the ancillary targets) is biased toward 
bluer colors over the full wavelength range. Spectrophotometry of these objects will preferentially 
exhibit excess flux relative to the SDSS imaging data at $\lambda<  4000$ \AA\ and a flux decrement at longer wavelengths.
Because the fiber offsets are intended to account for atmospheric differential refraction, data will show 
larger offsets in spectrophotometric fluxes relative to imaging photometry for observations performed at higher airmass.
\cite{dawson2012} discuss in details the quality of the BOSS spectrophotometry and reports that stellar contaminants in the quasar sample (i.e. quasar candidates that are actually stars) have $g-r$ colors 0.038 magnitudes bluer than the photometry with an RMS dispersion of 0.158 magnitudes.

\begin{figure}[htbp]
	\centering{\includegraphics[angle=-90,width=\linewidth]{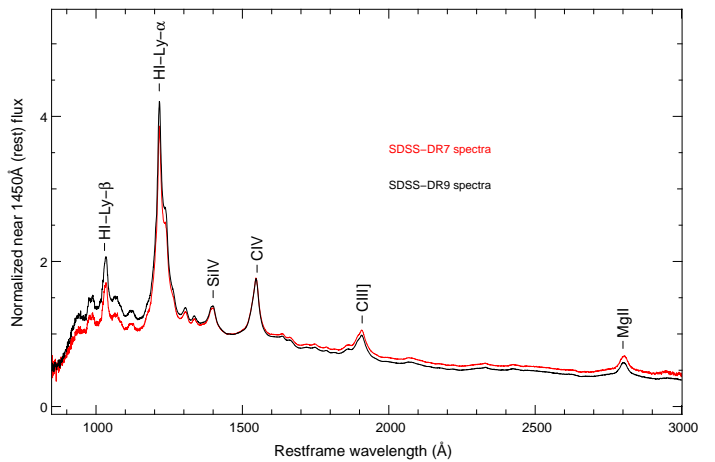}}
	\caption{ 
Composite spectra of 6,459 quasars observed both by SDSS-DR7 and BOSS 
for (i) SDSS-DR7 spectra (red) and, (ii) SDSS-DR9 spectra (black).
The slope of the two composite spectra should be similar
(as any variability should be averaged out). This is not the case because of the difference in focus
of the BOSS quasars and standard stars.
Note that this flux mis-calibration is different from object to object.
}
\label{fig:Composite}
\end{figure}

This problem is illustrated in \Fig{Composite} where the median composite spectra 
of quasars observed by both SDSS-I/II and BOSS are plotted together.
The resulting SDSS-DR7 spectrum is in red and BOSS spectrum in black. 
The BOSS composite spectrum is bluer than the same composite from SDSS-DR7 spectra.
Note that this flux mis-calibration is different from object
to object so that \Fig{Composite} shows only the mean difference between DR7 and DR9 spectra.

\subsubsection{Identified quasars with bad spectra}

During the course of the first two years of BOSS, different versions of
the SDSS spectroscopic pipeline were used after some systematic problems 
had been fixed, thus improving the overall quality of the data.
The visual inspection described below is performed on the fly, 
within a few days after the data are obtained, qualified 
and reduced by the version of the pipeline that is available at the time
the data are obtained. Once an object is positively identified
as a quasar, a galaxy, or a star from visual inspection, it keeps its identification in our catalog
unless an apparent mistake has been committed and is corrected in the course of some 
check performed afterwards by the scanners or by a user of the data.

When a new version of the pipeline is made available, all the data
are re-reduced. We then reinspect objects with uncertain identifications
({\tt QSO\_?}, {\tt QSO\_Z?}, {\tt Star\_?}, see Section~3.2) or spectra that are not qualified ({\tt Bad})
but we do not reinspect the objects with firm identifications.

It can happen that the spectra of a few objects are of lesser quality
with the new version of the pipeline. 
These objects are still in the catalog.

Even the most thorough work of the kind described here cannot be absolutely flawless. 
We encourage the reader to signal any mistake
to the first author of this paper in order to ensure highest quality
of the information provided in the catalog.

%
\section{Construction of the DR9Q catalog}
\label{s:Construction_Catalog}

In order to optimally measure the BAO clustering signal in the IGM, we must have as pure a catalog of
quasars as possible. In this catalog, peculiar features such as broad absorption lines (BAL)
or Damped Lyman-$\alpha$ systems (DLA) that may dilute the signal, should be identified.
We therefore designed quality control of the data based on a visual inspection of 
the spectra of all BOSS objects that might be a quasar.
During commissioning and the first year of the survey this quality control was also very 
useful to report problems with  the pipeline, which helped
improve the overall quality of the data reductions.

The catalog lists all the visually confirmed quasars.
About 10\% of these quasars have been observed several times (Dawson et al. 2012),
either because a particular plate has been re-observed (e.g. to increase the SNR for a particular scientific
project), or because a particular region in the sky has been reobserved at different epochs (e.g.  Stripe 82),
or, because plates overlap. Now, and throughout BOSS, overlapping plates are used as an
opportunity to
increase the SNR on a few objets (e.g. CORE objects).
These repeat observations are often useful to confirm the nature of objects with
low SNR spectra. However we did not attempt to co-add these data
mostly because they are often of quite different SNR.

	\subsection{A tool for the visual inspection}

Immediately after the processing by the BOSS pipeline, the reduced data (spectra and pipeline classification) are copied to the 
IN2P3 (Institut National de Physique Nucl\'eaire et de Physique des Particules) computing center\footnote{CC-IN2P3, http://cc.in2p3.fr}. 
A Java program gathers meta-information and saves it into an Oracle database. 

All spectra are matched to target objects, imaging and photometry information, and SDSS-II spectroscopy. 
They are processed by a Java program that computes basic statistics from the spectra and
fits a power law continuum and individual emission lines to each spectrum. 
The spectra are then made available online through a collaborative web 
application, from which human scanners can flag objects and decide classifications. 

This tool and the visual inspection procedure described in the next Section evolved with 
time during commissioning and the first six months of the survey. The whole procedure was repeated
at the end of the first two years to guarantee the homogeneity of the catalog.

	\subsection{Visual inspection procedure} 
	\label{s:VI}

\begin{figure}[htbp]
	\centering{\includegraphics[width=75mm]{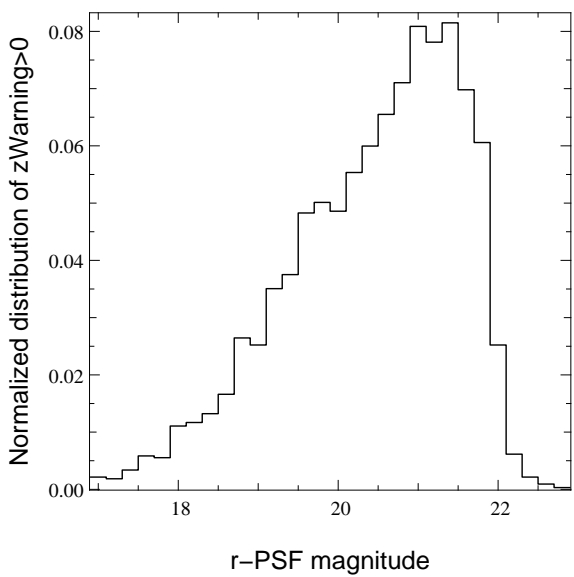}}
	\caption{ 
	Fraction of visually confirmed quasars
with a non-zero {\tt ZWARNING} flag as a function of the $r$-PSF magnitude (after correcting for
Galactic extinction). A positive {\tt ZWARNING} means that the pipeline considers 
its redshift estimate to be unreliable. This fraction increases at faint magnitudes.
	}
	\label{fig:zWarn_fraction}
\end{figure}

The identifications provided by the BOSS pipeline are already very good.
Nevertheless about 12\% of all quasar targets have a non-zero {\tt ZWARNING} flag, i.e. their redshift 
is not considered to be reliable by the pipeline. After visual inspection, 4\% of all confirmed quasars
have a non-zero {\tt ZWARNING} flag.
Not surprisingly, the fraction of these objects increases with magnitude (see \Fig{zWarn_fraction}).
%

%

\begin{figure*}[htbp]
		\centering{\includegraphics[angle=-90,width=\linewidth]{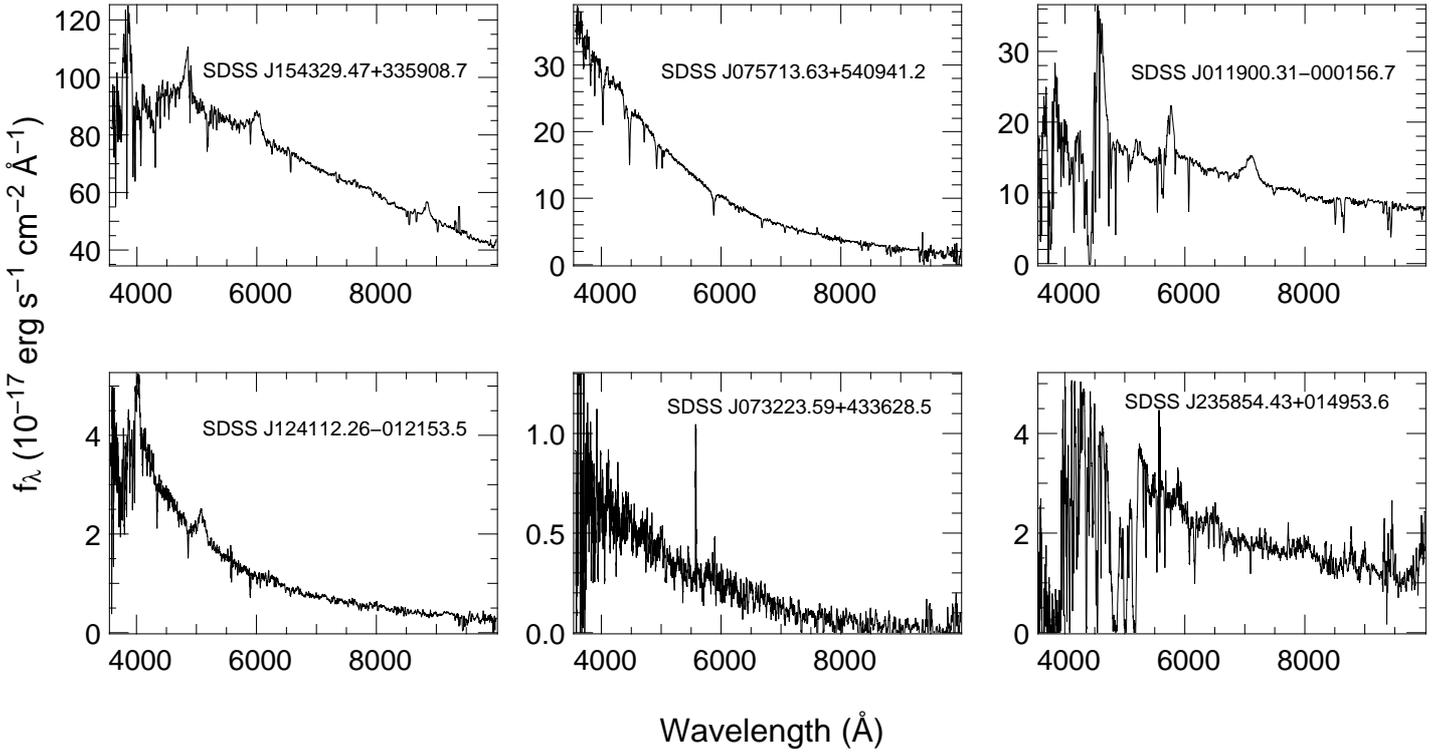}}
	\caption{
	\textit{First column: }
	Examples of $z > 2$ quasars classified as {\tt STAR} by the BOSS pipeline. The overall shape of the
spectrum is similar to the spectrum of F stars.
	\textit{Second column: }
	Examples of stars identified as {\tt QSO} by the BOSS pipeline. Strong absorption lines
or wiggles in the spectrum can mimic quasar features.
	\textit{Third colum: }
	Examples of $z>2$ quasars for which the BOSS pipeline provides an inaccurate redshift estimate 
that must be corrected during the visual inspection. The pipeline is confused by the strong absorption lines.
The spectra were boxcar median smoothed over 5 pixels.
	}
	\label{fig:ExMisID}
\end{figure*}

We visually inspected all quasar candidates and objects from quasar ancillary programs (see Section~2.2)
to (i) secure the identification of the object and, (ii) reliably estimate the systemic redshift of the quasar.
We manually confirmed or modified the identification of the object
and, when needed, corrected the redshift provided by the BOSS pipeline, i.e. when it was wrong 
(when e.g. an emission line is misidentified or a bad feature was considered an emission line)
or inaccurate (when emission lines are correctly identified but not properly centered). 
Examples of misidentified objects or inaccurate redshift estimates are displayed in \Fig{ExMisID}.

%

All the information on the objects  is stored in a database which is updated in real time as 
new data arrives from the telescope. 
Modifications from the visual inspection are stored also in the database.
For each plate, the objects classified by the pipeline as star, QSO with $z<2$, and QSO with $z \geq 2$ are 
made available to the scanner in three different lists. The cut in redshift corresponds to the Lyman-$\alpha$ emission line 
entering the BOSS spectrum. It also corresponds to a strong gap 
in the BOSS quasar redshift distribution due to target selection (see \Fig{DistributionRedshift}).

Most of the objects classified as stars by the pipeline are indeed stars and most of the objects
classified as quasar with $z<2$ are either quasars with $z<2$ or stars (see below). 
The objects classified as quasars at $z \geq 2$ are ranked by decreasing SNR.
This organizes the visual inspection and minimizes the risk of errors.
Most of the quasars with $z \geq 2$, the most valuable for the survey, are inspected by two different individuals. 

Objects that cannot be firmly identified by visual inspection are labeled in several categories.
Some spectra cannot be recognized because either the SNR is too low, or the spectrum has been badly extracted;
such objects are classified as {\tt Bad}. For others, the classification is not considered to be robust,
but there is some indication that they are stars ({\tt star\_?})
or quasars ({\tt QSO\_?}). For some objects both scanners were unable to give a firm
identification, such objects are labeled as `{\tt ?}'. Other objects are galaxies ({\tt Galaxy}). Finally some
objects are recognized as quasars but their redshifts are not certain ({\tt QSO\_Z?}).

\begin{figure}[htbp]
		\centering{\includegraphics[angle=-90,width=\linewidth]{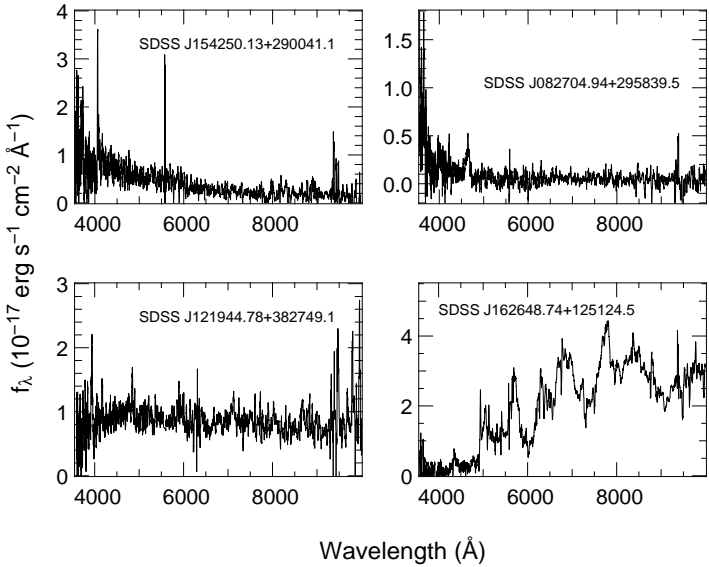}}
\caption{Examples of {\tt QSO\_?} (top panels) and {\tt QSO\_Z?} (lower panels).
The spectra were boxcar median smoothed over 5 pixels.}
\label{fig:exUnsecuredID}
\end{figure}

The output of the visual classification is provided as fields {\tt class\_person} and {\tt z\_conf\_person} in the
specObjAll table of the SDSS Catalog Archive Server (CAS) or the specObjAll.fits file from
the Science Archive Server (SAS).
The correspondence between the visual inspection classification we describe in this paper
({\tt QSO}, {\tt QSO\_BAL}, {\tt QSO\_Z?}, {\tt QSO\_?}, 
{\tt Star}, {\tt Star\_?}, {\tt Galaxy}, {\tt ?}) and the values of {\tt z\_conf\_person} and 
{\tt class\_person} 
is given in \Tab{VI_PIPE}.
Each time a new version of the BOSS pipeline becomes available, the data are reprocessed and objects in the
categories {\tt bad}, {\tt ?}, {\tt QSO\_?} and {\tt QSO\_Z?} are inspected again.
Examples of objects classified as {\tt QSO\_Z?} and {\tt QSO\_?} are displayed in \Fig{exUnsecuredID}.
Only objects classified as {\tt QSO} or {\tt QSO\_BAL} are listed  in the official DR9Q catalog.
Objects classified as  {\tt QSO\_Z?} are included in the supplemental list of quasars (see \Sec{VAC6}).
Objects classified as  {\tt QSO\_?} are also given for information in a separate list.
%

%
\begin{table}
\centering                        
\begin{tabular}{c c c c c}        
\hline\hline                 
z\_conf\_person & 0             & 1                  & 2             & 3        \\
 /              &               &                    &               &          \\
class\_person   &               &                    &               &          \\
\hline
0               & Not inspected & {\tt ?}            & -             & -        \\
1               & -             & -                  & {\tt Star\_?} & {\tt Star}     \\
3               & -             & {\tt QSO\_?}       & {\tt QSO\_Z?} & {\tt QSO}      \\
4               & -             & -                  & -             & {\tt Galaxy}   \\
30              & -             & -                  & -             & {\tt QSO\_BAL} \\
\hline   
\end{tabular}
\caption{The table gives the classification from the visual inspection corresponding to 
the combination of {\tt class\_person} (first column) and {\tt z\_conf\_person} (first row)
values provided in the headers of the SDSS-DR9 spectra 
available from the SDSS Catalog Archive Server. 
}
\label{t:VI_PIPE}
\end{table}

Of the 180,268 visually inspected targets corresponding to the DR9Q catalog, 87,822 were classified as unique quasars, 81,307 as stars and 6,120 as galaxies.
1,362 objects are likely quasars ({\tt QSO\_?}), 112 are quasars with an uncertain redshift ({\tt QSO\_Z?}) and
578 are likely stars ({\tt Star\_?}).
2,599 targets have bad spectra ({\tt Bad}) while we were not able to identify 368 objects ({\tt ?}).
Therefore 97.5\% of the objects are successfully classified. 
Only 27 true quasars were mis-identified by the BOSS pipeline as {\tt Star}, while 11,523 stars were classified 
as {\tt QSO}, most of them misidentified, however, as low redshift quasars, and only 1,241 have ZWARNING~=~0.
\Tab{DR9nb} gives a summary of these numbers.

Note that \cite{palanque2012} have obtained deeper MMT QSO data of some of the 
BOSS targets classified as {\tt QSO\_?} and confirmed that essentially all of these 
objects are true quasars. 

\begin{figure}[htbp]
	\centering{\includegraphics[width=.8\linewidth]{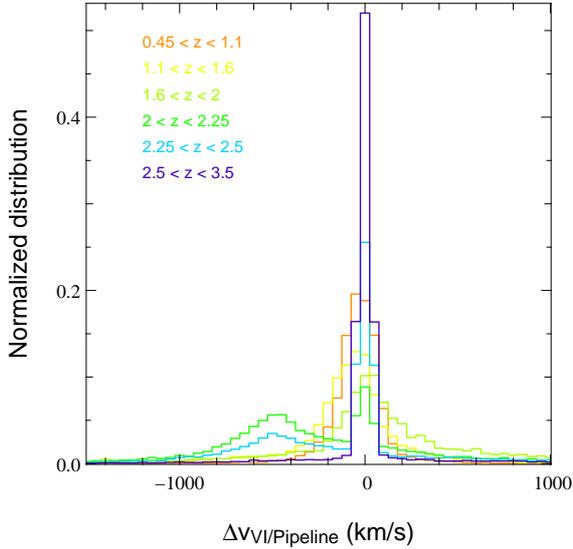}}
	\caption{Normalized (to unit integral) distribution of the velocity difference between the pipeline and visual inspection redshift estimates for different redshift bins.
	About half of the pipeline redshifts are corrected during the visual inspection.
	Most of the corrections are for quasars with $2 < z < 2.5$ where the \MgII\ emission line is available and where the pipeline redshift estimate does not correspond to the peak of the \MgII\ emission line.
	}
	\label{fig:dv_Scan2Pipe}
\end{figure}

During the visual inspection, a redshift is determined that will be refined further by an automatic procedure
(see \Sec{AutoZ}). The redshift of identified quasars provided  by the visual inspection is obtained applying 
the following procedure:
\begin{itemize}
	\item The first guess for the redshift is given by the BOSS pipeline and is not modified except if inaccurate or wrong.
The redshift from the pipeline can be wrong in cases where an emission line is misidentified. The presence of strong absorption 
at or near the emission, and especially a strong DLA,
is also a source of error. Often the redshift is just inaccurate because either it misses the 
peak of the \MgII\ emission line (and we consider that this line is the most robust indicator of the redshift)
or it is defined by the maximum of the \CIV\ emission line when we know that this line is  often
blueshifted  compared to \MgII\ \citep[][]{gaskell1982,mcintosh1999,VB01,richards2002,shen2008,hewett2010}.
	\item If the \MgII\ emission line is present in the spectrum, clearly detected, and not affected by sky subtraction, the visual 
inspection redshift is set {\sl by eye} at the maximum of this line. 
The typical uncertainty is estimated to be $\Delta z < 0.003$. The redshift is refined further, as described below.
	\item In other cases and for $z>2.3$ quasars, such that Mg~{\sc ii} is redshifted into the noisy part
of the red spectrum where sky subtraction errors make it unreliable,
the redshift is estimated using the positions of the red wing of the \CIV\ emission line which is known to be often blueshifted compared to \MgII\
and of the peak  of the Lyman emission line.
The precision is estimated to be $\Delta z  < 0.005$.
\end{itemize}
The visual redshift is not accurate to better than $\Delta z\sim 0.003$.
but can be used as a reliable guess for further automatic redshift determination (see \Sec{AutoZ}).
\Fig{dv_Scan2Pipe} displays the distribution of the velocity difference between the visual inspection redshift estimate 
and the redshift provided by the BOSS pipeline. At $z \leq 2$ the pipeline estimate is usually good and does 
not require significant adjustments. In the redshift range 2.0$-$2.3, about half of the redshifts are modified because the 
\MgII\ emission line is available and defines clearly the visual inspection redshift while the pipeline finds often a slightly 
lower redshift. 
At $z\gtrsim2.3$, 10\% of the redshifts are corrected.
Only 1,116 quasars ($\sim$2\%), regardless of ZWARNING flags, have a difference between the pipeline and visual redshifts larger than 0.1. 

%

In addition, peculiar spectral features are flagged:
\begin{itemize}
	\item When a  damped \lya\ absorption line is present in the forest, the object is 
assigned a flag ``DLA". This flag can be used to check
automatic Damped Lyman-$\alpha$ detections \citep[see ][]{noterdaeme2009,noterdaeme2012}.
	\item Broad absorption lines in \CIV\ and/or \MgII\ are also flagged.   At this point there is no
estimate of the width of the lines and we stay conservative.
This flag can be used to check automatic BAL detections (see \Sec{AutoBAL}).
	\item Problems such as the presence of artificial breaks in the spectrum, obviously wrong flux calibration, 
or bad sky subtraction are flagged as well whatever the identification of the object is. 
	These quality flags are pipeline-version dependent and are not meant to be released with the catalog. 
They are mainly useful for feedback to the pipeline team.
\end{itemize}

%

\begin{table*}
\centering                          
\begin{tabular}{l c c c}        
\hline                
\hline
 Classification & \# pipeline & \# pipeline           & \# visual inspection\\   
                &             & with {\tt ZWARNING}=0 &                     \\
\hline 
\hline                       
{\tt QSO}             & 102,696     & 86,855                 & 87,822   \\
{\tt QSO} with $z>2$  & (69,975)    & (64,004)               & (65,185) \\
{\tt QSO\_?}          & -           & -                      & 1,362    \\
{\tt QSO\_Z?}         & -           & -                      & 112      \\
{\tt Galaxy}          & 10,563      & 6,812                  & 6,120    \\
{\tt Star}            & 67,009      & 49,475                 & 81,307   \\
{\tt Star\_?}         & -           & -                      & 578      \\
{\tt Bad}             & -           & -                      & 2,599    \\
{\tt ?}               & -           & -                      & 368       \\
\hline
Total           & 180,268     & 143,142                & 180,268  \\
\hline                                
\end{tabular}
\caption{
	Number of objects identified as such by the pipeline with any  {\tt ZWARNING} value  (second column) and 
with {\tt ZWARNING}~=~0 (third column), and after the visual inspection (fourth column).
}           
\label{t:DR9nb}      
\end{table*}

%
%
\subsection{A note on Damped Lyman-$\alpha$ systems}

\begin{figure}[htbp]
	\centering{\includegraphics[width=75mm]{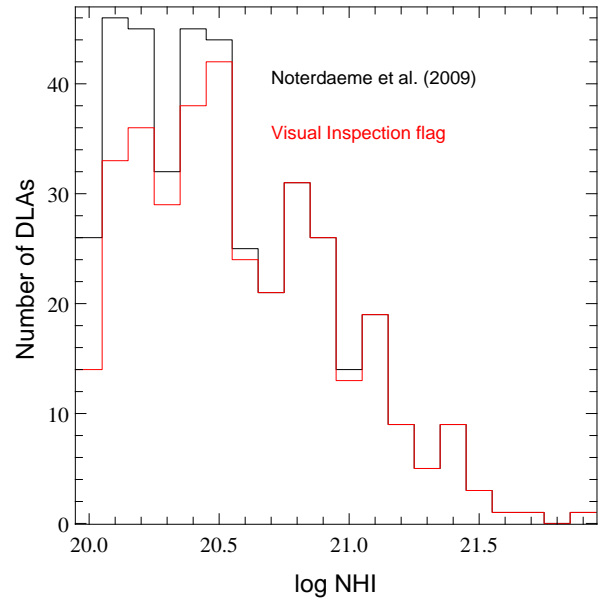}}
	\caption{
	  \HI\ column density distribution for DLAs and sub-DLAs detected by \cite{noterdaeme2009} 
in quasars observed 
both by SDSS-DR7 and BOSS (black histogram).
	The red histogram displays the same distribution but for DLAs flagged after visual inspection of BOSS spectra. This shows
that the visual inspection is robust for log~$N$(H~{\sc i})~$>$~20.3, the standard definition of DLAs.	
	}
\label{fig:DLA}
\end{figure}

In the course of the visual inspection, we flag the spectra with strong H~{\sc i} absorption 
(DLAs) in the Lyman-$\alpha$ forest. At this point we do not try to
measure the column density or to determine the redshift of the DLA. Flagging these lines
of sight can be useful to complement the search for DLAs by automatic procedures 
since this is a notoriously difficult task. 
\Fig{DLA} shows the number of DLAs we flag along SDSS-DR7 lines of sight
reobserved by BOSS, versus the $N$(H~{\sc i}) column density.
It can be seen that we visually recover most of the DLAs (log~$N$(H~{\sc i})~$>$~20.3)
identified in the SDSS-DR7 by Noterdaeme et al. (2009).
Only 11 such DLAs are missed by the visual inspection out of 257. 
The detection and analysis of DLAs in BOSS spectra is beyond the scope of this paper and 
will be described in \cite{noterdaeme2012}. 
%

%

%
\section{Automatic redshift estimate}
\label{s:AutoZ}

The visual inspection provides a reliable and secure redshift estimate for each quasar.
Nevertheless, it is somewhat subjective and the accuracy of such an estimate is limited and cannot be better than 500~km~s$^{-1}$.
In principle, it is possible to estimate the redshift of a quasar using a linear combination of principal components to fit the spectrum: 
the well known systematic shifts between emission lines are intrinsically imprinted in the components and the method can take into account 
the  variations from quasar to quasar \citep[see e.g.][]{paris2011}.  This should be a reliable procedure providing:
\begin{itemize}
	\item The reference sample used to derive the principal components is representative of the whole quasar population;
	\item The redshift of each quasar in the reference sample is reliable.
\end{itemize}
We will derive PCA components in order to reproduce the quasar spectrum between 
1,410 and 2,900 \AA, in the quasar rest frame, so that most of the prominent emission lines are covered,
especially C~{\sc iv} and Mg~{\sc ii}. This will yield an automatic estimate of the  quasar redshifts.
These components will be also used to fit emission lines individually to estimate a redshift for each 
emission line from the peak of the fit model. To derive these PCA components, 
we will use a reference sample of quasars for which the two main emission lines are well observed.
The redshifts of the quasars in the reference sample have also to be chosen carefully.    
The technique to derive PCA components of quasar spectra has been described in detail in several
papers \citep[e.g.][]{francis92,Yipal04,Suzukial05}. 
We refer the reader in particular to Section 2.3 of P\^aris et al. (2011).

		\subsection{Selection of the reference sample}
		\label{s:refsample}

To compute a set of principal components from a sample as representative as possible of the whole quasar population, we selected  
quasar spectra in SDSS-DR7 meeting the following requirements:
\begin{itemize}
	\item The rest frame wavelength range 1,410 - 2,900 \AA\ is redshifted into the observed wavelength range 
3,900 - 9,100 \AA\ (i.e. $1.77 < z < 2.13$). This observed wavelength range is chosen to avoid the flux-excess issue in 
the very blue portion of the spectra \citep[Section 2.4.1 and  ][]{paris2011} and bad sky line subtraction at the red end.
	\item The median squared SNR per pixel over the full wavelength range is higher than 5.
	\item The spectra do not display BAL troughs as listed in the  \cite{allen2011} catalog.
\end{itemize}
In SDSS-DR7, 8,986 quasar spectra meet these requirements. 
They all were visually inspected to remove spectra with obvious reduction issues (missing pixels, continuum breaks or very bad flux calibration).
We finally used 8,632 spectra.

The low-SNR cut we use here maximizes the number of quasars used for the PCA decomposition
and makes our sample as representative as possible of the BOSS quasars. 
		
	\subsection{Computing principal components}
	\label{s:PCz}

We now need an accurate redshift for each quasar before we calculate the PCA eigenvectors.
We first describe the use of Hewett \& Wild (2010) redshifts and then an improved approach 
using the peak of the Mg~{\sc ii} emission line in individual spectra.\\

\textit{\underline{Using Hewett \& Wild (2010) redshifts:}}
 
We first consider the redshifts provided by Hewett \& Wild (2010; HW10).
They have performed a systematic investigation of the relationship between different 
redshift estimation schemes and have derived empirical relationships between redshifts based
on different emission lines. 
They generated a high-SNR quasar template covering the UV and optical 
bands to be used to calculate cross-correlation redshifts.
They estimate and correct for the quasar 
luminosity-dependence of systematic shifts between quasar emission lines.
They are thus able to reduce systematic effects dramatically,
correcting redshifts for the {\sl mean} systematic shifts between emission lines.
Note however that this does not fully account for intrinsic quasar-to-quasar variation among the population.

Using these redshifts for the sample of representative quasars defined in \Sec{refsample}, we derive the 
PCA eigenvectors.  We then use the set of principal components 
to fit a linear combination of 4 principal components to the whole spectrum of $z \geq 2.2$ SDSS-DR7 quasar spectra 
and estimate their redshifts. This number of components has been chosen after several trials
in order to be able to derive a robust redshift for the maximum of objects.
Note that the samples used to compute the principal components and to which 
we apply the procedure are disjoint.

The median of the distribution of the velocity differences between the redshift given by HW10 and our redshift estimate is less 
than 30 ${\rm km \ s^{-1}}$. However, the rms of this distribution is about 1,200 ${\rm km \ s^{-1}}$ which is undesirably large
and is presumably due to quasar to quasar variations in emission-line shifts.

We can try to overcome this drawback  by using a redshift that is more representative of the individual characteristics
of the quasars in the reference sample. 
This is why we will derive a redshift from the observed \MgII\ emission line in each quasar spectrum.
Indeed this line has been recognized as a reliable indicator of the actual redshift
of the quasar (Shen et al. 2007, HW10).\\

\textit{\underline{Using \MgII\ emission line redshifts:}}

Using the set of PCA components previously described, we fit the \MgII\ emission line of each quasar in the 
same SDSS-DR7 reference sample. 
From this fit, we define the \MgII\ redshift using the peak-flux position of the emission line fit.
Using a combination of principal components to fit an emission line avoids the need to assume  a line profile 
(e.g. Gaussian, Lorentzian or Voigt).

To estimate the quality of each emission line fit:
\begin{itemize}
	\item We compute the amplitude of the emission line (expressed in units of the median error pixel of the 
spectrum in the window we use to fit the line) 
from the maximum flux relative to a fitted power-law continuum.
	\item We measure the FWHM of the emission line in ${\rm km \ s^{-1}}$.
\end{itemize}
The amplitude-to-FWHM ratio (expressed in s~km$^{-1}$)
provides an estimate of the prominence of the emission line. 
In particular, a weak and broad emission line will display a very low 
value of the amplitude-to-FWHM ratio.

To confirm the quality of the \MgII\ line measurement, we also fit \CIV\ emission lines using the same procedure. 
The \CIV\ emission line is easier to fit since it is stronger and the region of the spectrum where it is redshifted is
cleaner. If \CIV\ could not be fit, we also considered the \MgII\ fit to be unreliable. 

We then used the 7,193 spectra with both \CIV\ and \MgII\ amplitude-to-FWHM ratios larger than $8 \times 10^{-4}$~s~km$^{-1}$
to compute the new PCA components to be applied to the whole spectra.

\begin{figure}[htbp]
	\centering{\includegraphics[angle=-90,width=75mm]{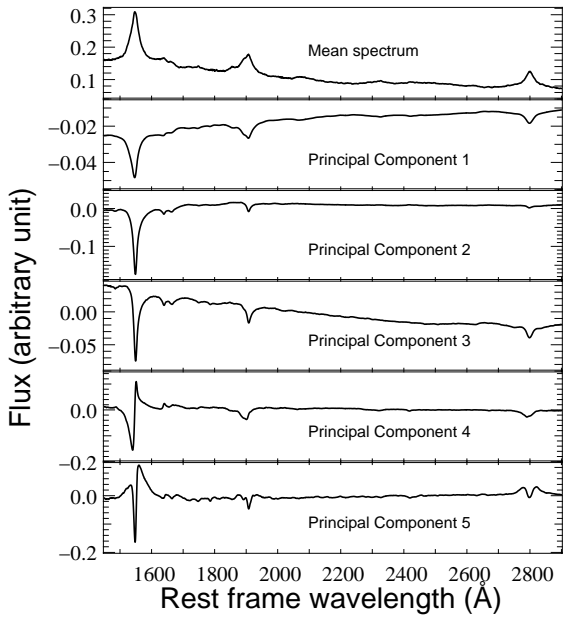}}
	\caption{
		Mean spectrum and the first five principal components derived in \Sec{PCz}.
		A linear combination of the first four principal components is used to estimate the global redshift of the quasar, 
while five components are used to fit emission lines locally.
		}
\label{fig:zPC}
\end{figure}

We use the set of principal components derived with the Mg~{\sc ii} redshifts in the following. 
\Fig{zPC} displays the mean spectrum together with the first five principal components.

%
	
	\subsection{Redshift estimates for BOSS quasars}

For each quasar in the DR9Q catalog, we use four principal components to fit the overall spectrum
after having subtracted the mean spectrum.
Four components are enough to reproduce the overall shape of the spectrum
and derive the redshift  \citep[]{Yipal04, paris2011}.
However, in order to avoid poor fitting due to the presence of strong absorption lines and especially
BAL troughs, 
we first obtain a fit with only two principal components.
This number is chosen because it provides a reasonable estimate of the amplitude of emission lines.
Using this first guess, we remove pixels below 2$\sigma$ and above 3$\sigma$ of the continuum where $\sigma$ is defined as 
the median flux error in an 11 pixel window.
We are thus able to remove broad absorption lines and badly subtracted sky emission lines 
(especially at the very red end of the spectra). We then increase the number of principal components iteratively to
three and four, removing narrow absorption lines, keeping the same detection thresholds.

Then, taking the visual inspection redshift estimate as an initial guess, it is possible to determine a redshift 
for each quasar by fitting a linear combination of four principal components to the spectrum, in which the redshift becomes 
a free parameter. We call this redshift the PCA redshift.

In addition, and in the same way as described in the previous subsection, we used five principal components to fit the Mg~{\sc ii} emission line
in BOSS spectra when possible and derived a redshift from the peak flux of the fit model. Using PCA allows to recover the line
without a priori assumptions about the line profile in a region of the spectrum affected by sky subtraction.
In the following we will call this redshift the PCA Mg~{\sc ii} redshift estimate.

\begin{figure}[htbp]
	\centering{\includegraphics[width=75mm]{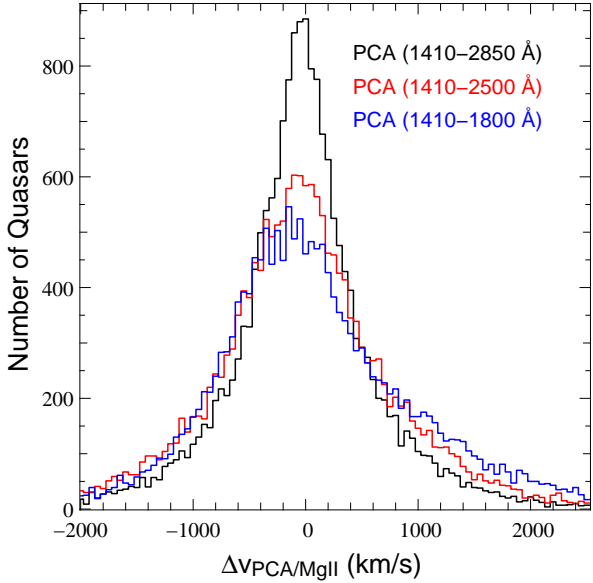}}
	\caption{
	Distributions of the velocity difference between PCA redshift estimates derived using different rest frame wavelength 
ranges and the PCA \MgII\ redshift estimate.
	}
\label{fig:CompPCA2MgII}
\end{figure}

We compare in \Fig{CompPCA2MgII} the distributions of the velocity difference between PCA and PCA Mg~{\sc ii}  redshift estimates.
The PCA was applied to all BOSS quasars with 1.57~$<$~$z_{\rm visual}$~$<$~2.3 so that both C~{\sc iv} and
Mg~{\sc ii} emission lines are in the observed redshift range and are not strongly affected by sky subtraction.
We considered three PCA estimates, varying the rest frame wavelength range over which the PCA was applied : (i) 1,410~$-$ 2,850~\AA~ (full range);
(ii) 1,410~$-$ 2,500~\AA~ (Mg~{\sc ii} is not included) and (iii) 1410~$-$~1800~\AA~ (only C~{\sc iv} is in the range).
There are 18,271 objects. It can be seen in \Fig{CompPCA2MgII} that the distributions are very similar. The median 
and rms of the distributions 
are ($-35.3$, 642), ($-52.2$, 780) and ($-30.3$, 851)
km~s$^{-1}$ respectively for the three wavelength ranges. 
The rms is dominated by low SNR spectra and slightly increases when the amount of information decreases.
The similarity of the distributions clearly shows that the PCA redshift estimate is 
consistent with the Mg~{\sc ii} estimate even when Mg~{\sc ii} is not included in the fit.

		\subsection{Comparison to HW10}

In order to compare the HW10 redshift estimates to ours, we selected SDSS-DR7 quasars re-observed by BOSS in the 
redshift range 2.00~$<$~$z$~$<$~2.30. We also restricted the sub-sample to quasars for which we were able to  
fit the \MgII\ emission line reliably and required the amplitude-to-FWHM of this line be larger than $8 \times 10^{-4}$~s~km$^{-1}$.
Even though the \MgII\ emission line is still detectable up to $z~=~2.5$, we restrict the redshift range to 
below $z~=~2.3$ to avoid the red end of the spectra where sky lines can be badly subtracted. 746 quasar spectra remain 
for the comparison.
%

\begin{figure}[htbp]
	\centering{\includegraphics[width=75mm]{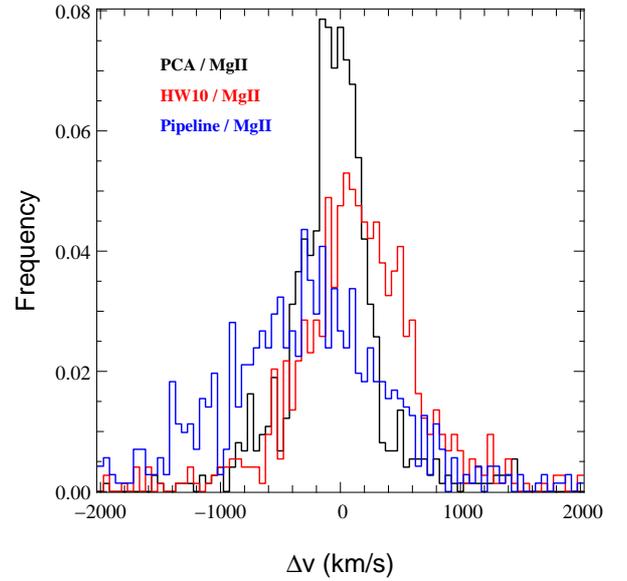}}	
	\caption{
	Normalized distributions of the velocity difference between our global PCA redshift estimate (black histogram), the pipeline 
redshift estimate (blue histogram) or \cite{hewett2010} redshifts (red histogram) with the redshift derived from a PCA 
fit of the \MgII\ emission line (see text). 
	}
	\label{fig:dv_HW10}
\end{figure}

\Fig{dv_HW10} displays the distributions of velocity differences between the PCA \MgII\ redshift estimate and our PCA global estimate 
(black histogram) or \citetalias{hewett2010} redshift (red histogram). Both distributions were normalized in the same manner and we 
also took into account the difference in the rest frame wavelength used by the different authors.

The median \citetalias{hewett2010} redshift estimate is shifted by $+$136.9 ${\rm km \ s^{-1}}$ 
(with positive velocity indicating redshift) compared to our 
median \MgII\ redshift with an rms of 467 ${\rm km \ s^{-1}}$. 
Both HW10 and Shen et al. (2011) find that the median shift of the \MgII\ emission line 
relative to the [OIII] doublet is smaller than 30 ${\rm km \ s^{-1}}$.
These discrepancies in the median velocity shift may not be very significant as different fitting recipes for any of these lines 
(\MgII, [OIII], C~{\sc iv}) can potentially cause systematic velocity differences of this order.

The median shift of our global estimate compared to the \MgII\ redshift is $-$49.9 ${\rm km \ s^{-1}}$, with an rms of 
389 ${\rm km \ s^{-1}}$.  The rms of the distribution is smaller than previously because we restrict our 
comparison here to spectra with high SNR.  
It is more peaked and the number of outliers is lower.
This is not surprising as we use the same components to fit the overall spectrum. 
However this illustrates the intrinsic dispersion between the results from the two methods.
 
The overall conclusion is that  our PCA estimate is very close to the Mg~{\sc ii} emission line redshift. 
And we are confident that the application of the procedure using  PCA components
to quasars for which the Mg~{\sc ii}  emission line is redshifted beyond
the observed wavelength range, will give robust redshift estimates.

%

		\subsection{Emission line redshifts}

Following the procedure described above, it is possible to reproduce the shape of each emission line  
with a linear combination of principal components. This combination can therefore fit
the individual lines without any a priori assumption about the line profile.
In the case of individual lines we have more flexibility to use more components because the fit  is more stable 
over a smaller wavelength range. We will use five PCA components and define the position (redshift) of the line 
as the position of the maximum of this fit.
\Tab{window_definition} displays the definition of each window used to fit emission lines together with the 
vacuum rest frame wavelengths taken from the NIST database\footnote{http://physics.nist.gov/PhysRefData/ASD/lines\_form.html}
used to compute the redshift. 
For multiplets (e.g. C~{\sc iv} and Mg~{\sc ii}), the rest frame wavelength used is the 
average wavelength over the transitions in the multiplet weighted by the oscillator strengths.
Together with the redshift estimate of each line, we also retrieve information on the symmetry of the line.
We compute the blue (red) HWHM (half width at half maximum) from the PCA fit,
bluewards (redwards) of its maximum.
The total FWHM is the sum of the blue and the red HWHMs. The continuum is provided by the fit of a power law
over the rest frame wavelength windows $1450-1500$, $1700-1850$ and $1950-2750$~\AA.

\begin{figure}[htbp]
	\centering{\includegraphics[width=75mm]{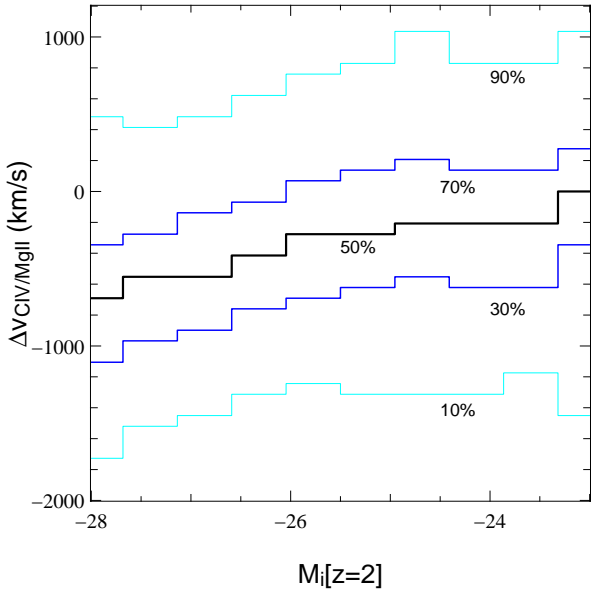}}
	\caption{
	Velocity difference between \CIV\ and \MgII\ emission line redshifts as a function of the absolute $i$ magnitude of the quasar.
	The solid black line shows the median velocity shift in 0.2 mag bins.
	Blue and cyan histograms display the 10$^{\rm th}$, 30$^{\rm th}$, 70$^{\rm th}$ and 90$^{\rm th}$ percentiles.
	The mean shift between the two emission lines increases with the quasar luminosity.
	}
	\label{fig:dv_vs_iMag}
\end{figure}

In \Fig{dv_vs_iMag} we plot the velocity of C~{\sc iv} relative to Mg~{\sc ii} versus the absolute magnitude
of the quasar. The more luminous the quasar, the more blueshifted is the C~{\sc iv} emission line.
Errors in the fit are less than 200~km~$^{-1}$.
These measurements can be useful to understand the relative shifts between 
different emission lines and discuss the structure of the broad line region \citep[see][]{shen2007,shang2007}.

The C~{\sc iii}]$\lambda$1909 line is blended with Si~{\sc iii}]$\lambda$1892 and to 
a lesser extent with Al~{\sc iii}$\lambda$1857. We do not attempt to deblend these lines. This means that
the redshift and red HWHM derived for this blend should  correspond to C~{\sc iii}]$\lambda$1909,
but the blue HWHM is obviously affected by the blend.

%

%
\begin{table}
\centering                          
\begin{tabular}{l c c}        
\hline                
Transition & Window & Rest frame wavelength (\AA ) \\   
\hline \hline                       
\CIV & 1450$-$1700 & 1549.061  \\     
\CIII & 1800$-$2000 & 1908.734  \\
\MgII & 2600$-$2850 & 2798.778  \\
\hline                                
\end{tabular}
\caption{Window and rest frame wavelength used to fit each emission line.}           
\label{t:window_definition}      
\end{table}

\section{Broad absorption line quasars}
\label{s:BAL}

Broad absorption troughs 
are flagged as BAL during the visual inspection. This flag 
means that an absorption feature broader than a usual intervening absorption 
(those arising in galaxies lying along the line of sight to the quasar)
is seen. These BALs may affect the Lyman-$\alpha$ forest and should be removed
from its analysis. We flag mostly C~{\sc iv} BALs but also Mg~{\sc ii} BALs.
Since during the visual inspection we do not measure the width of the trough,  
there is no a priori limit on the strength of the absorption.

We also implemented an automatic detection of C~{\sc iv} BALs.
We describe in \Sec{AutoBAL} the method used to detect BALs and estimate their 
properties automatically. We then test the robustness of the visual inspection in \Sec{visBAL} and
the results of the automatic detection in \Sec{AutoBALres}.

In the following subsections, we will concentrate on C~{\sc iv} BALs with $z>1.57$.
The quasar redshift limit is chosen so that the Si~{\sc iv} emission line is 
included in the spectra. This ensures that C~{\sc iv} BALs can be measured 
across the full range of velocities in balnicity index, e.g. up to 25,000~km~s$^{-1}$.
 
	\subsection{Method used to estimate BAL properties automatically}
	\label{s:AutoBAL}

In order to detect BALs and to characterize the strength of the troughs using an objective procedure, 
we compute the balnicity 
\citep[BI,][]{weymann1991} and the absorption indices \citep[AI,][]{hall2002} of the \CIV\ troughs.
In addition, we introduce a new index, the detection index, DI, which is a slight modification of BI.
In \Sec{AutoBALres}, we will measure these indices for all quasars regardless of visual inspection.

\begin{figure*}[htbp]
		\centering{\includegraphics[angle=-90,width=\linewidth]{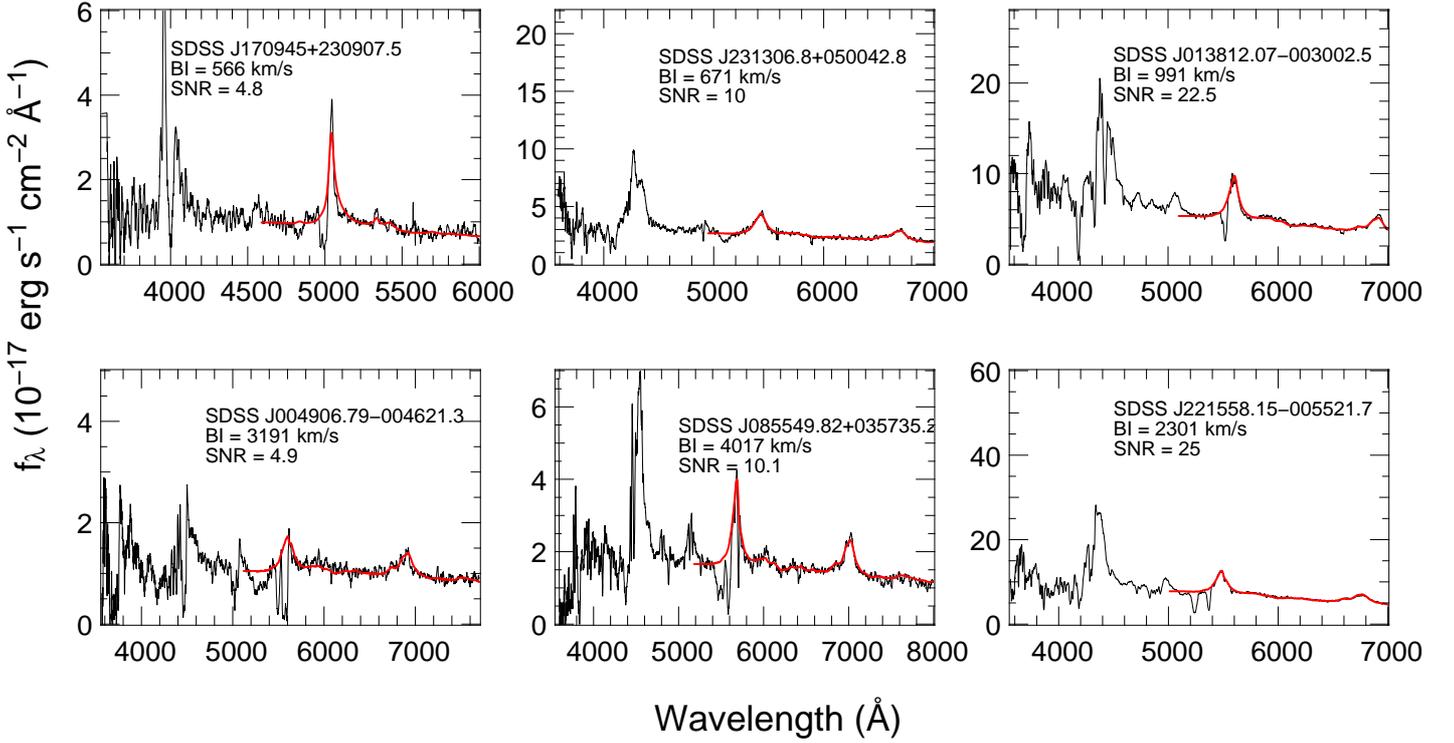}}
	\caption{
	Examples of high-redshift BAL quasar spectra in different ranges of signal-to-noise ratio
	and balnicity indices. The fit of the continuum is overplotted.
	}
	\label{fig:ExampleFitBAL}
\end{figure*}

The continuum has to be estimated first. 
For this, we use the same linear combination of four principal components described in \Sec{AutoZ}.
The resulting continuum covers the region from the \SiIV\ to the \MgII\ emission lines
(see examples in \Fig{ExampleFitBAL}). 
As described in Section~4.5, the procedure iteratively avoids absorption features and especially the BALs.
During the automatic procedure, we smoothed the data with a five pixel boxcar median.

%

With this continuum, we compute the balnicity index (BI) in the blue of the C~{\sc iv} emission line
using the definition introduced by \cite{weymann1991}:
\begin{equation}
BI ~=~ -  \int_{25,000}^{3,000}\left[ 1 - \frac{f(v)}{0.9} \right] C(v) \, \mathrm dv ,
\label{eq:BI}
\end{equation}
where $f(v)$ is the flux normalized to the continuum as a function of velocity displacement from the line center.
$C(v)$ is initially set to 0 and can take only two discrete values, 0 or 1. 
It is set to 1 whenever the quantity $1-f(v)/0.9$ is continuously positive over an
interval of at least 2,000 km/s.
It is reset to zero whenever the quantity in brackets becomes negative.
Therefore BI~=~0 does not mean that no trough is present. It means that, if a trough
is present, the absorption
does not reach 0.9 times the estimated continuum over a continuous window of
2,000~km~s$^{-1}$.

We will also define a detection index, DI, giving C a value 1 over the whole trough
{\sl if the criterion of a continuous trough over 2000~km~s$^{-1}$ is fulfilled}. This index
has the advantage of measuring the strength over the {\sl whole} trough. This index
will be useful to apply cuts in the analyses of the Lyman-$\alpha$
forest. Indeed these analyses need an estimate of the total strength of the trough
in order to avoid lines of sight spoiled by a strong BAL.

To study weaker troughs, \cite{hall2002} introduced the AI measurement defined as
\begin{equation}
AI ~=~ -  \int_{25,000}^{0}\left[ 1 - \frac{f(v)}{0.9} \right] C(v) \, \mathrm dv ,
\label{eq:AI}
\end{equation}
where $f(v)$ is the normalized flux and $C(v)$ has the same definition as for the DI
except that the threshold to set $C$ to 1 is reduced to 450 ${\rm km \ s^{-1}}$.
The AI index was introduced in order to take into account weaker troughs
and to measure troughs that are located close to the quasar rest velocity.
It is however more sensitive to the continuum placement than the BI.
Note that Trump et al. (2006) used a modified version of the AI wherein
the factor of 0.9 was removed from the integral to make the AI an
equivalent width measured in km~s$^{-1}$,
where 1,000~km~s$^{-1}$ was the threshold instead of 450~km~s$^{-1}$,
and where the integral extended to 29,000~km~s$^{-1}$.
In this work we use the original Hall et al. (2002) definition of the AI.

Following Trump et al. (2006), we calculate the reduced $\chi^2$ for each trough:
\begin{equation}
	\chi^2_{\rm trough} ~=~ \sum \frac{1}{N} \left(\frac{1-f(v)}{\sigma}\right)^2,
\label{eq:chi2trough}
\end{equation}
where N is the number of pixels in the trough, $f(v)$ is the normalized flux and $\sigma$ the 
rms of the pixel noise. The greater the value of $\chi^2_{\rm trough}$, the more likely
the trough is not due to noise.

We apply the automatic detection to all quasars in the DR9Q catalog and provide
values of DI, AI and BI. We estimate also an error on the indexes.
The error squared is obtained  by applying the same
formula as for the indexes replacing $(1-f/0.9)$ by $(\sigma/0.9)^2$ with $\sigma$ the rms of the 
noise in each pixel. Note however that the error on the strength of the trough is most of the time
dominated by the placement of the continuum. To estimate the latter we have
displaced the fitted continuum by 5\% and applied Eq.(2) of \cite{Kaspi2002}.

\subsection{Robustness of the visual detection of BALs}
	\label{s:visBAL}

\begin{figure}[htbp]
	\centering{\includegraphics[width=75mm]{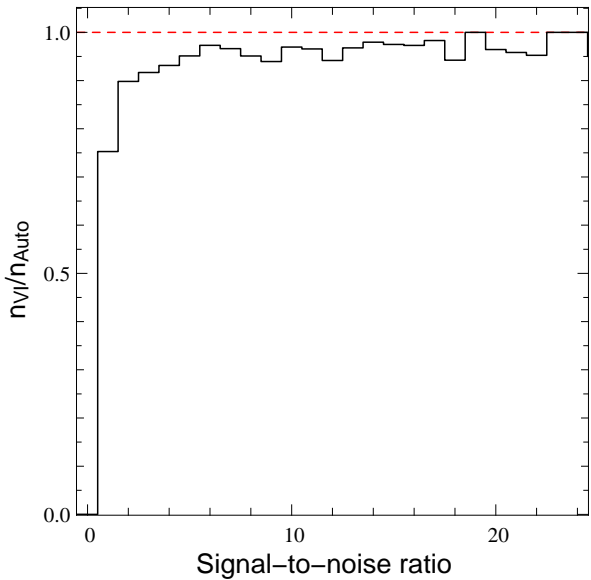}}
	\caption{
	Ratio of the numbers of BAL visual and 
automatic detections as a function of spectral SNR per pixel  at $\lambda_{\rm rest}$~=~1,700~\AA~ for BI~$>$~500~km~s$^{-1}$.
	As expected, this ratio decreases with decreasing SNR. 
	}
	\label{fig:DetectionRateBAL}
\end{figure}

During the visual inspection, we are conservative and flag a BAL only if the trough
is apparent. In addition, the automatic detections rely on the position of the continuum while
the visual inspection lacks this problem. This means that the BAL sample from the visual inspection
is purer than those from automatic detection. It is however unavoidable that,
as the strength of the absorption or the spectrum SNR decreases, the visual inspection will start to be
subjective. On the other hand the fraction of false BALs detected
by the automatic procedure will be higher.
\Fig{DetectionRateBAL} shows the ratio of the number of visual BALs to that 
of the automatic detections as a function of SNR per pixel at $\lambda_{\rm rest}$~=~1700~\AA~
for BI~$>$~500~km~s$^{-1}$.

%

Out of the whole DR9Q catalog, 7,533 quasars have been flagged visually as BAL.
Out of the 69,674 quasars with $z>1.57$, 
7,228 are flagged as BAL by visual inspection.
If we restrict the latter sample to quasars with SNR$>$10 at 1,700~\AA~in the rest frame
we have flagged 1,408 BALs  out of 7,317 quasars, a fraction of 19.2\%
which compares well with what was found by \cite{gibson09}. 

%
\begin{figure}[htbp]
	\centering{\includegraphics[width=75mm]{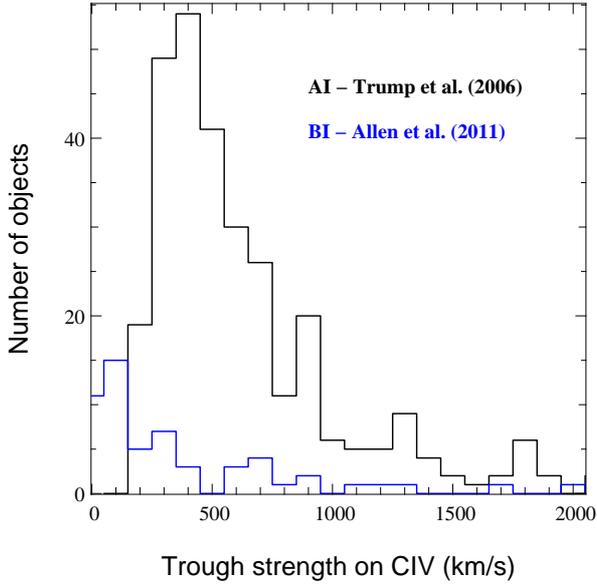}}
	\caption{
	Distribution of BI and AI for quasars detected as BAL by automatic procedures in previous 
        SDSS releases, and that were not flagged by the visual inspection of the BOSS spectra.
	The black histogram shows the distribution of AI as measured by \cite{trump2006} 
for 296 such quasars (all have BI~=~0). About half of them are not real BALs (see text).
	From the automatic detection by \cite{allen2011} (blue histogram), 57 quasars were missed by the visual inspection.
	Here again, only a handful of these objects actually display BAL troughs.
	}
	\label{fig:MissedBAL}
\end{figure}

\cite{trump2006} measured BAL troughs (BI and AI) in the SDSS-DR3 release.
We compare their detections and BI measurements with ours for
quasars in common between BOSS and SDSS-DR3.
Out of the 477 BALs (BI~$>$~0) that are detected by \cite{trump2006}, we
flag 425. 
We checked the BOSS spectra of these quasars individually. 
About half of them are not BALs and a handful, all with BI~$<$~500~km~s$^{-1}$,  are real BALs 
that were missed by the visual inspection. For the rest, it is hard to decide if they are real or not because of poor SNR.
Note that, in general, BOSS spectra are of higher SNR than previous SDSS spectra.

There are an additional 296 quasars in \cite{trump2006} that have C~{\sc iv} troughs that we do not flag as BALs.
These all have AI~$>$~0 but BI~=~0. 
The histogram of AI measurements from \cite{trump2006} for these objects 
is plotted as well in \Fig{MissedBAL} (black histogram). 
Most of the missing troughs have AI smaller than 1,000 ${\rm km \ s^{-1}}$. 
A visual inspection of the BOSS spectra reveals that most of the AIs have been overestimated and about half are 
not real mainly because the continuum in the red side of the C~{\sc iv}
emission line has been overestimated.

\begin{figure}[htbp]
	\centering{\includegraphics[angle=-90,width=\linewidth]{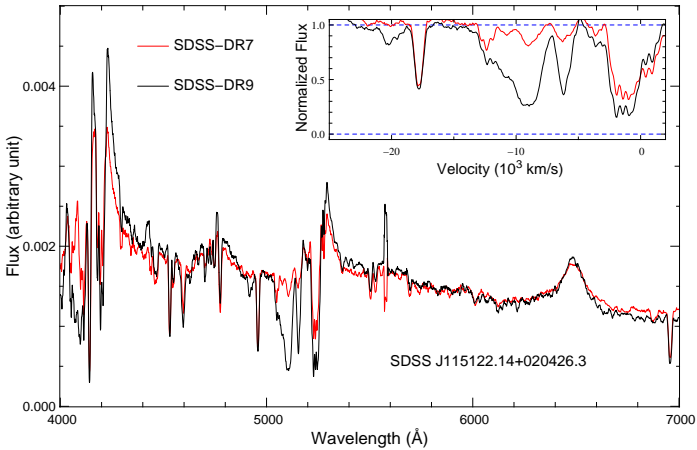}}
	\caption{
	Example of appearing BAL troughs. This quasar has been observed in SDSS-DR7 (red curve) and in SDSS-DR9 (black curve). 
The two spectra have been scaled to have a surface unity between 5,600 and 6,200\AA .
	The normalized flux in the \CIV\ region expressed in velocity is displayed in the inset.
	This quasar had was not detected as BAL in SDSS-DR7 (i.e. BI~=~0) while it has
${\rm BI} = 1,826 \ {\rm km \ s^{-1}}$ in the SDSS-DR9 spectrum.
	}
	\label{fig:BALappear}
\end{figure}

\cite{allen2011} searched for BALs in SDSS-DR6; they measured only BI.
Out of the 7,223 quasars with $z>1.57$ in common with BOSS, they find
722 quasars with BI~$>$~0. Of these 7,223 objects, we flag 1,259 as BALs of which 853 have BI$>$0. 
We checked the 131 objects for which we measured BI~$>$~0 but \cite{allen2011} found BI~=~0 individually.
Some of the additional BALs are identified because of better SNR in BOSS, some were missed 
by \cite{allen2011} because of difficulties in fitting the emission line
correctly, a handful are explained by the disappearance of the BAL between
the two epochs \citep{filizak2012}
and also by some appearances (see \Fig{BALappear}). We also find 57 objects that are detected by \cite{allen2011} and 
are missing in our visual detection. 
The BI distribution of these objects is shown as the blue histogram in \Fig{MissedBAL}.
About half of them are not BALs upon re-inspection and a handful 
are real BALs missed by the visual inspection. The nature of the rest of the objects is unclear. 

%

We conclude from this comparison that our catalog of BAL quasars flagged by 
visual inspection is pure at the 95\% level, but is probably incomplete below 
BI$\sim$500~${\rm km \ s^{-1}}$. This results from the conservative approach
we adopted when flagging the troughs implying that the number of detections in the visual inspection
is decreasing with decreasing SNR.

It is difficult to estimate the incompleteness especially at low SNR
because none of the previously published samples is reliable at small BI values.
Therefore we caution the reader against blind uses of the catalog.
SNR at rest wavelength $1700$~\AA~  (SNR\_1700) is provided in the catalog. This can be used
to identify the more reliable spectra.

\begin{figure}[htbp]
	\centering{\includegraphics[width=75mm]{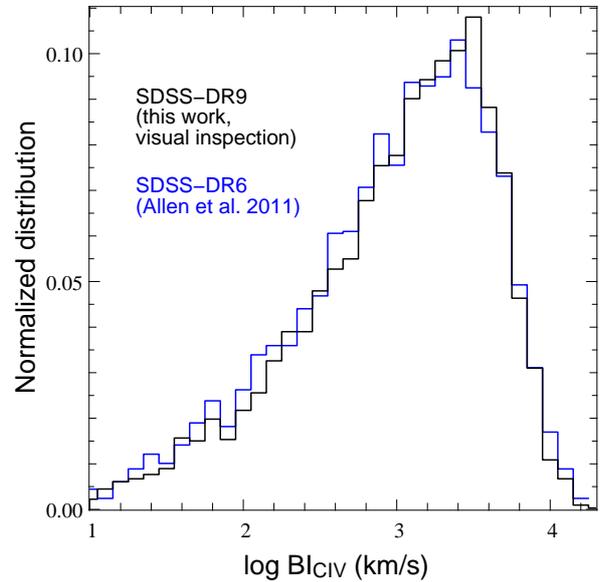}}
	\caption{
		Normalized distributions of the logarithm of balnicity indices measured from \CIV\ troughs.
		The BI distribution from the present catalog (black histogram) computed from 7,227 
visually flagged BAL quasars is very similar to the distribution from 
Trump et al. (2006, red histogram) obtained from 1,102 BAL quasars from the SDSS-DR3 quasar catalog \citep{schneider2005}.
		The distribution is also very similar to the BI distribution from Allen et al. (2011, blue histogram)
based on the SDSS-DR6 quasar catalog.
	}
	\label{fig:BIdistri}
\end{figure}

The BI distributions normalized by the total number of quasars with
BI$>$500 ${\rm km \ s^{-1}}$ in each sample for visually flagged BALs in  DR9 (this work) and DR6 \citep{allen2011} are compared in \Fig{BIdistri}.
We find  3,130 BALs with BI$>$500 ${\rm km \ s^{-1}}$ out of 69,674 BOSS quasars
with $z>1.57$  (4.5~\%). If we restrict ourselves to quasars with SNR~$>$~10, these numbers are
813 BALs out of 7,317 quasars, 
corresponding to a rate of 11.1\%. 
This compares well with the $\sim$10\% uncorrected observed fraction of BAL found  
by Allen et al. (2011) at $z\sim 2.5$.

%

\subsection{Automatic detection}
	\label{s:AutoBALres}
	
We also performed an automatic detection of the C~{\sc iv} troughs using the 
continua in the wavelength range between the Si~{\sc iv} to C~{\sc iv} emission lines
computed as described in \Sec{AutoZ}. 
BIs, AIs and DIs are calculated for all quasars with $z>1.57$ using \Eq{BI} and \Eq{AI}.
The values are given in the catalog together with 
the number of troughs both with width $>$~2000 and 450~km~s$^{-1}$. We also give, for
quasars with BI~$>$~0, the minimum and maximum velocities relative to $z_{\rm em}$,
$v_{\rm min}$ and $v_{\rm max}$, spanned by the whole absorption flow.

Out of the 69,674 (resp. 7,317) quasars with $z>1.57$ (resp. and SNR\_1700$>$10), 
8,124 (resp. 3,499)  BALs, with $\chi^2_{\rm trough}$~$>$~10,
have AI~$>$~0  ${\rm km \ s^{-1}}$. 
A visual inspection of spectra with small values of AI indicates
that a number are due to inadequate continuum fitting. We advise to be careful with AI values
smaller than 300~km~s$^{-1}$ (see also below). 

Out of the 69,674 (resp. 7,317) quasars with $z>1.57$ (resp. and SNR\_1700$>$10), 
4855 (resp. 1,196)  BALs have BI~$>$~0  ${\rm km \ s^{-1}}$. This corresponds
to 7\% (resp. 16.3\%). 821 BALs (11.2\%) have BI$>$500~km~s$^{-1}$.
While the overall detection rate is larger than for the visual inspection, it is
important to note that the automatic detection finds only 8 more objects with BI$>$ 500~km~s$^{-1}$
in spectra with SNR$>$10 than the visual inspection. Upon reinspection, we found that half of them are not real 
and are due either to a peculiar continuum or to the presence of strong metal lines from a DLA at 
$z_{\rm abs}\sim z_{\rm em}$. Three are real, but shallow BALs.
This shows that the automatic and visual detections give nearly identical results
for BI$>$500~km~s$^{-1}$. At lower BI and lower SNR, and consistently with what was found
by comparison with previous surveys, the number of unreliable detections is large.

%
\begin{figure}[htbp]
	\centering{\includegraphics[width=75mm]{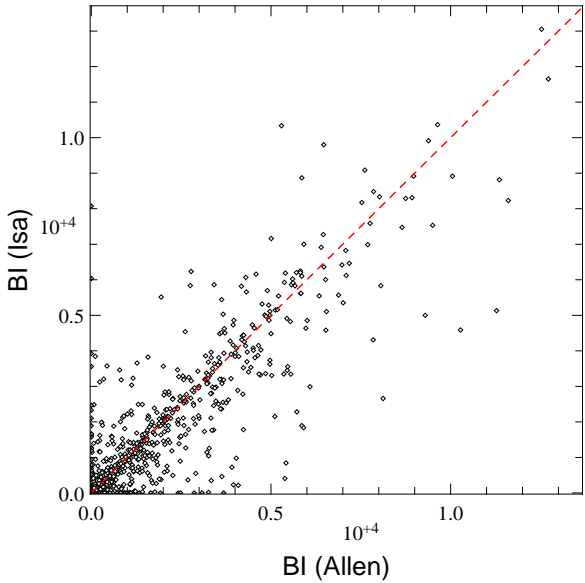}}
	\caption{
	Balnicity index (BI) from this work against BI measured by \cite{allen2011} for SDSS-DR6 objects 
re-observed by BOSS. 
	}
	\label{fig:BIAllvsDR9}
\end{figure}

\begin{figure*}[htbp]
	\centering{
		\begin{minipage}{.45\linewidth}
			\centering{\includegraphics[width=\linewidth]{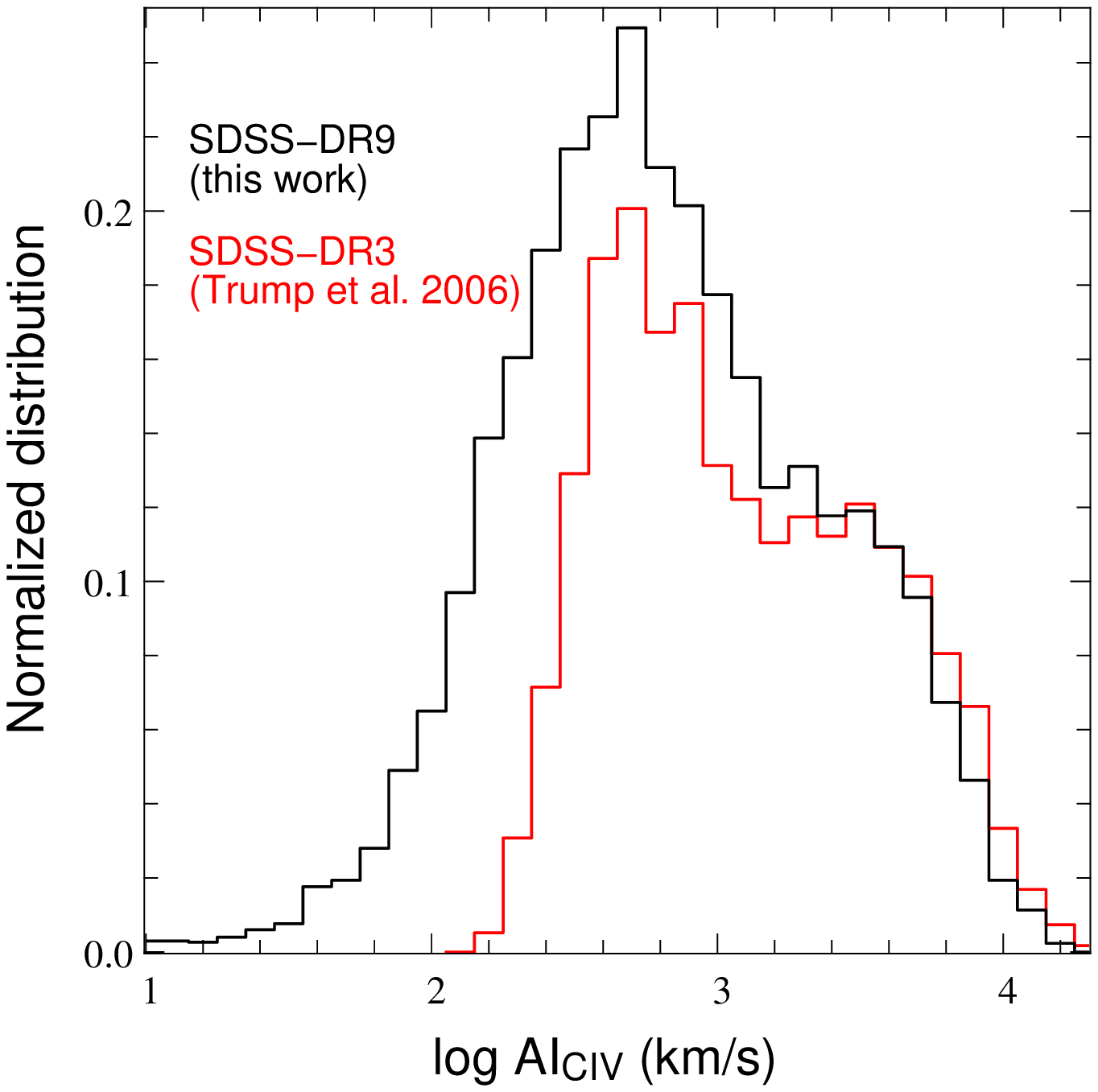}}
		\end{minipage}
		\hfill
		\begin{minipage}{.45\linewidth}
			\centering{\includegraphics[width=\linewidth]{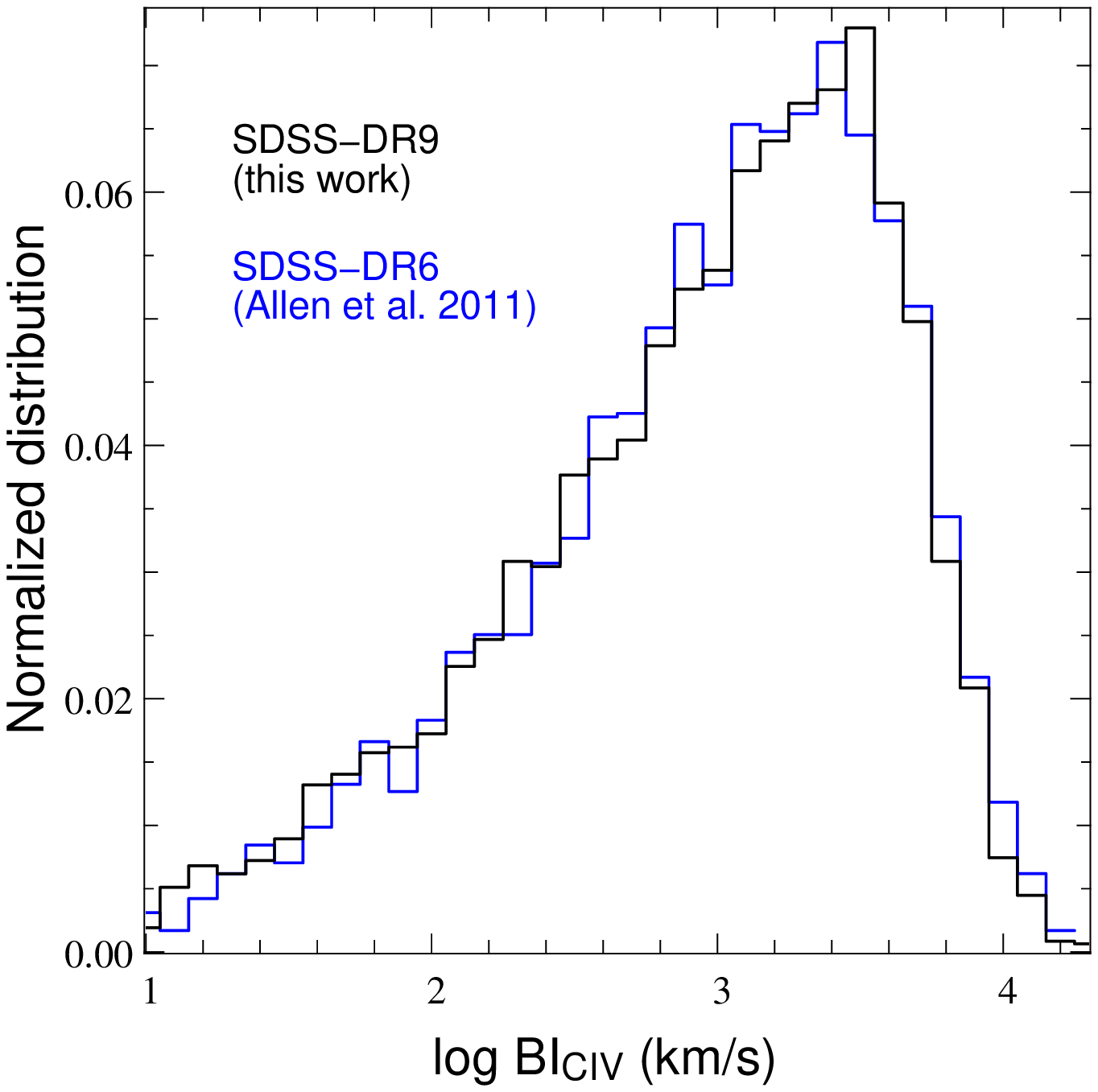}}
		\end{minipage}
		}
	\caption{
	\textit{Left panel:} Distribution of absorption indices (AI) from our automatic detection (black histogram) and from SDSS-DR3 
\citep[red histogram,][]{trump2006}. The distributions are normalized for $\log {\rm AI} > 3$.
 The difference between the two results at low AI is a consequence of slightly different formula used to measure AI.
	\textit{Right panel: }	Distribution of balnicity indices (BI) from our automatic detection (black histogram) and from 
SDSS-DR6 \citep[blue histogram,][]{allen2011}.
	}
	\label{fig:BIdistri_auto}
\end{figure*}

We compare in \Fig{BIAllvsDR9} the BI values measured by Allen et al. (2011)
for SDSS-DR6 spectra with BI values measured by our automatic procedure
using BOSS spectra for the same quasars. Although the scatter is large,
the median difference is only $\sim$30~km~s$^{-1}$.  Note that part of the scatter
is probably due to BAL variability \citep[see][]{Gibson2008,Gibson2010,filizak2012}.
In \Fig{BIdistri_auto} left (resp. right) panel, we compare the frequency distribution of AI (resp. BI) values
in our BAL sample detected automatically with that of previous studies.
The distributions are normalized in the same manner.
It can be seen that the shape of the distributions are very similar.
They peak around AI~=~300~km~s$^{-1}$ which is the lower limit we set for robust
detection.

We have shown here that BI measurements provided in the catalog are 
robust for SNR~$>$~5 (see \Fig{DetectionRateBAL}) and BI~$>$~500~km~s$^{-1}$. 
Any statistical analysis should be restricted to the corresponding sample.
The catalog gives a few properties of detected C~{\sc iv} troughs and of
Si~{\sc iv} and Al~{\sc iii} troughs but only in cases where BI(C~{\sc iv})~$>$~500~km~s$^{-1}$
and SNR$>$5. These troughs have been measured by Gibson et al. (2009) in SDSS-DR5.

%

%
\section{Description of the DR9Q catalog}
\label{s:Catalog_description}
The DR9Q catalog is available both as a standard ASCII file and 
a binary FITS table file at the
SDSS public website  http://www.sdss3.org/dr9/algorithms/qso\_catalog.php.
The files contain the same number of columns, 
the FITS headers contain all of the required documentation
(format, name, unit of each column).  
The following description applies to the standard ASCII file. 
\Tab{CatalogFormat} provides a summary of the information
contained in each of the columns in the ASCII catalog.
The supplemental list of quasars (see Section 8) together with the
list of objects classified as {\tt QSO\_?} are also available at the same
SDSS public website.

\par\noindent
Notes on the catalog columns:\\
\smallskip
\noindent
1. The DR9 object designation, given by the format \hbox{SDSS Jhhmmss.ss+ddmmss.s}; only the final~18
characters  are listed in the catalog (i.e., the \hbox{``SDSS J"} for each entry is dropped).
The coordinates in the object name follow IAU convention and are truncated, not rounded.

\noindent
2-3. The J2000 coordinates (Right Ascension and Declination) in decimal degrees.  
The astrometry is from DR9 \citep[see][]{DR9}.

\noindent
4.  The 64-bit integer that uniquely describes the spectroscopic observation
that is listed in the catalog (Thing\_ID). 

\noindent
5-7. Information about the spectroscopic observation (Spectroscopic plate number, 
Modified Julian Date, and spectroscopic fiber number) used to
determine the characteristics of the spectrum.
These three numbers are unique for each spectrum, and
can be used to retrieve the digital spectra from the public SDSS database.

\noindent
8. Redshift from the visual inspection (see \Sec{VI}).

\noindent
9. Redshift from the BOSS pipeline \citep[see \Sec{survey} and][]{bolton2012}.

\noindent
10. Error on the BOSS pipeline redshift estimate.

\noindent
11. ZWARNING flag from the pipeline. ZWARNING~$>$~0 indicates bad fits in 
the redshift-fitting code.

\noindent
12.  Automatic redshift estimate from the fit of the quasar continuum over the
rest frame wavelength range 1,410$-$2,000~\AA~ with
a linear combination of four principal components (see \Sec{AutoZ}).
When the velocity difference between automatic PCA and visual inspection redshift estimates is larger than 
3000~${\rm km \ s^{-1}}$, this PCA redshift is set to $-1$.
The inaccuracy in the PCA estimate is often due to difficulties in the fit of the continuum.
In that case no automatic measurements are made on these objects and BI is set to $-1$. 

\noindent
13. Error on the automatic PCA redshift estimate.
If the PCA redshift is set to $-1$, the associated error is also set to $-1$.

\noindent
14. Estimator of the PCA continuum quality (between 0 and 1). See Eq.(11) of 
P\^aris et al. (2011).

\noindent
15-17. Redshifts measured from \CIV , \CIII\ complex and \MgII\ emission lines 
from a linear combination of five principal components (see \Sec{AutoZ}).

\noindent
18. Morphological information. If the SDSS photometric pipeline classified the image of the quasar
as a point source, the catalog entry is~0; if the quasar is extended, the catalog entry is~1.

\noindent
19-21. 
Quasars targeted by BOSS are tracked with the BOSS\_TARGET1 flag bits \citep[19; see details of selection method in][]{ross2012}. 
In addition, 5\% of fibers on each plate are dedicated to ancillary programs tracked 
with the ANCILLARY\_TARGET1 (20) and ANCILLARY\_TARGET2 (21) flag bits.
The bit values and the corresponding program names are listed in \cite{dawson2012}. 

\noindent
22. A quasar known from SDSS-DR7  has an entry equal to 1, and 0 otherwise

\noindent
23-25. Spectroscopic plate number, Modified Julian Date, and spectroscopic fiber number
in SDSS-DR7.

\noindent
26. {\tt Uniform }flag. See \Sec{Uniformflag}.

\noindent
27. The absolute magnitude in the $i$ band at $z=2$ calculated after correction for
Galactic extinction and assuming
\hbox{$H_0$ = 70 km s$^{-1}$ Mpc$^{-1}$,}
$\Omega_{\rm M}$~=~0.3, $\Omega_{\Lambda}$~=~0.7, and a power-law (frequency)
continuum index of~$-0.5$.
The K-correction is computed using Table~4 from \cite{richards2006}.

\noindent
28. The $\Delta (g-i)$ color, which is the difference in the Galactic
extinction corrected $(g-i)$ for the quasar and that of the mean of the
quasars at that redshift .  If $\Delta (g-i)$ is not defined for the quasar,
which occurs for objects at either \hbox{$z < 0.12$} or \hbox{$z > 5.12$}
the column will contain~``$-9.000$".  See \Sec{DR9summary} for a description of this
quantity.

\noindent
29. Spectral index $\alpha _{\nu}$ (see \Sec{Spectralindex}).

\noindent
30. Median signal-to-noise ratio computed over the whole spectrum.

\noindent
31. Median signal-to-noise ratio computed over the window 1,650-1,750 \AA\ in the quasar rest frame.

\noindent
32. Median signal-to-noise ratio computed over the window 2,950-3,050 \AA\ in the quasar rest frame.

\noindent
33. Median signal-to-noise ratio computed over the window 5,100-5,250 \AA\ in the quasar rest frame.

\noindent
34-37. FWHM (${\rm km \ s^{-1}}$), blue and red half widths at half-maximum (HWHM;
the sum of the latter two equals FWHM), and amplitude (in units of the median rms pixel noise,
see \Sec{AutoZ}) of the \CIV\ emission line. 

\noindent
38-39. Rest frame equivalent width and corresponding uncertainty in \AA\ of the \CIV\ emission line.

\noindent
40-43. 
Same as 34-37 for the  \CIII\ emission complex. 
It is well known that C~{\sc iii}]$\lambda$1909 is blended with Si~{\sc iii}]$\lambda$1892
and to a lesser extend with Al~{\sc iii}$\lambda$1857. We do not attempt
to deblend these lines. Therefore the redshift and red
HFHM derived for this blend correspond to
C~{\sc iii}]$\lambda$1909. The blue HFWM is obviously affected by the blend

\noindent
44-45. Rest frame equivalent width and corresponding uncertainty in \AA\ of the \CIII\ emission complex. 

\noindent
46-49.  Same as 34-37
for the \MgII\ emission line. 

\noindent
50-51. Rest frame equivalent width and corresponding uncertainty in \AA\ of the \MgII\ emission line.

\noindent
52. BAL flag from the visual inspection. It is set to 1 if a BAL feature was seen during the visual inspection. It is set to 0 
otherwise. Note that  BAL quasars are flagged during the visual inspection at any redshift.

\noindent
53-54. Balnicity index (BI) for \CIV\ troughs, and its error, expressed in ${\rm km \ s^{-1}}$. See definition in \Sec{AutoBAL}.
The Balnicity index is measured for quasars with $z > 1.57$ only.
If the BAL flag from the visual inspection is set to 1 and the BI is equal to 0, this means either that there is no 
\CIV\ trough (but a trough is seen in another transition) or that the trough seen during the visual inspection does 
not meet the formal requirement of the BAL definition. 
In cases with bad fits to the continuum, the balnicity index and its error are set to -1.

\noindent
55-56. Absorption index, and its error, for \CIV\ troughs expressed in ${\rm km \ s^{-1}}$. See definition in \Sec{AutoBAL}.
In cases with bad continuum fit, the absorption index and its error are set to -1.

\noindent
57-58. Detection index, and its error, for \CIV\ troughs expressed in ${\rm km \ s^{-1}}$. See definition in \Sec{AutoBAL}.
In cases with bad continuum fit, the detection index and its error are set to -1.

\noindent
59. Following Trump et al. (2006), we calculate the reduced $\chi^2$ for each trough
from \Eq{chi2trough}.
We require that troughs have $\chi^2_{\rm trough}$~$>$~10 to be considered as
true troughs (see \Sec{AutoBAL}).

\noindent
60. Number of troughs of width larger than 2,000~km~s$^{-1}$

\noindent
61-62. Full velocity range over which \CIV\ troughs are at least 10\% below the continuum 
for troughs of width larger than 2,000~km~s$^{-1}$.  

\noindent 
63. Number of troughs of width larger than 450~km~s$^{-1}$

\noindent
64-65. Full velocity range over which \CIV\ troughs are at least 10\% below the continuum
for troughs of width larger than 450~km~s$^{-1}$.

\noindent
66-68. Rest frame equivalent width in \AA\ of \SiIV, \CIV\ and \AlIII\  troughs detected in BAL quasars
with BI $>$ 500 ${\rm km \ s^ {-1}}$ and SNR\_1700 $>$~5.
They are set to 0 otherwise or in cases where no trough is detected and to -1 if the continuum is not reliable.

\noindent
69-70 The SDSS Imaging Run number and the Modified Julian Date (MJD) of the
photometric observation used in the catalog.  The MJD is given as an integer;
all observations on a given night have the same integer MJD
(and, because of the observatory's location, the same UT date). For example,
imaging run 94 has an MJD of 51075; this observation was taken on
1998 September~19~(UT).

\noindent
71-74 Additional SDSS processing information: the
photometric processing rerun number; the camera column (1--6) containing
the image of the object, the field number of the run containing the object,
and the object identification number
\citep[see][for descriptions of these parameters]{stoughton2002}.

\noindent
75-84. DR9 flux and errors (not corrected for Galactic extinction) in the five SDSS filters. 

\noindent
85-89. TARGET photometric flux
in the five SDSS filters.

\noindent
90. Galactic extinction in the $u$ band based on the maps of
\cite{schlegel1998}. For an $R_V = 3.1$ absorbing medium,
the extinctions in the SDSS bands can be expressed as

$$ A_x \ = \ C_x \ A_u $$

\noindent
where $x$ is the filter ($ugriz$), and values of $C_{g, r, i, z}$ are
 0.736, 0.534, 0.405, and 0.287. See \cite{Schlafly2011} however.

\noindent
91. The logarithm of the Galactic neutral hydrogen column density along the
line of sight to the quasar. These values were
estimated via interpolation of the 21-cm data from \cite{stark1992},
using the COLDEN software provided by the {\it Chandra} X-ray Center.
Errors associated with the interpolation are typically expected to
be less than $\approx 1\times 10^{20}$~cm$^{-2}$ 
\citep[e.g., see \S5 of][]{elvis1994b}.

\noindent
92. The logarithm of the vignetting-corrected count rate (photons s$^{-1}$)
in the broad energy band \hbox{(0.1--2.4 keV)} in the
{\it ROSAT} All-Sky Survey Faint Source Catalog \citep{voges2000} and the
{\it ROSAT} All-Sky Survey Bright Source Catalog \citep{voges1999}.
The matching radius was set to 30$''$ (see \Sec{ROSAT});

\noindent
93. The SNR of the {\it ROSAT} measurement.

\noindent
94. Angular Separation between the SDSS and {\it ROSAT} All-Sky Survey
locations (in arcseconds).

\noindent
95-98. UV fluxes and errors from GALEX, aperture-photometered from the original GALEX images
in the two bands FUV and NUV (see \Sec{GALEX}).

\noindent
99-100. The $J$ magnitude and error from the Two Micron All Sky Survey
All-Sky Data Release Point Source Catalog \citep{cutri2003} using
a matching radius of ~2.0$''$ (see \Sec{2MASS}).  A non-detection by 2MASS is indicated by a ``0.000" in these
columns.  Note that the 2MASS measurements are Vega-based, not AB,
magnitudes.  

\noindent
101-102. SNR in the $J$ band and corresponding 2MASS jr\_d flag.

\noindent
103-106. Same as 98-101 for the $H$-band.

\noindent
107-110. Same as 98-101 for the $K$-band.

\noindent
111. Angular separation between the SDSS and 2MASS positions (in arcseconds).

\noindent
112-113. The $w1$  magnitude and error from the Wide-field Infrared Survey Explorer
\citep[WISE;][]{wright2010} All-Sky Data Release Point Source Catalog  using a matching radius of 2$"$
(see \Sec{WISE}).

\noindent
114-115 SNR and $\chi^2$ in the WISE $w1$ band.

\noindent
116-119. Same as 111-114 for the $w2$-band.

\noindent
120-123. Same as 111-114 for the $w3$-band.

\noindent
124-127. Same as 111-114 for the $w4$-band.

\noindent
128. Angular separation between SDSS and WISE positions (in arcseconds).

\noindent
129. If there is a source in the FIRST catalog (version July 2008) within~2.0$''$ 
of the quasar position, this column contains the FIRST peak flux density
(see \Sec{FIRST}).
\noindent
An entry of ``0.000" indicates no match to a FIRST source; an entry of
``$-1.000$" indicates that the object does not lie in the region covered by
the final catalog of the FIRST survey. 

\noindent
130. The SNR of the FIRST source whose flux is given in column~128.

\noindent
131. Angular separation between the SDSS and FIRST positions (in arcseconds).

\addtocounter{table}{1}

%
\section{Summary of sample}
\label{s:DR9summary}
\subsection{Broad view}

The DR9Q catalog contains 87,822 unique, visually confirmed quasars, of which 65,205 and 61,931 have, respectively, $z\geq2$ and $z>2.15$. 
91\% of these quasars were discovered by BOSS.
The first two years of operations cover an area of approximately 3,275 ${\rm deg^2}$
leading  to a mean density of $>$15 quasars with $z>2.15$  per square degree.
In the following, we describe the properties of the quasar population drawn from
the whole sample. However, we also provide a {\tt uniform} flag (see \Sec{Uniformflag}). 
A sample of quasars with {\tt uniform > 0} is a sufficiently statistical sample for, e.g., 
clustering measurements on some scales \citep[e.g.,][]{white2012} and luminosity function demographics.
 
\begin{figure}[htbp]
	\centering{\includegraphics[width=80mm]{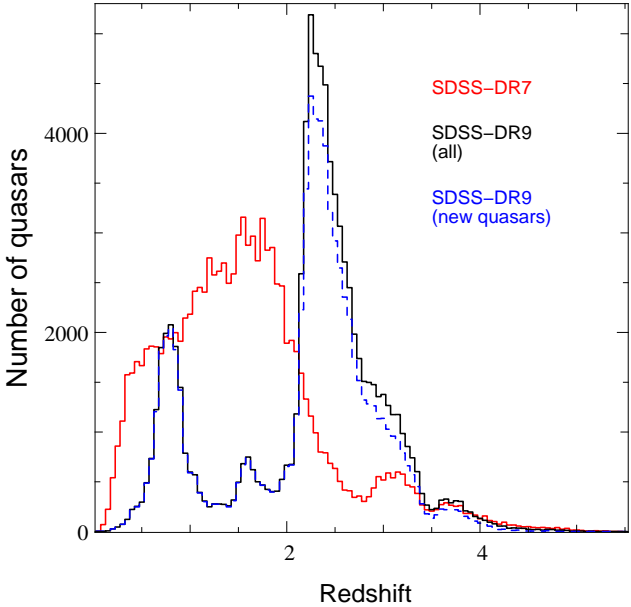}}
	\caption{ 
	The redshift distribution of quasars from this catalog is displayed in black.
	The same distribution is shown for newly discovered quasars only (dashed blue histogram).
	Most of the SDSS-DR9 quasars have a redshift greater than 2.
	The redshift distribution of quasars from the SDSS-DR7 catalog \citep{schneider2010} is shown for comparison in red.	
	The latter is dominated by quasars at low redshift.
	The present catalog contains 2.6 times more quasars at $z>2.15$ than the DR7 catalog.
	}
	\label{fig:DistributionRedshift}
\end{figure}

Quasars from the present catalog span a range of redshift from $z~=~0.058$ to $z~=~5.855$.
The redshift distribution is given in  \Fig{DistributionRedshift} together with that from  SDSS-DR7
\citep[red histogram,][]{schneider2010}. It is apparent from the figure that BOSS primarily targets
$z>2.15$ quasars as it was designed.  
Only 7,932 of those quasars were previously known, e.g. detected by previous surveys and the majority
of those were previous SDSS discoveries.
The DR9Q catalog thus contains about 2.6 times more high-redshift quasars than the whole SDSS-I/II survey.
The two peaks in the redshift distribution at $z \sim 0.8$ and $z \sim 1.6$ are due to known degeneracies 
in the SDSS color space. Six objects have $z<0.1$. These are Seyfert galaxies 
that were classified as quasars in order to differentiate them from normal galaxies. 
%

\begin{figure}[htbp]
	\centering{\includegraphics[width=80mm]{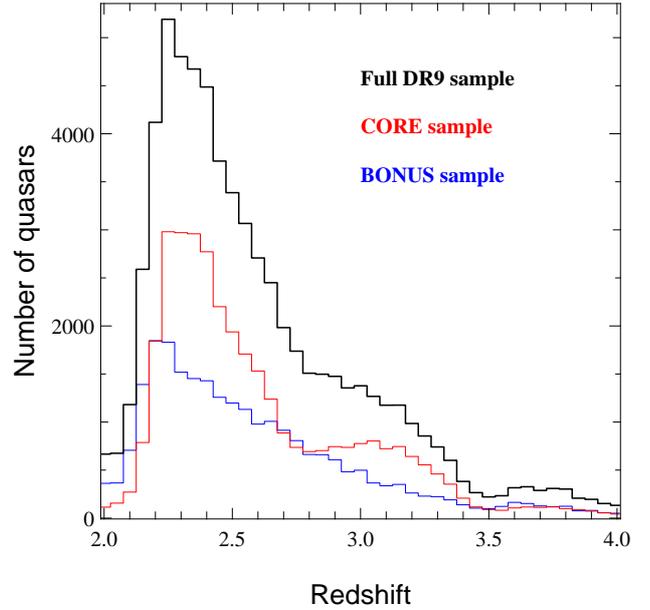}}
	\caption{ 
	Redshift distribution of SDSS-DR9 quasars in the range 2.00-4.00 for the whole distribution (black histogram), 
the CORE sample (red histogram) and the BONUS sample (blue histogram).
	The CORE sample was uniformly selected through the likelihood method \citep{kirkpatrick2011} during most of the first 
year of operation and the XDQSO method \citep{bovy2011} for the second year.
	The BONUS sample was selected through a combination of four target selection algorithms to maximize the number 
of high-redshift quasars in the sample.
	}
	\label{fig:DistributionRedshiftZoom}
\end{figure}

\begin{figure*}[htbp]
		\centering{\includegraphics[angle=-90,width=\linewidth]{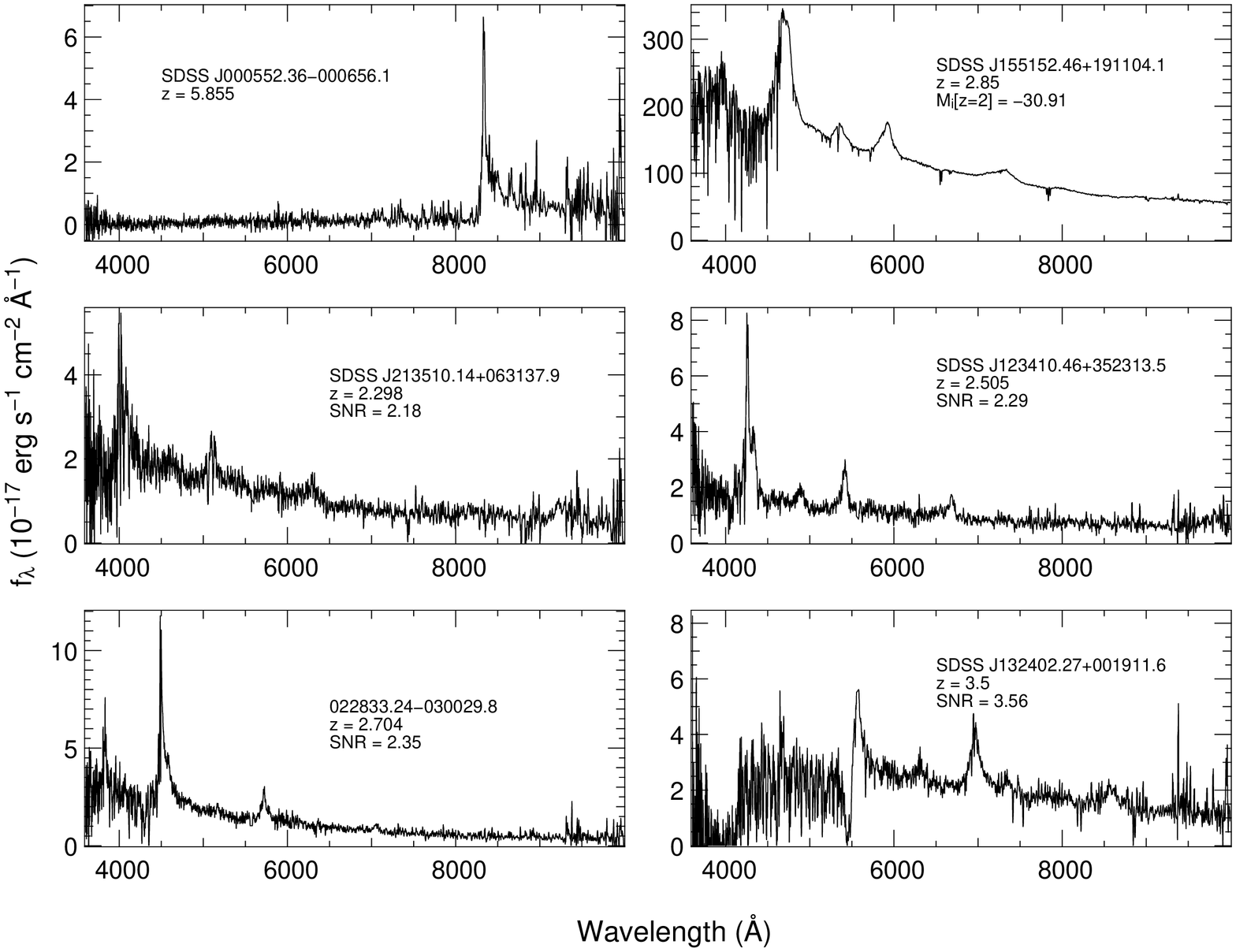}}
	\caption{ 
	{\it First row, left:} spectrum of the highest redshift quasar ($z = 5.855$) observed by BOSS ;
this quasar was discovered by \cite{Fan2004}; the quasar with highest redshift {\it discovered} by BOSS is
SDSS J222018.50$-$010147.0 at $z=5.605$. {\it First row, right:} spectrum of the most luminous ($M_{\rm i}\left[z=2\right] = -30.91$) quasar available 
in this catalog. {\sl Middle and bottom rows:}
	 Four typical quasar spectra selected to be representative 
in terms of SNR at different redshifts ($z\sim 2.3, \ 2.5, \ 2.7, \ 3.5$). The SNR listed is the median SNR per pixel over the whole spectrum.
	 All the spectra were boxcar median smoothed over 5 pixels.
	}
	\label{fig:Examples}
\end{figure*}

\Fig{DistributionRedshiftZoom} displays the redshift distributions in the redshift range of interest for BOSS
for the whole sample (black histogram), the CORE sample (red histogram) and the BONUS sample (blue histogram). 
The CORE sample is selected via uniform target selection
\citep[see details in][]{ross2012}, and is designed for statistical studies of the quasar population
(see Section 7.4).
On the other hand, the BONUS sample is the result of the combination of four target selection algorithms.
This sample was designed to maximize the number of high-redshift quasars.
Typical spectra are shown in \Fig{Examples}.

\Tab{nbQTS} gives the number of objects targeted by the various selection methods
and visually inspected (column \#2)
as described in \cite{ross2012} together with the number of objects classified by  visual 
inspection as quasars (column \#3), 
quasars with $z > 2.15$ (column \#5), stars (column \#6)
or galaxies (column \#7). Column \#8 and \#9 give, respectively, the number of objects with good spectra but 
uncertain identification and the number of objects with data of too low SNR to allow for identification (see Section~3 for 
a detailed description of the different categories). 
Note that a single object can be selected by several methods.

%
\begin{table*}
\centering                          
\begin{tabular}{l c c  c c c c c c}
\hline                      
\hline                
Selection      & Maskbits    & \# Objects   & \# QSO     & \# QSO $z>2.15$ & \# STAR    &  \# GALAXY   & \# ?     & \# BAD\\
\hline
                    & BOSS\_TARGET1 & & & & & & & \\
\hline
QSO\_CORE                 &  10 & 3,468        & 1,3        & 1,084          & 1,975      & 64           & 41       & 11 \\
QSO\_BONUS               & 11  & 4,259        & 803        & 437            & 3,319      & 89           & 30       & 18 \\
QSO\_KNOWN\_MIDZ  & 12    & 9,927        & 9,775   & 9,121            & 36         & 3            & 56       & 57 \\
QSO\_KNOWN\_LOHIZ & 13  & 24           & 24            & 0              & 0          & 0            & 0        & 0\\
QSO\_NN                      & 14 & 72,365       & 45,319   & 34,864         & 25,541     & 569          & 407      & 529\\
QSO\_UKIDSS              &  15 & 48           & 27             & 20             & 19         & 2            & 0        & 0\\
QSO\_KDE\_COADD     & 16 & 1,362        & 305         & 202            & 921        & 56           & 50       & 30\\
QSO\_LIKE                   & 17 & 90,762       & 54,313    & 35,869         & 32,909     & 1,699        & 938      & 903\\
QSO\_FIRST\_BOSS     &  18  & 3,348        & 2,507     & 1,629          & 433        & 142          & 174      & 92\\
QSO\_KDE                    &  19  & 92,203       & 47,564  & 34,252         & 42,248     & 1,021        & 647      & 723\\
QSO\_CORE\_MAIN     &   40 & 67,677       & 41,817   & 32,355         & 23,930     & 799          & 461      & 670\\
QSO\_BONUS\_MAIN   &  41 &  148,085      & 76,660  & 53,000         & 65,320     & 2,931        & 1,467    & 1705\\
QSO\_CORE\_ED         &  42  & 22,715       & 15,019   & 12,387         & 7,055      & 198          & 169      & 274\\
QSO\_CORE\_LIKE       &  43 & 23,951       & 17,635    & 12,522         & 5,591      & 319          & 188      & 218\\
QSO\_KNOWN\_SUPPZ & 44 &  24           & 24              & 0              & 0          & 0            & 0        & 0\\
\hline
                    & ANCILLARY\_TARGET1 & & & & & & & \\
\hline
QSO\_AAL               & 22 & 174          & 172             & 1              & 0          & 1            & 0        & 1\\
QSO\_AALS             & 23 & 281          & 277             & 2              & 0          & 0            & 1        & 3\\
QSO\_IAL                & 24 & 80           & 80                & 0              & 0          & 0            & 0        & 0\\
QSO\_RADIO          & 25 & 72           & 71                & 0              & 0          & 0            & 1        & 0\\
QSO\_RADIO\_AAL & 26 &      58           & 58           & 0              & 0          & 0            & 0        & 0\\
QSO\_RADIO\_IAL  & 27 & 31           & 30                & 0              & 0          & 0            & 0        & 1\\
QSO\_NOAALS        & 28 & 32           & 31                & 0              & 0          & 0            & 1        & 0\\
QSO\_GRI               & 29 & 1,117        & 354            & 343            & 373        & 177          & 26       & 187\\
QSO\_HIZ               & 30 & 335          & 0                  & 0              & 272        & 4            & 3        & 56\\
QSO\_RIZ               & 31 & 728          & 47               & 42             & 545        & 78           & 11       & 47\\
\hline
                    & ANCILLARY\_TARGET2 & & & & & & & \\
\hline
HIZQSO82                    &   0  & 62           & 2            & 2              & 55         & 1            & 0        & 4\\
HIZQSOIR                    &   1  & 28           & 0            & 0              & 25         & 0            & 0        & 3\\
KQSO\_BOSS               &    2  & 183          & 81         & 39             & 89         & 4            & 9        & 0\\
QSO\_VAR                   &    3  & 1,380        & 856     & 296            & 431        & 85           & 5        & 3\\
QSO\_VAR\_FPG          &   4  & 576          & 549       & 263            & 6          & 3            & 14       & 4\\
RADIO\_2LOBE\_QSO  &   5  &  332          & 149      & 15             & 131        & 29           & 7        & 16\\
QSO\_SUPPZ                &    7  & 208          & 208      & 0              & 0          & 0            & 0        & 0\\
QSO\_VAR\_SDSS        &   8  & 1,887        & 568      & 166            & 1,185      & 39           & 35       & 60\\
\hline                                
\end{tabular}
\caption{Number of visually inspected DR9 BOSS quasar targets (third column) and identifications in the DR9Q catalog for each target 
selection method (first column; see Table~4 of  Ross et al. 2012, and Tables~6 and 7 in the Appendix of Dawson et al., 2012). 
These categories overlap because many objects are selected by multiple algorithms.
}
\label{t:nbQTS}      
\end{table*}

%
%

\begin{figure}[htbp]
	\centering{\includegraphics[width=75mm]{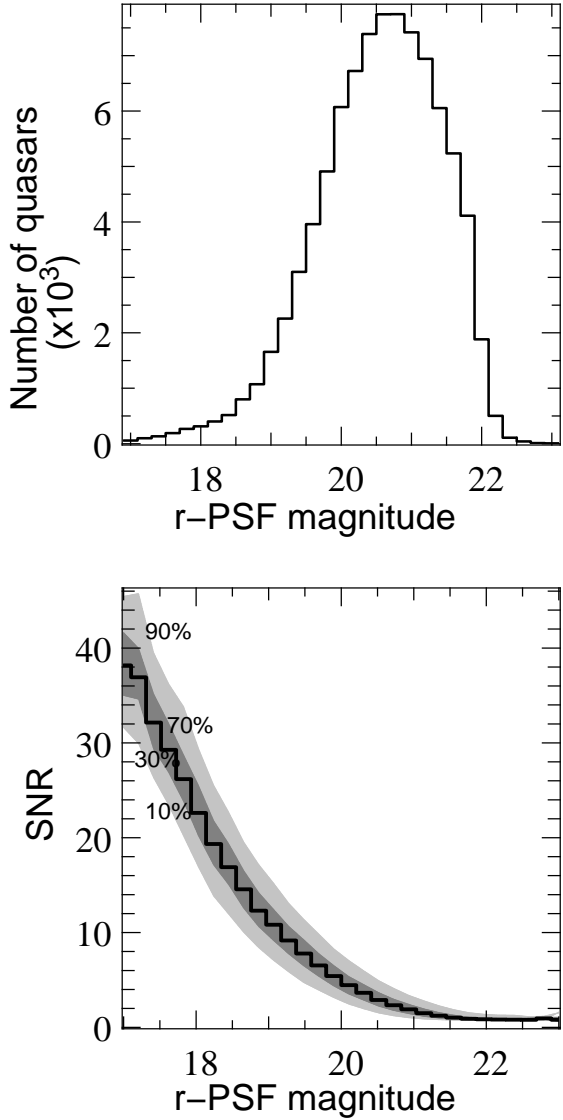}}	
	\caption{ 
	\textit{Top panel: }	
	Distribution of $r$ magnitude of SDSS-DR9 quasars (PSF; corrected for Galactic extinction).
	\textit{Bottom panel: }
	Median SNR per pixel over the whole spectrum with respect to the $r$-PSF magnitude
(black histogram). 
Percentiles are indicated in grey.
	}
	\label{fig:gmag_distribution}
\end{figure}

\begin{figure}[htbp]
	\centering{\includegraphics[width=70mm]{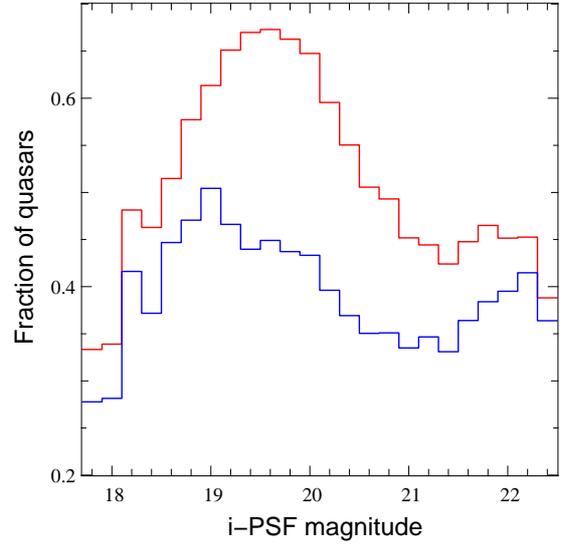}}
	\caption{ 
Fraction of quasar candidates confirmed as quasars (red histogram) and $z > 2.15$ quasars (blue histogram)
versus the $i$-band PSF magnitude (corrected for Galactic extinction).
	}
	\label{fig:SuccessQTS}
\end{figure}

BOSS targets fainter objects than SDSS-I/II.
The $r$-PSF magnitude distribution (corrected for Galactic extinction) of SDSS-DR9 quasars is shown in \Fig{gmag_distribution} 
(top panel), and peaks at $\sim$20.8. The median signal-to-noise ratio computed over 
the whole spectrum versus the  $r$-PSF magnitude is shown  in the 
bottom panel of \Fig{gmag_distribution}. Percentiles are indicated in grey.
The $i$-PSF magnitude distribution of quasar candidates, spectroscopically confirmed quasars and $z > 2.15$ confirmed quasars
are shown in \Fig{SuccessQTS}. There is no  drop of the success rate at high magnitude,
indicating again that the SNR threshold used to define a survey quality plate is well chosen.

%

%
%
\begin{figure}[htbp]
	\centering{\includegraphics[width=75mm]{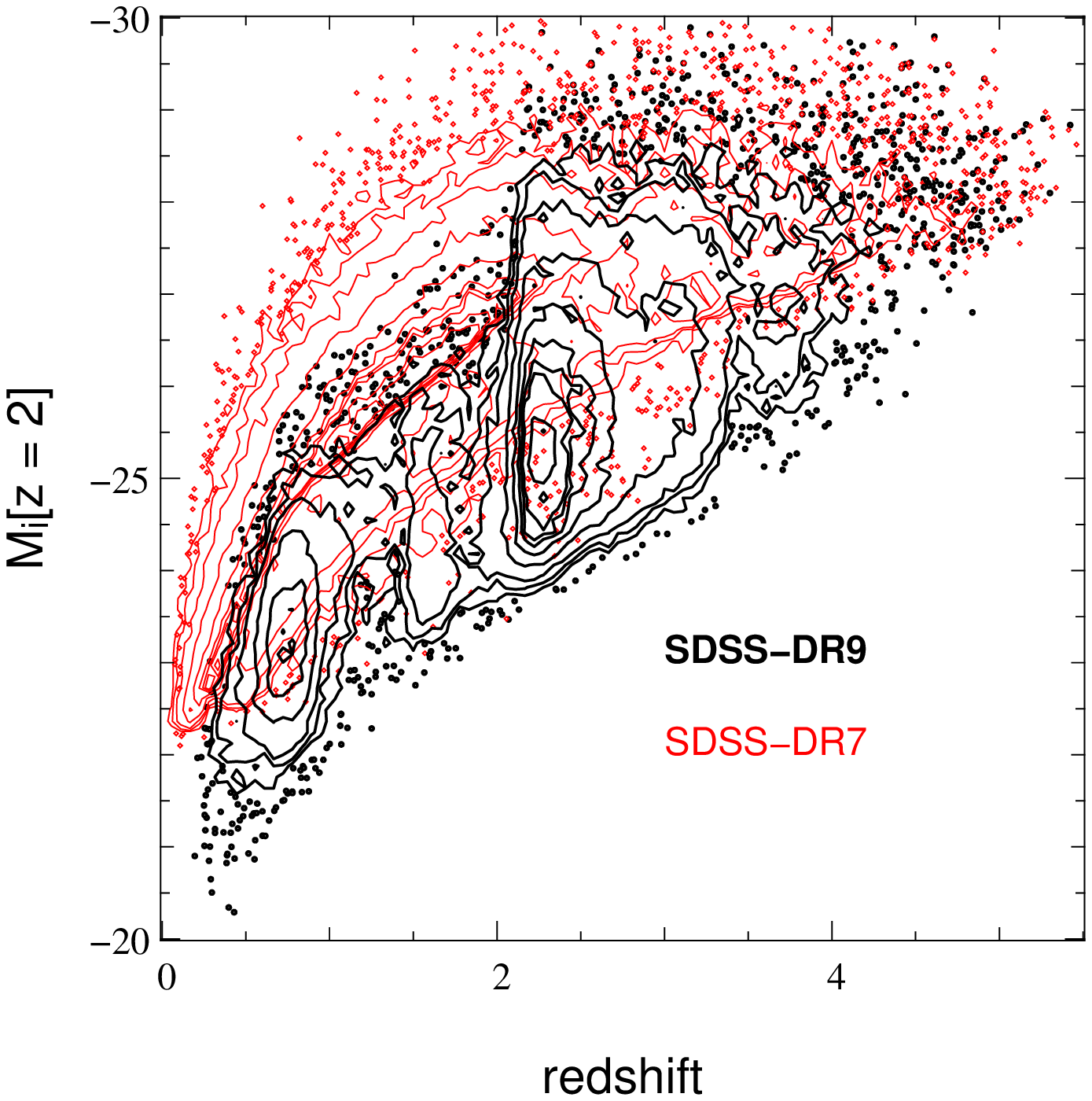}}
	\caption{ 
	$L-z$ plane for SDSS-DR9 quasars (black contours and points) and SDSS-DR7 quasars \citep[red contours and points;][]{schneider2010}.
	The luminosity assumes $H_{0}$ = 70~${\rm km \ s^{-1}\  Mpc^ {-1}}$ and the K-correction is given by \cite{richards2006}
who consider $K(z=2)=0$.
        Contours are drawn at constant point density.
	}
	\label{fig:iMag_vs_z}
\end{figure}
\Fig{iMag_vs_z} shows the distribution of objects in the redshift-luminosity ($L$ versus $z$) plane for 
the BOSS survey (black contours and points) together with the same quantities for the SDSS-DR7 \citep[red contours and points;][]{schneider2010}. 
We calculate the absolute $i$-band (at $z=2$) magnitudes, $M_{\rm i}$, using the observed
$i$-band PSF magnitudes and the K-corrections given in Table~4 of \cite{richards2006}.  
This shows the coverage available for calculating the evolution of the faint end of the quasar luminosity function, 
and for placing constraints on the luminosity dependence of quasar clustering \citep{white2012}.

\begin{figure*}[htbp]
	\centering{\includegraphics[width=140mm]{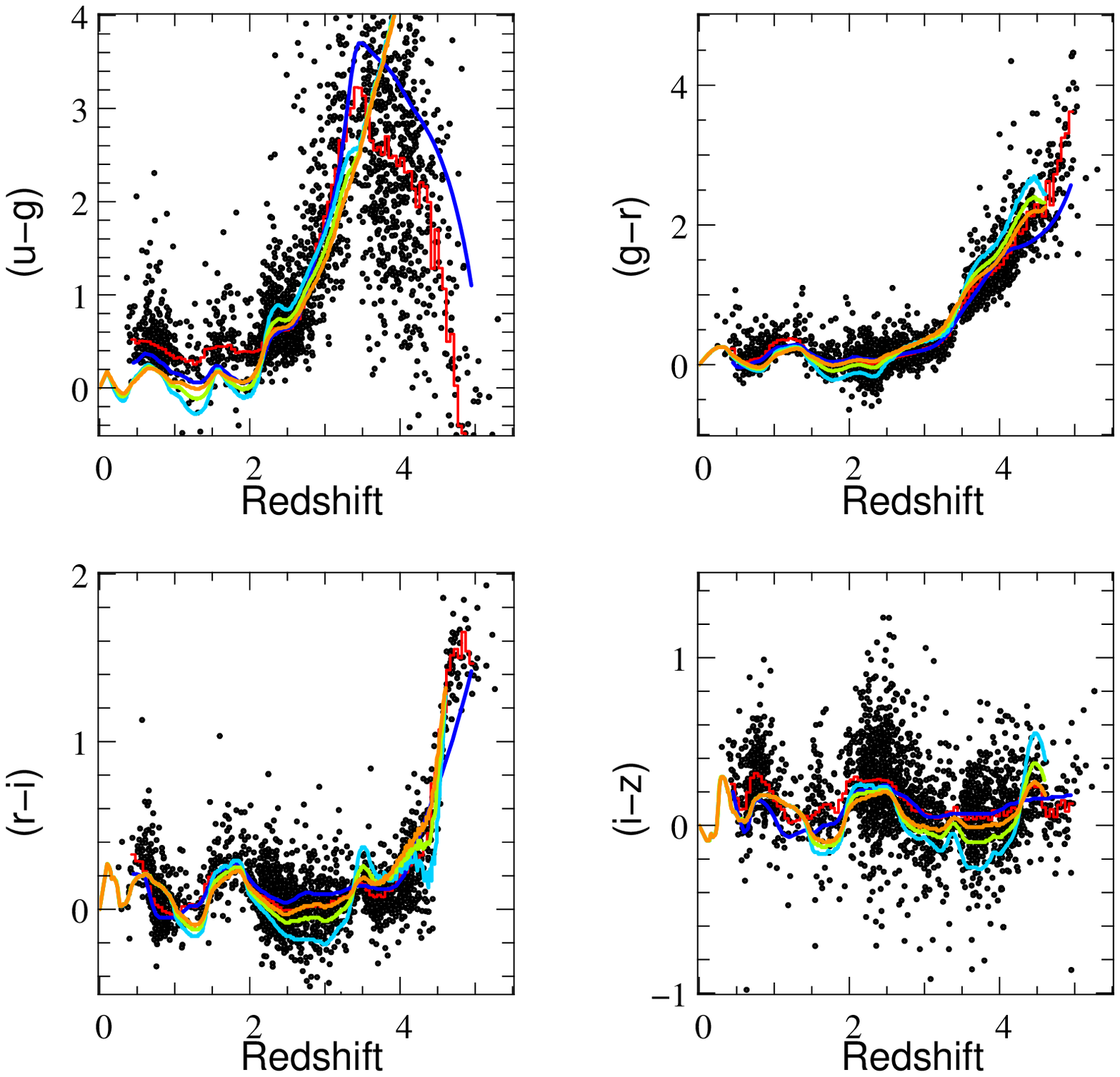}}
	\caption{ SDSS colors vs. redshift for quasars in the DR9Q catalog. 
The thin solid red line is the median color in bins of redshift.
The thick color lines are models from simulations used to determine the BOSS quasar completeness \citep[McGreer et al., in prep.; see also][]{ross2012b} for three different quasar luminosities: $M_i[z=2] = -22.49$ (cyan), $M_i[z=2] = -24.99$ (green) and $M_i[z=2] = -27.49$ (orange); and empirical tracks
for the DR7 quasars \citep[blue][]{bovy2011}.
These simulations include the Baldwin effect \citep{baldwin1977}.
Therefore the colors depend on the quasar luminosity.
The model is systematically bluer than the data
at low redshift because BOSS systematically excludes UV-excess
sources.
	}
	\label{fig:ColorRedshift}
\end{figure*}

\Fig{ColorRedshift} shows the SDSS ($u$-$g$), ($g$-$r$), ($r$-$i$), and ($i$-$z$) colors as a function of redshift for the DR9Q catalog. 
Also shown are the mean color in redshift bins (thin red solid line), and the models described in \citet[thick colored lines]{ross2012b}.
This model is systematically bluer than the data at low redshift; BOSS target selection systematically excludes 
UV-excess quasars.
The trends with redshift are due to various emission lines moving in
and out of the SDSS broadband filters, and the onset of the Lyman-$\alpha$ forest and Lyman-limit systems 
\citep[e.g.,][]{fan1999,hennawi2010,richards2002,richards2003,Bovy2012,peth2011};  see also \cite{Prochaskaal09}
and \cite{worseck10} for biases in the SDSS target selection.

\begin{figure*}[htbp]
	\centering{\includegraphics[width=140mm]{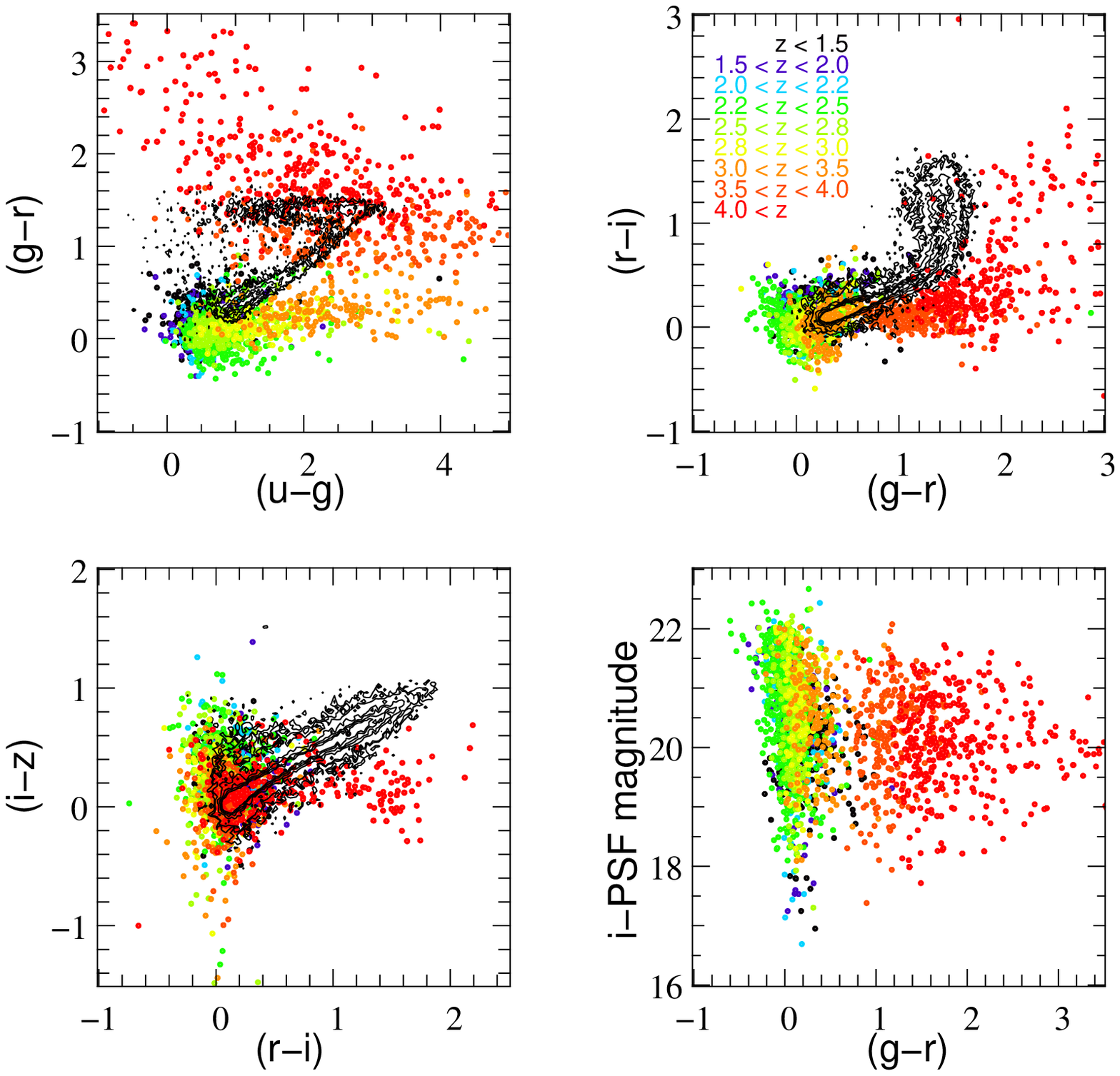}}
	\caption{ 
Color-color diagrams for all quasars in the DR9Q catalog. Colors of points encode their redshifts
(see top right panel).
	The stellar locus is represented with black contours.
	}
	\label{fig:ColorColor}
\end{figure*}

\Fig{ColorColor} shows the SDSS color-color diagrams for the quasars in the
DR9Q  catalog. This figure illustrates the redshift dependence of quasar colors \citep[see also \Fig{ColorRedshift}; ][]{fan1999}.
The quasars at $z\sim 2.7$ are located in the stellar locus (black contours).
%

%

\subsection{Spectral index and composite spectra}
\label{s:Spectralindex}

The quasar continuum can be expressed as $f_{\rm cont} \propto \nu_{{\rm rest}}^{\alpha_{\nu}}$, 
where $\alpha_{\nu}$ is the spectral index. 
This index is obtained by fitting a power law over wavelength
ranges outside the Lyman-$\alpha$ forest and devoid of strong emission lines.
The regions of the fits are 1450-1500, 1700-1850 and 1950-2750~\AA~ in the rest frame.
The continuum is iteratively fitted to remove absorption lines and to limit the impact 
of the iron emission blends on the $\alpha_{\nu}$ measurement.

\begin{figure}[htbp]
	\centering{\includegraphics[width=.8\linewidth]{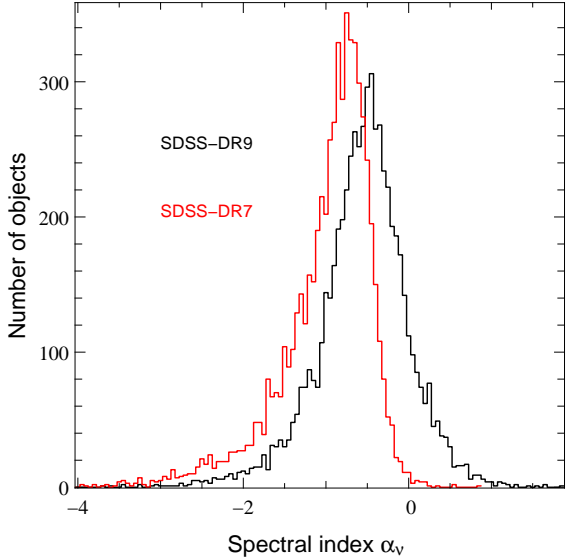}}
	\caption{ 
	Distribution of the spectral index $\alpha _{\nu}$ of $z > 2$ of SDSS-DR7 quasars (red histogram) re-observed by BOSS (black histogram).
	The spectral index was measured using the rest frame wavelength ranges 1450-1500, 1700-1850 and 1950-2750 \AA .
	As seen already from the composite spectrum shown in \Fig{Composite}, the spectral indices measured using SDSS-DR9 
quasar spectra are bluer than those obtained using DR7 spectra (see Section~7.2).
	}
	\label{fig:distri_al}
\end{figure}

The distribution of the quasar spectral index of SDSS-DR7 quasars re-observed by BOSS is shown in \Fig{distri_al}.
The median spectral index measured for BOSS spectra (black histogram) is $\alpha_{\nu , DR9} = -0.517$
while the median value measured with SDSS-DR7 spectra is $\alpha_{\nu , DR7} = -0.862$.
This discrepancy is mainly the consequence of 
the inaccuracy of the BOSS flux calibration in the blue (see \Fig{Composite}).
This may explain as well the fact that the distribution is more symmetric than 
previously measured \citep[e.g.][]{richards2003} lacking the red tail. 
Therefore the reader should be careful of this measurement using BOSS quasar spectra.

\begin{figure}[htbp]
		\centering{\includegraphics[angle=-90,width=\linewidth]{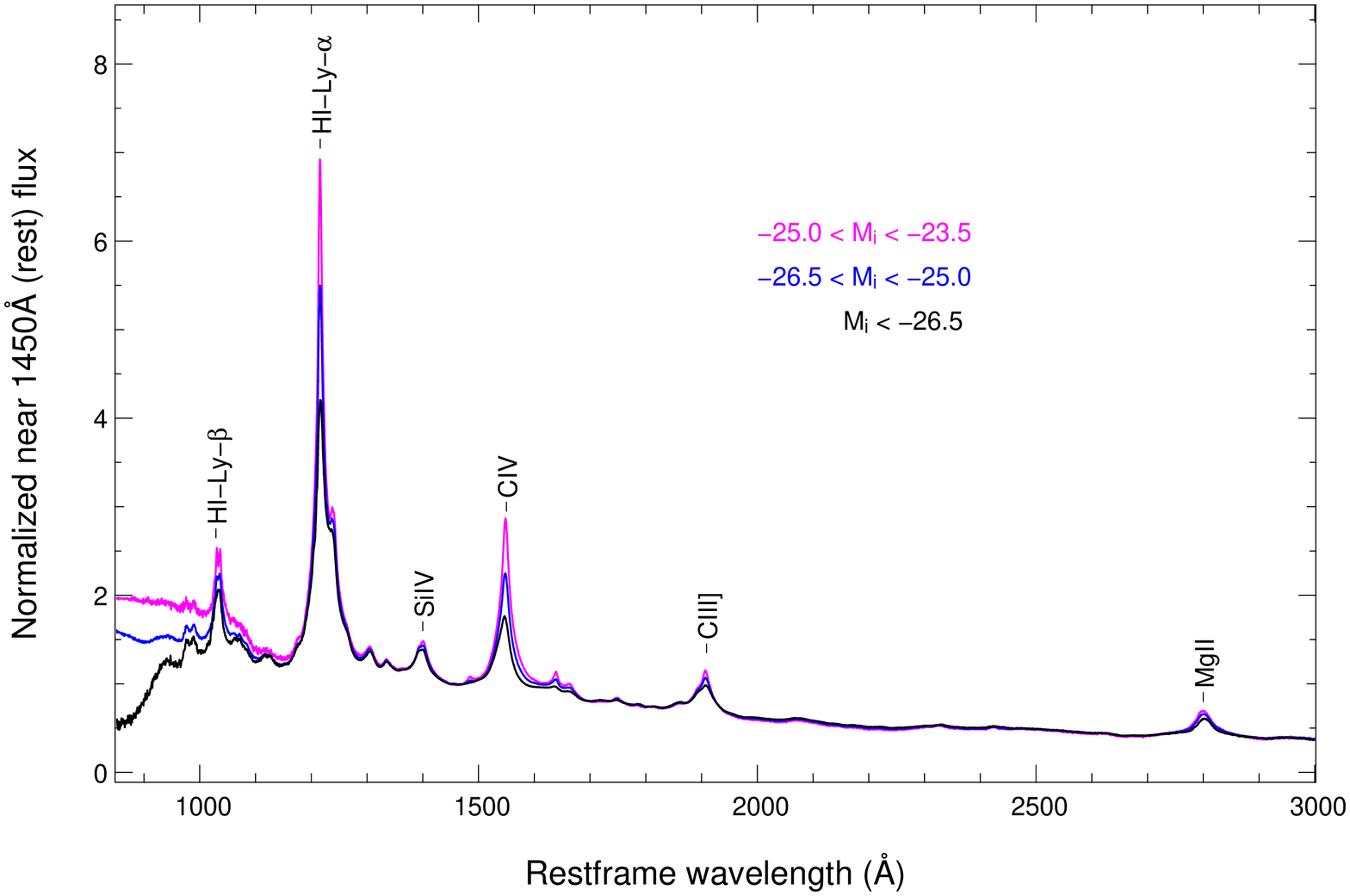}}
	\caption{ 
Composite spectra of BOSS quasars in different ranges of absolute 
magnitude: $-25.0 < M_{\rm i} < -23.5$ (magenta), $-26.5 < M_{\rm i} < -25.0$ (blue) and $M_{\rm i} < -26.5$ (black).
	All the spectra were normalized to have a flux unity near 1450~\AA\ in the quasar rest frame. The
Baldwin effect is apparent (see Table~6).
	}
\label{fig:Compositebis}
\end{figure}

Although the absolute flux calibration is in error,
it is interesting to compare the composite spectra in different absolute magnitude
bins. They are displayed in \Fig{Compositebis} for the absolute magnitude bins $-25.0 < M_{\rm i} < -23.5$ (magenta), 
$-26.5 < M_{\rm i} < -25.0$ (blue) and $M_{\rm i} < -26.5$ (black).
The equivalent widths of the emission lines decreases with increasing luminosity.
This is the well-known Baldwin effect \citep{baldwin1977}. The 
rest equivalent widths of the most important equivalent emission lines
is given for different absolute magnitude bins in \Tab{EWcomposite}.

\begin{table}
\centering                         
\begin{tabular}{l c c c c}     
\hline\hline              
${\rm M_i \left[z=2 \right]}$ & \multicolumn{4}{c}{Restframe equivalent width (\AA )} \\
       & \SiIV\ & \CIV\ & \CIII\ & \MgII\ \\
\hline                       
$ -25.0 < {\rm M_i} < -23.5$ & 10.6  &  65.8  & 31.3 & 44.2  \\
$ -26.5 < {\rm M_i} < -25.0$ &   9.5  &  48.6  & 27.4 & 37.0 \\
${\rm M_i} < -26.5              $ &    8.3 &  34.0  & 23.4 & 29.8 \\
\hline                                  
\end{tabular}
\caption{Rest frame equivalent widths measured on the composite spectra displayed in \Fig{Compositebis}}.
\label{t:EWcomposite}
\end{table}

\subsection{Rest equivalent widths in individual spectra}
\label{s:Eq_width}
As explained in Section~4.5, 
we used five PCA components to fit the emission lines and
derive their redshift. We used these fits to measure also the rest equivalent width and
widths (FWHM and half widths at half maximum) of the emission lines.
The continuum is fitted as a power law to the best PCA component fit 
over the windows 1,450-1,470~\AA~ and 1,650-1,820~\AA.
We modified the windows used by Shen et al. (2011) (1,445-1,465~\AA~and
1,700-1,705~\AA)  to minimize the fraction of bad fits, especially for emission lines narrower
than the mean.
 
\begin{figure}[htbp]
	\centering{\includegraphics[width=75mm]{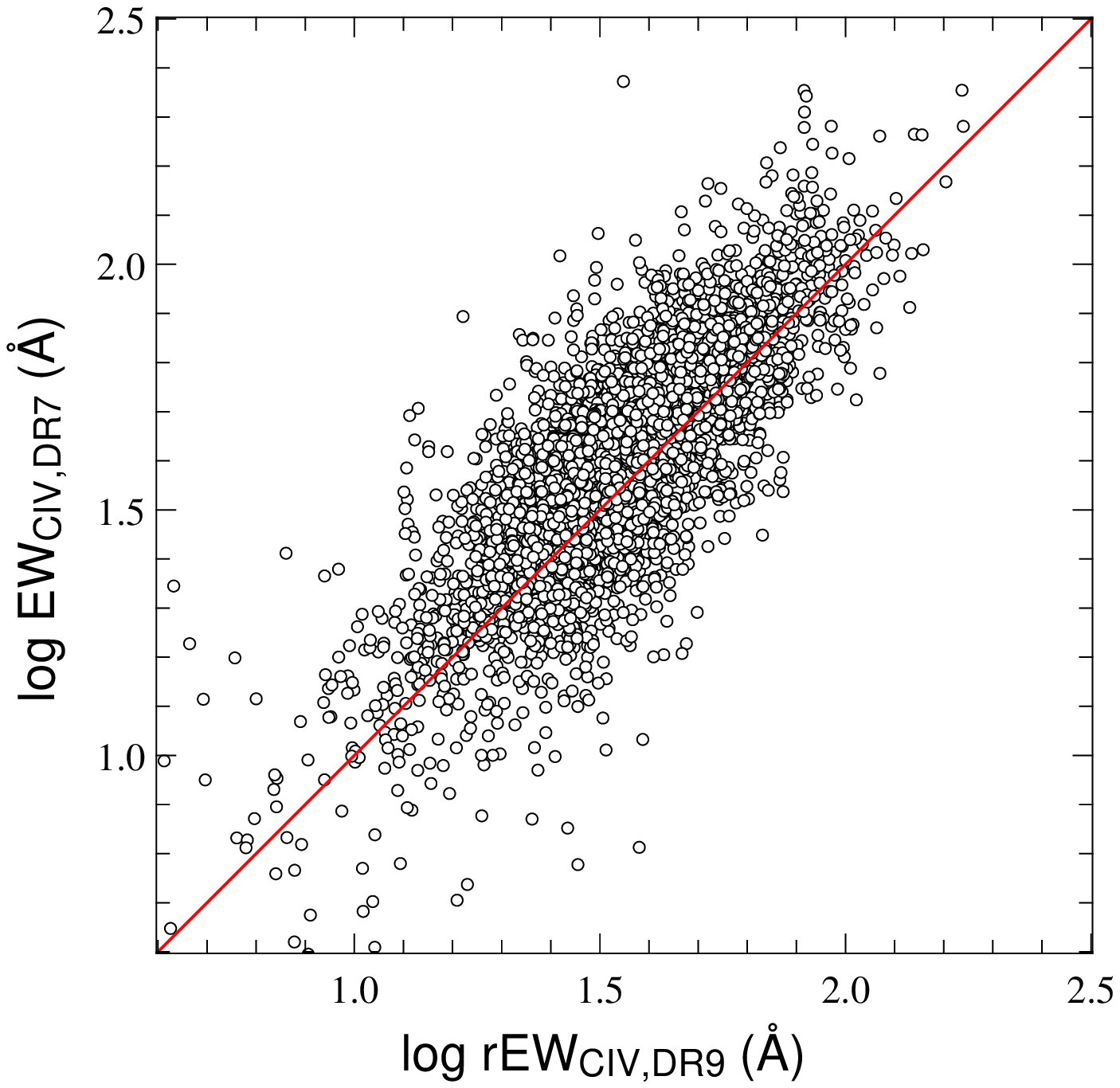}}
	\caption{
	 Rest frame equivalent width of the \CIV\ emission line of SDSS-DR7 quasars,
	measured from the SDSS-DR7 spectra \citep{shen2011} and spectra obtained  by BOSS (this work).
	}
	\label{fig:CIV_rEW}
\end{figure}

\Fig{CIV_rEW} shows the comparison between the rest equivalent width measured
on SDSS\_DR7 spectra by \cite{shen2011} and that measured on BOSS spectra of the
same quasars. Our rest equivalent widths are about 10~\% smaller on average. 
This systematic shift is likely related to a difference in the rest frame wavelength range used to compute the rest frame equivalent width.
While we strictly limited the equivalent width computation to the 1,500-1,600 \AA\ range, Shen et al. (2011) used this range to fit the line but accounted for the extra wings to estimate the rest frame equivalent width. 
The rms scatter is about 33~\%.   
We checked by hand some of the largest discrepancies and found that our
procedure seems to behave well.
We applied our procedure to both SDSS-DR7 data and BOSS data from the same
quasars. The mean difference is 4\% and rms 25\%. 
Statistical errors are of the order of 15\%, 
and variability can account for another 15\%
\citep[see e.g.][]{Bentz2009,Willhite2006}. BOSS spectra are also of better quality.

The rest equivalent widths are listed in the catalog for C~{\sc iv}, the C~{\sc iii}] complex and Mg~{\sc ii}.
A value of $-1$ indicates that the PCA failed to fit the emission line.
The variance was computed as the integral over the width of the line of the variance in each pixel.
Note however that errors are mostly due to the position of the continuum.

%

%

\subsection{Uniform sample}
\label{s:Uniformflag}
We provide a similar {\tt uniform} flag in our catalog to previous versions of the SDSS quasar catalogs 
(e.g., Schneider et al. 2007). 
Quasars in our catalog with {\tt uniform = 1} are CORE targets that were 
selected with the XDQSO technique (Bovy et al. 2011) {\em after} XDQSO became the CORE targeting algorithm 
of choice for BOSS \citep[e.g., in or after Chunk 12;][]{ross2012}. XDQSO will remain the BOSS quasar target 
algorithm for the rest of the survey, so this {\tt uniform = 1} sample will grow significantly in subsequent releases.

Quasars with {\tt uniform = 2} would have been selected by XDQSO {\em if} it had been the CORE algorithm prior 
to Chunk 12. {\tt uniform = 2} objects are quite complete to what XDQSO would have selected \citep[e.g.,][]{ross2012}, 
so {\tt uniform > 0} is a sufficiently statistical sample for, e.g., clustering measurements on some scales 
\citep[e.g.,][]{white2012}. Quasars in our catalog with {\tt uniform = 0} are not homogeneously selected CORE targets. 
Finally, the very few (30) quasars in our catalog with {\tt uniform = -1} have no chunk information, but are a sufficiently 
small sample to be discarded for the purposes of statistical analyses.

\subsection{Multiwavelength matching}
\subsubsection{ROSAT all sky survey}
\label{s:ROSAT}
We cross-correlate the DR9Q catalog with the ROSAT all sky survey 
catalogues listing the sources detected in the energy band 0.1$-$2.4~keV.
The matching radius is set to 30$"$.

We report the logarithm of the vignetting-corrected count rate (photons s$^{-1}$)
from the {\it ROSAT} All-Sky Survey Faint Source Catalog \citep{voges2000} and the
{\it ROSAT} All-Sky Survey Bright Source Catalog \citep{voges1999}.
An entry of~"$-9.000$" in the column RASS\_COUNTS indicates no X-ray detection.
We also report the SNR at the position of the quasar and the separation between the
quasar and the X-ray source.

There are 16 matches with the Bright Source Catalog and 298 with the Faint Source Catalog.
It never happened to find more than one source within the matching radius.
No DR9 quasar is detected both in the Bright and Faint Source catalogues.
Only the most reliable detections were included in our catalog: X-ray counterparts for 13 quasars 
were flagged for possible detection quality issues and therefore are not included in the present quasar catalog.

%
\subsubsection{The Galaxy Evolution Explorer (GALEX)}
\label{s:GALEX}
The GALEX space mission \citep{martin2005} has performed an all-sky imaging survey in two UV bands 
(FUV: 1350 to 1750 \AA~; NUV: 1750 to 2750 \AA)
down to $m_{\rm AB} \sim 20.5$ and a medium-deep imaging survey that
reaches $m_{\rm AB} \sim 23$ (e.g.,  \cite{Bianchi2011}). Both surveys are used here.
 
GALEX  images are force photometering
GALEX images (from GALEX Data Release 5) at the SDSS-DR8 centroids (Aihara et al. 2011), such that 
low signal-to-noise point-spread function (PSF) fluxes of objects
not detected by GALEX is obtained.

A total of 77,236 quasars lie in the GALEX FUV footprint, 78,062 lie in
the NUV footprint, and 77,197 are covered by both bandpasses.

\subsubsection{The two micron all sky survey (2MASS)}
\label{s:2MASS}

We cross-correlate the DR9Q catalog with the All-Sky Data
Release Point Source Vatalog (Cutri et al., 2003) using a matching radius
of 2.0''.  

Together with the Vega magnitudes in the J, H and K-bands (xMAG with x~=~J, H or K) 
and their errors (ERR\_xMAG), we report
the SNR (xSNR) since the errors on the magnitude
do not differentiate the 2$\sigma$ upper limits (in a 4'' radius aperture) from detections. 
We also give for each band the value of the 2MASS flag rd\_flg[1] (entry xRDFLAG)  
which gives the meaning of the peculiar values of xMAG and ERR\_xMAG
(see http://www.ipac.caltech.edu/2mass/releases/- allsky/doc/explsup.html)
 
 There are 1,441 matches in the catalog.

\subsubsection{The Wide-Field Infrared Survey (WISE)}
\label{s:WISE}

        We take the DR9Q catalog, and match to the 
        Wide-Field Infrared Survey \citep[WISE; ][]{wright2010} All-Sky Data
        Release\footnote{http://wise2.ipac.caltech.edu/docs/release/allsky/}, 
        asking for all quasars in the DR9Q catalog that are in the All-Sky Source Catalog. 
        WISE photometry covers four bands, 3.4, 4.6, 12 and 22~$\mu$m, where the
        angular resolution of WISE is $\approx$6'' for 3.4, 4.6, 12~$\mu$m, and
        $\approx$12'' for 22~$\mu$m \citep{wright2010}. 
        After testing for various matching radii (1'', 2'', 6'', 12'', 18''), we use a 
        matching radius of 2.0'', and a total of 45,987 rows
        of WISE photometry data are returned, along with the separation in 
        arcseconds between the SDSS and WISE source (stored in SDSSWISE\_SEP). 

        In the DR9Q catalog, we report the photometric quantities, w$x$mpro,  w$x$sigmpro,  
        w$x$snr,  w$x$rchi2, where $x={1-4}$ and represents the four WISE bands
        centered at wavelengths of 3.4, 4.6, 12 and 22~$\mu$m. These magnitudes 
        are in the Vega system, and are measured with profile-fitting photometry
        (see e.g. http://wise2.ipac.caltech.edu/docs/release/- allsky/expsup/sec2\_2a.html
        and http://wise2.ipac.caltech.edu/-  docs/release/allsky/expsup/sec4\_4c.html\#wpro).         
        
        Formulae for converting WISE Vega magnitudes to flux density units (in Janskys) 
        and AB magnitudes are given in \citet{wright2010} and \citet{Jarret2011} and also here:
        http://wise2.ipac.caltech.edu/docs/release/- allsky/expsup/sec4\_4h.html\#conv2flux        

        Although the MIR WISE properties of the BOSS quasars will be 
        valuable for many scientific questions, we {\it strongly} urge the
        user to not only consider the various ``health warnings'' associated
        with using the BOSS quasar dataset (as given in Section 2) but also those 
        connected to the WISE All-Sky Release Data
Products\footnote{http://wise2.ipac.caltech.edu/docs/release/allsky/expsup/sec1\_4.html}.
                
        We do not investigate any of the 2MASS, or UKIDSS properties associated with WISE
matches here, but these are being investigated in \cite{ross2012b}. 

\subsubsection{FIRST}
\label{s:FIRST}
We cross-correlate the DR9Q catalog with objects that are detected in the
FIRST radio survey (Becker et al. 1995). We use the version of July 2008.

 If there is a source in the FIRST catalog within~2.0$''$ 
of the quasar position, we indicate the FIRST peak flux density and the SNR.
Note that extended radio sources may be missed by this matching.

Note that, as in SDSS-I/II, FIRST sources are automatically included in the
target selection. An additional cut in color ($u$-$g$~$>$~0.4) is added to avoid
as much as possible low-redshift sources \citep{ross2012}. 
The catalog contains 3,283 FIRST matches.

%
\section{Additional quasars}
\label{s:VAC6}

\begin{figure}[htbp]
	\centering{\includegraphics[width=75mm]{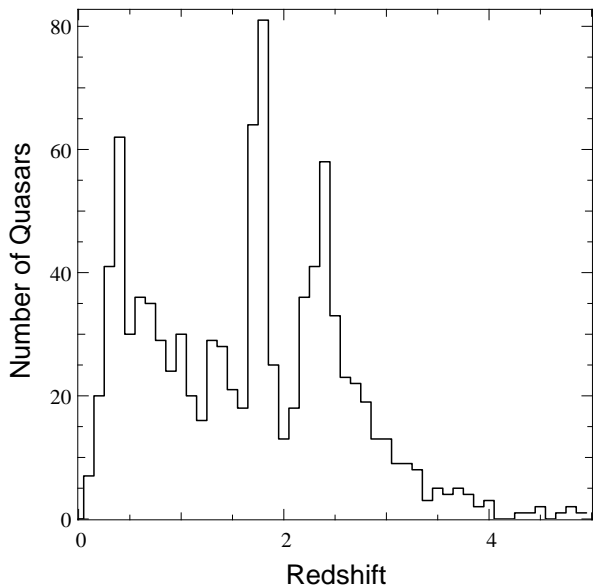}}
	\caption{ 
	Redshift distribution of the 949 additional quasars described in \Sec{VAC6}.
	}
	\label{fig:RedshiftSupList}
\end{figure}

We provide a supplemental list of 949 quasars, of which 318 at $z>2.15$,  that have been identified 
among quasar targets after DR9 was ``frozen" (\Sec{AQT}) or among galaxy targets (\Sec{GT}).
This supplemental list of quasars is provided in the same format as the DR9Q catalog 
but in a separate file and  is meant to be merged with the whole catalog for DR10.
\Fig{RedshiftSupList} gives the redshift distribution of these additional quasars.
The list is available together with the DR9Q catalog and the list of objects classified
as {\tt QSO\_?} at the
SDSS public website http://www.sdss3.org/dr9/algorithms/qso\_catalog.php.

\subsection{Additional quasars from the quasar target list}
\label{s:AQT}
The quasar catalog was frozen\footnote{By which we mean 
no additional quasar or change in the identifications
were intended to be included in the catalog} in February 2012, but we subsequently identified an additional
301 quasars (294 with $z>2.15$) that have been targeted as quasar candidates.
Some of these were identified with improvements of the pipeline. Others
are identified from a good spectrum taken on a plate which was not survey quality so was not
included in the first inspection.
A handful are objects that have been misidentified during the first inspection but were corrected during 
the checks. 

In addition, a few ancillary programs 
were not included in the first inspection. Therefore the supplemental list contains the quasars  that have been targeted
only by these programs. These comprise 500 quasars of which 20 have $z>2.15$.

Finally, only objects classified as {\tt QSO} or {\tt QSO\_BAL} are listed  in the official DR9Q catalog.
Objects classified as  {\tt QSO\_Z?} (126 in total; 122 corresponding to the DR9Q inspection) are also included in the supplemental list. 
Most of the latter are very peculiar BAL quasars.

\subsection{Galaxy targets}
\label{s:GT}

In order to be as complete as possible we also tried to identify serendipitous quasars.
For this, we visually inspected all objects from the BOSS galaxy target list that the pipeline 
reliably classifies ({\tt ZWARNING}~=~0) as quasars with $z>2$, 
and all objects classified as {\small\tt GALAXY/BROADLINE}. 
We also visually inspected 10\% of the galaxy targets classified as quasars with {\tt ZWARNING}
not equal to zero; none were in fact quasars, so we did not inspect the remaining such objects.
A large fraction (65\%) of the unclassifiable objects are attributed to the {\tt QSO} class, but with low significance. 
They include all sorts of unidentified objects and spectra with 
calibration problems but probably very few real quasars, if any. 
In total we identified 22 additional quasars, 4 of which were at $z>2.15$, out of 
more than 3,000 targets. There is one quasar classified as {\tt QSO\_Z?}.
%

%

%
\section{Conclusion}
\label{s:Summary}
The quasar catalog presented here contains 87,822 quasars,  61,931 having $z>2.15$, with robust identification 
from visual inspection and redshift derived from the fit of PCA components to the spectra.  
This catalog has been gathered during the first two years of the BOSS  operation covering 3,275~deg$^2$.
It will be the basis for studies of the luminosity function and the spatial
distribution of quasars as well as studies of the clustering properties of the Lyman-$\alpha$ forest.
In particular it will be used to measure for the first time the BAO clustering signal in the IGM 
at $z\sim 2.3$ from the Lyman-$\alpha$ forest.

For quasars with $z_{\rm em}>1.57$, 
the catalog  identifies 7,228 broad absorption line quasars from visual inspection, of which
3,130 have BI~$>$~ 500~km~s$^{-1}$. 
In the 7,317 spectra with SNR~$>$~10, we find 813 BALs with BI~$>$~500~km~s$^{-1}$
corresponding to a fraction of 11.1\%.  We implement a procedure 
to identify BALs automatically, fitting the quasar continuum with PCAs.
3,330 BALs with BI~$>$~500~km~s$^{-1}$ have been identified in this way, of which
821 in spectra of SNR~$>$~10.
The catalog gives their characteristics, balnicity and absorption indices.  
The list of BALs will be used for statistical analysis of this population
of quasars. Since SDSS-DR7 $z>2.15$ quasars are reobserved by BOSS, 
this will be a unique opportunity to study the variability of these troughs.
High redshift ($z>2$) quasar continua together with pixel masks, improved noise estimates,  and other products designed to aid in the 
BAO-Lyman-$\alpha$ clustering analysis will be released in \cite{lee2012}.

BOSS is a five year program and the next version of our quasar catalog, to be released as a part of  
SDSS-DR10  in July 2013, should contain about two times as many
quasars as the DR9Q catalog. Improvements in the pipeline will allow us to achieve 
identification of more objects. We will also perform multiple checks and  improve our 
procedures in order to place better constraints on the characterisitcs of the quasar spectra.

\begin{acknowledgements}
	  I.P. received partial support from Center of Excellence in Astrophysics and Associated Technologies (PFB 06).
      The French Participation Group to SDSS-III was supported by the Agence Nationale de la
Recherche under contract ANR-08-BLAN-0222.  
W.N.B. and N.F.-A. gratefully acknowledge support from NSF AST-1108604.
A.D.M. is a research fellow of the Alexander von Humboldt Foundation of Germany.

      Funding for SDSS-III has been provided by the Alfred P. Sloan
Foundation, the Participating Institutions, the National Science
Foundation, and the U.S. Department of Energy Office of Science.
The SDSS-III web site is http://www.sdss3.org/.

SDSS-III is managed by the Astrophysical Research Consortium for the
Participating Institutions of the SDSS-III Collaboration including the
University of Arizona,
the Brazilian Participation Group,
Brookhaven National Laboratory,
University of Cambridge,
Carnegie Mellon University,
University of Florida,
the French Participation Group,
the German Participation Group,
Harvard University,
the Instituto de Astrofisica de Canarias,
the Michigan State/Notre Dame/JINA Participation Group,
Johns Hopkins University,
Lawrence Berkeley National Laboratory,
Max Planck Institute for Astrophysics,
Max Planck Institute for Extraterrestrial Physics,
New Mexico State University,
New York University,
Ohio State University,
Pennsylvania State University,
University of Portsmouth,
Princeton University,
the Spanish Participation Group,
University of Tokyo,
University of Utah,
Vanderbilt University,
University of Virginia,
University of Washington,
and Yale University.

\end{acknowledgements}

\bibliographystyle{aa}
\bibliography{DR9Q}

\longtab{4}{
\begin{longtable}{clcl}
\caption{\label{t:CatalogFormat} DR9Q catalog format}\\
\hline\hline
Column & Name & Format & Description$^a$ \\
\hline
\endfirsthead
\caption{continued.}\\
\hline\hline
Column & Name  &  Format & Description \\
\hline
\endhead
\hline
\endfoot
%
1    & SDSS\_NAME                      &  A19      & SDSS-DR9 designation  hhmmss.ss+ddmmss.s (J2000)\\
2    & RA                                      &  F11.6   & Right Ascension in decimal degrees (J2000)\\
3    & DEC                                    & F11.6    & Declination in decimal degrees (J2000)\\
4    & THING\_ID                           &  I10       & Thing\_ID \\
5    & PLATE                                 & I5         & Spectroscopic Plate number \\
6    & MJD                                   & I6         & Spectroscopic MJD \\
7    & FIBERID                             & I5         & Spectroscopic Fiber number \\
\hline
%
%
8    & Z\_VI                                   &  F9.4    & Redshift from visual inspection \\
9    & Z\_PIPE                               &  F9.4   & Redshift from BOSS pipeline \\
10    & ERR\_ZPIPE                      &  F9.4     & Error on BOSS pipeline redshift \\
11   & ZWARNING                        & I4          & ZWARNING flag  \\
12  & Z\_PCA                               & F9.4      & Refined PCA redshift \\
13  & ERR\_ZPCA                        & F9.4       & Error on refined PCA redshift \\
14  & PCA\_QUAL                        & F9.4      & Estimator of the PCA continuum quality\\
15   & Z\_CIV                               & F9.4       & Redshift of \CIV\ emission line \\
16   & Z\_CIII                              & F9.4       & Redshift of \CIII\ emission complex \\
17   & Z\_MGII                             & F9.4      & Redshift of \MgII\ emission line\\
\hline
%
%
18   & SDSS\_MORPHO                 &   I2     & SDSS morphology flag 0 = point source 1 = extended \\
19   & BOSS\_TARGET1                 &  I20    & BOSS target flag for main survey  \\
20   & ANCILLARY\_TARGET1       &   I20   & BOSS target flag for ancillary programs \\
21   & ANCILLARY\_TARGET2       &   I20   & BOSS target flag for ancillary programs  \\
22   &SDSS\_DR7                           & I2      & 1 if the quasar is known from DR7\\ 
23  & PLATE\_DR7                         & I5       & SDSS-DR7 spectroscopic Plate number if the quasar is known from DR7 \\
24  & MJD\_DR7                           &  I6       & SDSS-DR7 spectroscopic MJD  if the quasar is known from DR7\\
25  & FIBERID\_DR7                     & I4       & SDSS-FR7 spectroscopic Fiber number  if the quasar is known from DR7\\
26   & UNIFORM                           & I2        & Uniform sample flag \\
27   & MI                                      &  F9.4   & $M_{\rm i}\left[{\rm z = 2} \right] \left( H_0 = 70 {\rm km \ s^{-1} \ Mpc^{-1}}, \ \Omega _M = 0.3, \ \Omega _{\Lambda} = 0.7, \ \alpha _{\nu} = -0.5 \right)$ \\
28   & DGMI                                 &  F9.4   & $\Delta (g-i) = (g-i) - \langle (g-i) \rangle _{\rm redshift}$ (Galactic extinction corrected) \\
29   & ALPHA\_NU                       &  F9.4    & Spectral index measurement $\alpha _{\nu}$ \\
\hline
%
%
30   & SNR\_SPEC                      &    F9.4    & Median signal-to-noise ratio over the whole spectrum\\
31   & SNR\_1700                      &    F9.4    & Median signal-to-noise ratio in the window 1,650 - 1,750\AA\ (rest frame)\\
32   & SNR\_3000                      &    F9.4    & Median signal-to-noise ratio in the window 2,950 - 3,050\AA\ (rest frame)\\
33   & SNR\_5150                      &    F9.4    & Median signal-to-noise ratio in the window 5,100 - 5,250\AA\ (rest frame)\\
34   & FWHM\_CIV                     &   F9.4     & FWHM of \CIV\ emission line in ${\rm km \ s^{-1}}$ \\
35   & BHWHM\_CIV                  &    F9.4     & Blue HWHM of \CIV\ emission line in ${\rm km \ s^{-1}}$ \\
36   & RHWHM\_CIV                  &    F9.4   & Red HWHM of \CIV\ emission line in ${\rm km \ s^{-1}}$ \\
37   & AMP\_CIV                        &   F9.4      & Amplitude of \CIV\ emission line in units of median rms pixel noise \\
38   & REWE\_CIV                     &   F9.4     & Rest frame equivalent width of \CIV\ emission line in \AA \\
39   & ERR\_REWE\_CIV           &  F9.4     & Uncertainty on the rest frame equivalent width of \CIV\ emission line in \AA \\
40   & FWHM\_CIII                    &     F9.4       & FWHM of \CIII\ emission complex in ${\rm km \ s^{-1}}$ \\
41   & BHWHM\_CIII                 &     F9.4       & Blue HWHM of \CIII\ emission line in ${\rm km \ s^{-1}}$ \\
42   & RHWHM\_CIII                 &     F9.4       & Red HWHM of \CIII\ emission line in ${\rm km \ s^{-1}}$ \\
43   & AMP\_CIII                       &     F9.4       & Amplitude of \CIII\ emission complex in units of median rms pixel noise\\
44   & REWE\_CIII                    &      F9.4       & Rest frame equivalent width of \CIII\ emission line in \AA \\
45  & ERR\_REWE\_CIII           &  F9.4     & Uncertainty on the rest frame equivalent width of \CIII\ emission complex in \AA \\
46   & FWHM\_MGII                  &      F9.4       & FWHM of \MgII\ emission line in ${\rm km \ s^{-1}}$ \\
47   & BHWHM\_MGII               &       F9.4       & Blue HWHM of \MgII\ emission line in ${\rm km \ s^{-1}}$ \\
48   & RHWHM\_MGII               &       F9.4       & Red HWHM of \MgII\ emission line in ${\rm km \ s^{-1}}$ \\
49   & AMP\_MGII                     &      F9.4       & Amplitude of \MgII\ emission line in units of median rms pixel noise \\
50   & REWE\_MGII                  &      F9.4       & Rest frame equivalent width of \MgII\ emission line in \AA \\
51  & ERR\_REWE\_MGII         &  F9.4     & Uncertainty on the rest frame equivalent width of \MgII\ emission in \AA \\
\hline
52   & BAL\_FLAG\_VI                  &       I2       & BAL flag from visual inspection \\
53   & BI\_CIV                              &      F9.4       & Balnicity index of \CIV\ trough in ${\rm km \ s^{-1}}$ \\
54   & ERR\_BI\_CIV                    &      F9.4       & Error on the Balnicity index of \CIV\ trough in ${\rm km \ s^{-1}}$ \\
55   & AI\_CIV                             &      F9.4       & Absorption index of \CIV\ trough in ${\rm km \ s^{-1}}$ \\
56   & ERR\_AI\_CIV                   &      F9.4       & Error on the absorption index of \CIV\ trough in ${\rm km \ s^{-1}}$ \\
57   & DI\_CIV                             &      F9.4       & Detection index of \CIV\ trough in ${\rm km \ s^{-1}}$ \\
58   & ERR\_DI\_CIV                   &     F9.4       & Error on the detection index of \CIV\ trough in ${\rm km \ s^{-1}}$ \\
59   & CHI2THROUGH                 &     F9.4        & $\chi^2$ of the trough from \Eq{chi2trough}\\
60   & NCIV\_2000                       &   I3         & Number of distinct \CIV\ troughs of width larger than 2,000~km~s$^{-1}$ \\
61   & VMIN\_CIV\_2000              &    F9.4       & Minimum velocity of the \CIV\ troughs defined in row 60 ${\rm km \ s^{-1}}$ \\
62   & VMAX\_CIV\_2000             &     F9.4       & Maximum velocity of the \CIV\ troughs defined in row 60  in ${\rm km \ s^{-1}}$ \\
63   & NCIV\_450                         &   I3         & Number of distinct \CIV\ troughs of width larger than 450~km~s$^{-1}$ \\
64   & VMIN\_CIV\_450               &    F9.4       & Minimum velocity of the \CIV\ troughs defined in row 63 in ${\rm km \ s^{-1}}$ \\
65   & VMAX\_CIV\_450              &     F9.4       & Maximum velocity of the \CIV\ troughs defined in row 63 in ${\rm km \ s^{-1}}$ \\
66   & REW\_SIIV                       &  F9.4           & rest frame equivalent width of the \SiIV\ trough \\
67   & REW\_CIV                         &  F9.4          & rest frame equivalent width of the \CIV\ trough  \\
68   & REW\_ALIII                      &  F9.4           & rest frame equivalent width of the Al~{\sc iii} trough \\
\hline
%
%
69   & RUN\_NUMBER                     & I6       & SDSS Imaging Run Number of photometric measurements \\
70   & PHOTO\_MJD                       & I6      & Modified Julian Date of imaging observation \\
71   & RERUN\_NUMBER                 & A4     & SDSS Photometric Processing Rerun Number \\
72   & COL\_NUMBER                      & I2      & SDSS Camera Column Number (1-6) \\
73  & FIELD\_NUMBER                   & I5      & SDSS Field Number \\
74   & OBJ\_ID                                & A20      & SDSS Object Identification Number \\
%
%
75   & UFLUX                              &      F9.4       & flux in the $u$-band (not corrected for Galactic extinction)\\
76   & ERR\_UFLUX                    &       F9.4      & Error in $u$ flux \\
77   & GFLUX                              &      F9.4       & flux in the $g$-band (not corrected for Galactic extinction) \\
78   & ERR\_GFLUX                    &      F9.4       & Error in $g$ flux \\
79   & RFLUX                              &      F9.4       & flux in the $r$-band (not corrected for Galactic extinction) \\
80   & ERR\_RFLUX                    &      F9.4       & Error in $r$ flux \\
81   & IFLUX                               &      F9.4       & flux in the $i$-band (not corrected for Galactic extinction) \\
82   & ERR\_IFLUX                     &      F9.4       & Error in $i$ flux \\
83   & ZFLUX                              &      F9.4       & flux in the $z$-band (not corrected for Galactic extinction) \\
84   & ERR\_ZFLUX                    &      F9.4       & Error in $z$ flux \\
85   & TARGET\_UFLUX             &       F9.4       & TARGET flux in the $u$-band (not corrected for galactic extinction) \\
86   & TARGET\_GFLUX             &       F9.4       & TARGET flux in the $g$-band (not corrected for galactic extinction) \\
87   & TARGET\_RFLUX             &       F9.4       & TARGET flux in the $r$-band (not corrected for galactic extinction) \\
88   & TARGET\_IFLUX              &       F9.4       & TARGET flux in the $i$-band (not corrected for galactic extinction) \\
89   & TARGET\_ZFLUX             &        F9.4       & TARGET flux in the $z$-band (not corrected for galactic extinction) \\
90   & U\_EXT                           &       F9.4       & $u$ band Galactic extinction (from \citep{schlegel1998}) \\
91   & HI\_GAL                          &       F9.4       & ${\rm log} N_{\rm H}$ (logarithm of Galactic \HI\ column density in ${\rm cm^{-2}}$)\\
\hline
%
%
92   & RASS\_COUNTS             &       F9.4       & log RASS full band count rate (counts s$^{-1}$)\\
93   & RASS\_COUNTS\_SNR   &       F9.4       & SNR of the RASS count rate \\
94   & SDSS2ROSAT\_SEP        &       F9.4       & SDSS-RASS separation in arcsec \\
95   & NUVFLUX                       &       F9.4       & $nuv$ flux (GALEX) \\
96   & ERR\_NUVFLUX             &       F9.4       & Error in $nuv$ flux \\
97   & FUVFLUX                       &       F9.4       & $fuv$ flux (GALEX) \\
98   & ERR\_FUVFLUX              &       F9.4       & Error in $fuv$ flux \\
%
99   & JMAG                             &      F9.4       & $J$ magnitude (Vega, 2MASS) \\
100   & ERR\_JMAG                   &      F9.4       & Error in $J$ magnitude \\
101   & JSNR                              &     F9.4        & J-band SNR \\
102   & JRDFLAG                       &     I2      & J-band photometry flag\\ 
103 & HMAG                            &   F9.4  & $H$ magnitude (Vega, 2MASS) \\
104  & ERR\_HMAG                   &      F9.4       & Error in $H$ magnitude \\
105  & HSNR                             &     F9.4        & H-band SNR \\
106  & HRDFLAG                       &     I2      & H-band photometry flag\\ 
107  & KMAG                            &  F10.6    & $K$ magnitude (Vega, 2MASS) \\
108  & ERR\_KMAG                  &      F10.6        & Error in $K$ magnitude \\
109  & KSNR                              &     F10.6        & K-band SNR \\
110  & KRDFLAG                       &      I2      & K-band photometry flag\\ 
111  & SDSS2MASS\_SEP         &   F10.6   & SDSS-2MASS separation in arcsec \\
112  & W1MAG                         &       F10.6       & $w1$ magnitude (Vega, WISE) \\
113  & ERR\_W1MAG               &       F10.6       & Error in $w1$ magnitude \\
114  & W1SNR                         &    F10.6          & SNR in w1 band    \\
115  & W1CHI2                         &   F10.6           & $\chi^2$ in w1 band \\
116  & W2MAG                        &       F10.6       & $w2$ magnitude (Vega, WISE) \\
117  & ERR\_W2MAG              &       F10.6       & Error in $w2$ magnitude \\
118  & W2SNR                         &    F10.6          & SNR in w1 band    \\
119  & W2CHI2                         &   F10.6           & $\chi^2$ in w1 band \\
120  & W3MAG                        &       F10.6       & $w3$ magnitude (Vega, WISE) \\
121  & ERR\_W3MAG              &       F10.6       & Error in $w3$ magnitude \\
122  & W3SNR                         &    F10.6          & SNR in w1 band    \\
123  & W3CHI2                         &   F10.6           & $\chi^2$ in w1 band \\
124  & W4MAG                        &       F10.6       & $w4$ magnitude  (Vega, WISE) \\
125  & ERR\_W4MAG               &       F10.6       & Error in $w4$ magnitude \\
126  & W4SNR                         &    F10.6          & SNR in w1 band    \\
127  & W4CHI2                         &   F10.6           & $\chi^2$ in w1 band \\
128  & SDSS2WISE\_SEP             &       F10.6       & SDSS-WISE separation in arcsec \\
129  & FIRST\_FLUX                     &      F10.6       & FIRST peak flux density at 20 cm expressed in mJy \\
130  & FIRST\_SNR                      &      F10.6       & SNR of the FIRST flux density \\
131  & SDSS2FIRST\_SEP              &       F10.6       & SDSS-FIRST separation in arcsec \\
\hline
\multicolumn{4}{l}{$^a$ All magnitudes are PSF magnitudes}
\end{longtable}
}

\end{document}